\PassOptionsToPackage{x11names, table}{xcolor}
\documentclass[11pt]{article}
\pdfoutput=1
\usepackage{dcolumn}
\usepackage{bm}

\usepackage[square,numbers,compress,sort]{natbib}
\usepackage{graphicx}
\usepackage{amssymb,amsmath}
\usepackage{multirow}
\usepackage{slashed,xfrac}
\usepackage[makeroom]{cancel}
\usepackage{rotating}
\usepackage{color,url}
\usepackage[colorlinks=true
,urlcolor=blue
,anchorcolor=blue
,citecolor=blue
,filecolor=blue
,linkcolor=blue
,menucolor=blue
,linktocpage=true
,pdfproducer=medialab
,pdfa=true
]{hyperref}

\usepackage{epsfig,psfrag,rotating,soul}
\usepackage{rotfloat}
\usepackage[normalem]{ulem}
\usepackage{framed,fancybox,xcolor,xfrac,geometry}
\usepackage[x11names,table]{xcolor}
\usepackage{hhline}

\renewcommand\eqref[1]{Eq.~(\ref{#1})}
\newcommand\eqrefs[1]{Eqs.~(\ref{#1})}
\newcommand\figref[1]{Fig.~\ref{#1}}

\newcommand\tabref[1]{Table~\ref{#1}}

\graphicspath{{figs/}}

\evensidemargin \oddsidemargin
\allowdisplaybreaks

\def\significance{\sigma^{\rm stat}}
\def\cosw{c_{\rm w}}
\def\sinw{s_{\rm w}}

\def\euno{\varepsilon_1}
\def\edos{\varepsilon_2}
\def\etres{\varepsilon_3^*}
\def\ecuatro{\varepsilon_4^*}
\newcommand{\nn}{\nonumber}

\newcommand{\be}{\begin{equation}}
\newcommand{\ee}{\end{equation}}
\newcommand{\bear}{\begin{eqnarray}}
\newcommand{\eear}{\end{eqnarray}}
\newcommand{\mL}{\mathcal{L}}
\newcommand{\mO}{\mathcal{O}}
\newcommand{\mX}{\mathcal{X}}
\newcommand{\Frac}[2]{\frac{\displaystyle #1}{\displaystyle #2}}

\def\thefootnote{\fnsymbol{footnote}}

\begin{document}
\thispagestyle{empty}

\begin{flushright}
IFT-UAM/CSIC-17-048  \\
FTUAM-17-09\\
\end{flushright}
~
~
\vspace{0.5cm}

\begin{center}

\begin{Large}
\textbf{\textsc{Production of vector resonances at the LHC via $\boldsymbol{WZ}$-scattering:\\ a unitarized EChL analysis}}
\end{Large}

\vspace{1cm}

{\sc
R. L. Delgado$^1$%
\footnote{\tt \href{mailto:rafael.delgado.lopez@ucm.es}{rafael.delgado.lopez@ucm.es}}%
, A. Dobado$^1$%
\footnote{\tt \href{mailto:dobado@fis.ucm.es}{dobado@fis.ucm.es}}%
, D. Espriu$^2$%
\footnote{\tt \href{mailto:espriu@icc.ub.edu}{espriu@icc.ub.edu}}%
, C. Garcia-Garcia$^3$%
\footnote{\tt \href{mailto:claudia.garcia@uam.es}{claudia.garcia@uam.es}}%
, M.J. Herrero$^3$%
\footnote{\tt \href{mailto:maria.herrero@uam.es}{maria.herrero@uam.es}}%
, X.~Marcano$^3$%
\footnote{\tt \href{mailto:xabier.marcano@uam.es}{xabier.marcano@uam.es}}%
, J.J. Sanz-Cillero$^1$%
\footnote{\tt \href{mailto:jusanz02@ucm.es}{jjsanzcillero@ucm.es}}%
}

\vspace*{.7cm}

{\sl

$^1$Departamento de F\'isica Te\'orica I, Universidad Complutense de Madrid, \\
Plaza de las Ciencias 1, 28040 Madrid, Spain

\vspace*{0.1cm}

$^2$Departament de ¬•F\'\i sica Qu\`antica i Astrof\'\i sica and Institut de Ci\`encies del Cosmos (ICCUB),
Universitat de Barcelona, Mart\'i i Franqu\`es 1, 08028 Barcelona, Catalonia, Spain

\vspace*{0.1cm}

$^3$Departamento de F\'{\i}sica Te\'orica and Instituto de F\'{\i}sica Te\'orica, IFT-UAM/CSIC,\\
Universidad Aut\'onoma de Madrid, Cantoblanco, 28049 Madrid, Spain

}

\end{center}

\vspace*{0.1cm}

\begin{abstract}
In the present work we study the production of vector resonances at the LHC by means of the vector  boson scattering $WZ \to WZ$ and explore the sensitivities to these resonances for the expected future LHC luminosities. We are assuming that these vector resonances are generated dynamically from the self interactions of the longitudinal gauge bosons, $W_L$ and $Z_L$,  and work under the framework of the electroweak chiral Lagrangian to describe in a model independent way the supposedly strong dynamics of these modes. The properties of the vector resonances, mass, width and couplings to the $W$ and $Z$ gauge bosons are derived from the inverse amplitude method approach.  We implement all these features into a single model, the IAM-MC, adapted for MonteCarlo, built in a Lagrangian language in terms of  the electroweak chiral Lagrangian and a chiral Lagrangian for the vector resonances,  which mimics the resonant behavior of the IAM and provides unitary amplitudes. The model has been implemented in MadGraph, allowing us to perform a realistic study of the signal versus background events at the LHC. In particular, we have focused our study on the  $pp\to WZjj$ type of events, discussing first on the potential of the hadronic and semileptonic channels of the final $WZ$, and next exploring in more detail the clearest signals. These are provided by the leptonic decays of the gauge bosons, leading to a final state with $\ell_1^+\ell_1^-\ell_2^+\nu jj$, $\ell=e,\mu$, having a very distinctive signature,  and showing clearly the emergence of the resonances with masses in the range of 1.5-2.5 TeV, which we have explored.  
\noindent
 \end{abstract}

\def\thefootnote{\arabic{footnote}}
\setcounter{page}{0}
\setcounter{footnote}{0}

\newpage

\section{Introduction}
\label{intro}
One of the most likely indications of the existence of physics beyond the standard model (SM)
could be the appearance of resonances in the scattering of longitudinally polarized $W$ and
$Z$ electroweak (EW) gauge bosons. This would be a formidable hint of the existence of new interactions
involving the electroweak symmetry breaking sector (EWSBS) of the SM. This possibility is indeed
contemplated in all composite Higgs scenarios, characterized by the existence of a scale $f \gg v = 246$ GeV where some new strong interactions trigger the dynamical breaking of a global symmetry group $G$ to a certain subgroup $H$. 
The Goldstone bosons that appear provide the longitudinal degrees of freedom of the weak gauge bosons, while the Higgs boson would be one of the leftover Goldstone bosons. 
A non-zero mass for the latter is often provided by electroweak radiative corrections, e.g., via some misalignment mechanism between the gauge group and the global unbroken subgroup~\cite{Agashe:2004rs}.

In the present work we will not assume any specific model for the strong dynamics underlying the
EWSBS nor for the above mentioned misalignment mechanism.
 Instead, we will work under the generic and minimal assumptions for the above global groups and the spontaneous symmetry breaking pattern given by $SU(2)_L \times SU(2)_R \to SU(2)_{L+R}$. 
 This involves the minimal set of Goldstone bosons that are needed to generate the EW gauge boson masses, $m_W$ and $m_Z$,  and also preserves the wanted custodial symmetry $SU(2)_C=SU(2)_{L+R}$. 
 This symmetry protects the SM tree level relation $m_W=\cos\theta_W m_Z$ from potentially dangerous strong dynamics corrections, keeping the values of the $m_{W,Z}$ masses close to each other.
Under these generic assumptions, the most convenient approach to study in a model independent way the phenomenology of the strongly interacting EWSBS is provided by the electroweak chiral Lagrangian that is based on the above EW chiral symmetry breaking pattern and has the same EW gauge symmetries as the SM.
 The use of these effective chiral Lagrangians in the context of the electroweak theory was initiated long ago in the eighties \cite{Appelquist:1980vg,Longhitano:1980iz,Longhitano:1980tm,Chanowitz:1985hj,Cheyette:1987jf,Dobado:1989ax,Dobado:1989ue} by following the guiding lines of the well established chiral perturbation theory (ChPT) of low energy
 QCD~\cite{Weinberg:1978kz,Gasser:1984gg, Gasser:1983yg}.
It was used in the early nineties for LEP phenomenology \cite{Dobado:1990zh,Espriu:1991vm}, and for LHC prospects~\cite{Dobado:1990jy,Dobado:1990am,Dobado:1995qy,Dobado:1999xb}, and it has received an important push and upgrade in the last years, mainly after the discovery of the Higgs particle. All this lead to the building of the EW chiral Lagrangian with a light
Higgs (EChL)~\cite{Alonso:2012px,Buchalla:2013rka,Espriu:2012ih, Delgado:2013loa, Delgado:2013hxa, Brivio:2013pma, Espriu:2013fia, Espriu:2014jya,  Delgado:2014jda, Buchalla:2015qju, Arnan:2015csa}. A great effort has also been done in exploring the main implications of the EChL for LHC phenomenology (see, for instance, \cite{deFlorian:2016spz} for a recent summary),
although  no strongly interacting signal from the EWSBS has been seen yet at the LHC. The absence of these signals  at present and past colliders is translated, within the EChL framework, into experimental bounds on the size of the a priori unknown chiral parameters of the EChL \cite{Falkowski:2013dza, Brivio:2013pma, Khachatryan:2014jba, Aad:2014zda, ATLAS:2014yka, Fabbrichesi:2015hsa, Buchalla:2015qju,Aaboud:2016uuk}.

One of the most characteristic features of strong dynamics is undoubtedly the appearance of resonances in the spectrum, thus one should also expect new resonances if the EWSBS is strongly interacting. The use of the EChL for the study of this strong dynamics suggests that the scale associated to these resonances is related to the parameter with dimension of energy controlling the perturbative expansion within this chiral effective field theory, given typically, in the minimal scenario that we work with, by $4 \pi v$.  Therefore, one expects resonances to appear with masses typically of a few TeV,  clearly in the range covered at the LHC.  The theoretical framework  for the description of such resonances is,  however, not universal and one has to rely on a particular (author dependent) approach. Once one chooses, as we do, the approach provided by the EChL, there are basically two main paths to proceed.  Either the resonances are introduced explicitly at the Lagrangian level and the new terms added to the EChL are required to share the same symmetries of this latter, in particular the EW chiral symmetry, or they are not explicitly included but they are instead dynamically generated from the EChL itself. The first approach has been followed in several
works~\cite{Pich:2012jv,Pich:2012dv,Pich:2013fea,Pich:2015kwa,Pich:2016lew} essentially along the lines of previous works within the context of low energy QCD \cite{Ecker:1989yg}.
This type of chiral resonances have also been studied at the LHC  \cite{Alboteanu:2008my}.
The second approach has been followed in a number of works that use
the inverse amplitude method (IAM) to impose the unitarity of the amplitudes predicted with the EChL~\cite{Espriu:2012ih, Espriu:2013fia,Delgado:2013loa, Delgado:2013hxa, Delgado:2014dxa, Espriu:2014jya, Arnan:2015csa, Dobado:2015hha, Corbett:2015lfa,BuarqueFranzosi:2017prc}.
Within this approach,  the self-interactions of the longitudinal  EW bosons, which are assumed to be strong, are the responsible of the dynamical generation of the resonances,  and these are expected to show up in the scattering of the longitudinal modes, $W_L$ and $Z_L$, essentially as it happens in the context of ChPT where the QCD resonances emerge in the scattering of pions \cite{Truong:1988zp, Dobado:1989qm, Dobado:1992ha, Hannah:1995si}. The IAM was indeed used long ago in the context of the strongly interacting EWSBS framework but without the Higgs particle, and the production of these IAM resonances at the LHC was also addressed \cite{Dobado:1989gr, Dobado:1990jy, Dobado:1990am}. The advantage of this second approach is that it provides unitary amplitudes, which are absolutely needed for a realistic analysis at the LHC, and it predicts the properties of the resonances, masses, widths and couplings, in terms of the chiral parameters of the EChL. The disadvantage of this method is that it does not deal with full amplitudes but with partial waves, which are not very convenient for a MonteCarlo analysis at the LHC.

The present work addresses the question of whether these IAM dynamically generated resonances of the EWSBS could be visible at the LHC by means of the study of the EW vector boson scattering (VBS). 
These VBS processes are the most relevant channels to explore at the LHC if the longitudinal gauge modes are really strongly interacting, since they involve the four point self-interactions of the EW gauge bosons. Moreover, the resonances should emerge more clearly in VBS processes as they are generated from this strong dynamics. Our study aims to quantify the visibility of these resonances and also to determine the integrated luminosities that would be required to this end.
 More concretely, our purpose here is to estimate the event rates at the LHC of the production of a $SU(2)_{L+R}$  triplet
vector resonance, $V$,  via $WZ\to WZ$ scattering, and the subsequent decays of the
final $W$ and $Z$. We have selected this particular subprocess because it has several appealing features
in comparison with other VBS channels.  In the presence of such dynamical
vector resonances, these  emerge/resonate (in particular, the charged $V^{\pm }$ones)
in the s-channel of $W Z  \to W Z$, whereas in other subprocesses like
$W^+W^+ \to W^+ W^+$, $W^+W^- \to ZZ$, $ZZ \to W^+W^-$ and $ZZ \to ZZ$ do not. Other interesting
cases like $W^+W^- \to W^+ W^-$ where the neutral resonance, $V^0$, could similarly emerge in
the s-channel have, however, severe backgrounds.  For this reason it is known to be very difficult to
disentangle the signal from the SM irreducible background at the LHC. In particular, the SM one-loop
gluon initiated subprocess, $g g \to W^+W^-$, turns out to be a very important background in this case
due to the huge gluon density in the proton at the LHC energies. Our selected process $WZ\to WZ$, in contrast,
does not suffer from this background, and therefore it provides one of the cleanest windows
to look for these vector resonances at the LHC.

Consequently, our theoretical framework will be: 1) the effective electroweak chiral theory
with a light Higgs boson in terms of the `chiral'
effective couplings, $a_{1,2,3,4,5}$, and $a$ and $b$ effective Higgs boson couplings (custodial
symmetry of the underlying strong dynamics will be assumed); 2) the unitarization
of $W_L Z_L \to W_L Z_L$ via the IAM, following the works~\cite{Delgado:2013loa, Delgado:2013hxa, Dobado:2015hha, Delgado:2014dxa,Espriu:2012ih, Espriu:2013fia, Espriu:2014jya, Arnan:2015csa} 
 and  making sure that the predictions at the LHC comply with the obvious requirement
of unitarity; 3) we work with EW gauge bosons in the external legs of the VBS amplitudes and not with Goldstone bosons. This means that we go beyond the
simpler predictions provided by the equivalence theorem (ET)~\cite{Cornwall:1974km, Vayonakis:1976vz, Lee:1977eg, Gounaris:1986cr}, and this will allow us to make
realistic predictions for massive $W$ and $Z$ gauge bosons production and
their decays
at the LHC; 4) out of the EChL we shall construct and effective Lagrangian including vector resonances, based on the Proca 4-vector 
formalism~\cite{Pich:2012jv,Pich:2012dv,Pich:2013fea,Pich:2015kwa,Pich:2016lew}, in order to introduce in a Lagrangian language the resonances that are dynamically generated by the IAM. This effective Lagrangian includes the proper resonance
couplings to the $W$ and $Z$ and have
the symmetries of the EChL, in particular the EW Chiral symmetry. With this Lagrangian we will mimic the resonant behavior of the IAM amplitudes, having the resonance masses and widths as predicted by the IAM. Indeed, we will make use of this vector Lagrangian to extract
the Lorentz structure of the $WZ$ scattering vertex to be coded in the MonteCarlo. 
The coupling itself will turn
out to be a momentum-dependent function that will be derived from the IAM
unitarization process
in the $IJ=11$ channel. This IAM-MC model presented here is proper for a MonteCarlo analysis and it is included in MadGraph5 \cite{Alwall:2014hca}
for this work. The corresponding UFO file for the present IAM-MC model
can be provided on demand.
We would like to emphasize that our IAM-MC model provides full $A(WZ\to WZ)$ amplitudes with massive external EW gauge bosons.
The corresponding cross section $\sigma(WZ\to WZ)$ is computed from these full amplitudes and not from the first partial waves that do not provide a sufficiently accurate result, as we have checked.

Finally, a careful study of the signal versus backgrounds for the full process
$pp \to WZ jj$, leading to events with two jets plus one $W^+$ and one $Z$ will be performed. We will first discuss on the potential of the hadronic and semileptonic channels of the final $WZ$. Then we will explore the cleanest channels leading to events with two jets and  
the three leptons and missing energy which come from
the leptonic decays of the final $W^+$ and $Z$. For that study we will  employ the well established
VBS selection cuts \cite{Haywood:1999qg, Doroba:2012pd, Szleper:2014xxa, Aad:2016ett} and some specific optimal cuts on the final particles, which will
eventually allow us to extract the emergent vector resonances from the SM background in
this kind of $\ell {\bar \ell} \ell \nu j j $ events at the LHC.  

The paper is organized as follows. In section~\ref{theEChL} we summarize the main features
of the EChL. In section~\ref{scenarios} we present the predictions for the 
$W Z \to W Z$ scattering process
within this EChL framework, we unitarize the corresponding amplitudes with the IAM, and we select specific EChL scenarios with emergent vector resonances in this $WZ$ scattering process. Section~\ref{sec-model} is devoted to the presentation of our IAM-MC model
and the description of how we deal with IAM vector resonances in $WZ$ scattering within a MonteCarlo framework.
In section~\ref{LHC} we present our numerical results for the production and sensitivity
to vector resonances in  $pp\to WZjj$ events at LHC. A dicussion on the extrapolated rates for the hadronic and semileptonic channels is also included. The leptonic channels leading to $\ell {\bar \ell} \ell \nu j j $ events are also explored  in this section. A comparative study of the signal and background events is included.
The final section summarizes our main conclusions.
The final appendices collect some of our analytical results and Feynman rules for the VBS amplitudes.

\section{The Effective Electroweak Chiral Lagrangian}
\label{theEChL}
Given that the possible physics existing beyond the minimal SM is model
dependent, even after restricting ourselves to the realm of strongly EWSBS, it is necessary
to employ a technology that is as model independent as possible. The appropriate tool
to do so is provided by the effective EChL. In this theory
the information about the underlying microscopic theory is encoded in a number of so-called low-energy constants, i.e., coefficients of local operators.

The EChL is a gauged non-linear effective field theory (EFT) coupled to a singlet scalar particle
that contains as dynamical fields the EW gauge bosons, $W^{\pm}$, $Z$ and $\gamma$,
the corresponding would-be Goldstone-bosons, $w^{\pm}$, $z$, and the Higgs scalar
boson, $H$. We will not discuss the fermion sector in this article.
The $w^{\pm}$, $z$ are described by a matrix field $U$ that takes values in
the $SU(2)_L \times SU(2)_R/SU(2)_{L+R}$
 coset, and transforms as $U \to g_L U g_R^\dagger$ under the
action of the global group  $SU(2)_L \times SU(2)_R$. We will assume here that
 the scalar sector of the EChL preserves the custodial symmetry, except
for the explicit breaking due to the gauging of the $U(1)_Y$ symmetry. We believe that this assumption is well justified, since experimental measurements involving the well known $\rho$ parameter, or the effective couplings that parametrize the interaction between the Higgs and the EW gauge bosons show no evidence of custodial breaking in the bosonic sector other than that induced from $g'\neq0$. 

The basic building blocks of the
$SU(2)_L\times U(1)_Y$ gauge invariant EChL are the following:
\begin{eqnarray}
U(w^\pm,z) &=& 1 + i w^a \tau^a/v + \mO(w^2)\;\in SU(2)_L \times SU(2)_R/SU(2)_{L+R}, \label{Umatrix}\\
{\cal F}(H)&=& 1+2a\frac{H}{v}+b \left(\frac{H}{v} \right)^2+\dots ,\label{polynomial}\\
D_\mu U &=& \partial_\mu U + i\hat{W}_\mu U - i U\hat{B}_\mu, \label{eq.cov-deriv} \\
\hat{W}_{\mu\nu} &=& \partial_\mu \hat{W}_\nu - \partial_\nu \hat{W}_\mu
 + i  [\hat{W}_\mu,\hat{W}_\nu ],\;\hat{B}_{\mu\nu} = \partial_\mu \hat{B}_\nu -\partial_\nu \hat{B}_\mu ,
\label{fieldstrength}\\
\hat{W}_\mu &=& g \vec{W}_\mu \vec{\tau}/2 ,\;\hat{B}_\mu = g'\, B_\mu \tau^3/2 ,\\
{\cal V}_\mu &=& (D_\mu U) U^\dagger .
\label{EWfields}
\end{eqnarray}

According to the usual counting rules, the $SU(2)_L \times U(1)_Y$ invariant terms
in the EChL are organized by means of their `chiral dimension', meaning that a term ${\cal L}_d$ with
`chiral dimension' $d$ will contribute to $\mO(p^d)$ in the corresponding power momentum expansion.
The chiral dimension of each term in the EChL can be found out by following the scaling
with $p$ of the various contributing basic functions. Derivatives and  masses are considered
as soft scales of the EFT and of the same order in the chiral counting, i.e. of ${\cal O}(p)$.
The gauge boson masses, $m_W$ and $m_Z$ are examples of these soft masses in the case of the EChL.
These are generated from the covariant derivative in
\eqref{eq.cov-deriv} once the $U$ field is expanded in terms of the $w^a$ fields as:
\be
D_\mu U = \Frac{i \partial_\mu \vec{w}\, \vec{\tau}  }{v} + i\,  \Frac{g v}{2}  \, \Frac{\vec{W}_\mu\, \vec{\tau}}{v} - i\,  \Frac{g' v}{2}  \, \Frac{B_\mu\,  \tau^3}{v} + \dots
\ee
where the dots represent terms with higher powers of $(w^a/v)$ and whose precise form will
depend on the particular parametrization of $U$.  Once the gauge fields are rotated to the
physical basis they get the usual gauge boson squared mass values at lowest order:
$m_W^2=g^2 v^2/4$ and $m_Z^2 =(g^2+g'^{2}) v^2 /4$.

In order to have a power counting consistent with
the loop expansion one needs all the terms in the covariant derivative above to be
of the same order. Thus, the proper   assignment is
 $\partial_\mu$, $(g v)$ and  $(g'v) \sim \mO(p)$ or, equivalently,
$\partial_\mu$, $m_W$,  $m_Z \sim \mO(p) $. In addition, we will also consider in this
work the Higgs boson mass $m_H$ as another soft mass in the EChL with a
similar chiral counting as $m_W$ and $m_Z$.  That implies, $m_H \sim
\mO(p)$, or equivalently $(\lambda v^2 ) \sim \mO(p^2)$, with $\lambda$ being the SM
Higgs self-coupling.

\begin{figure}[t!]
\begin{center}
\includegraphics[width=.8\textwidth]{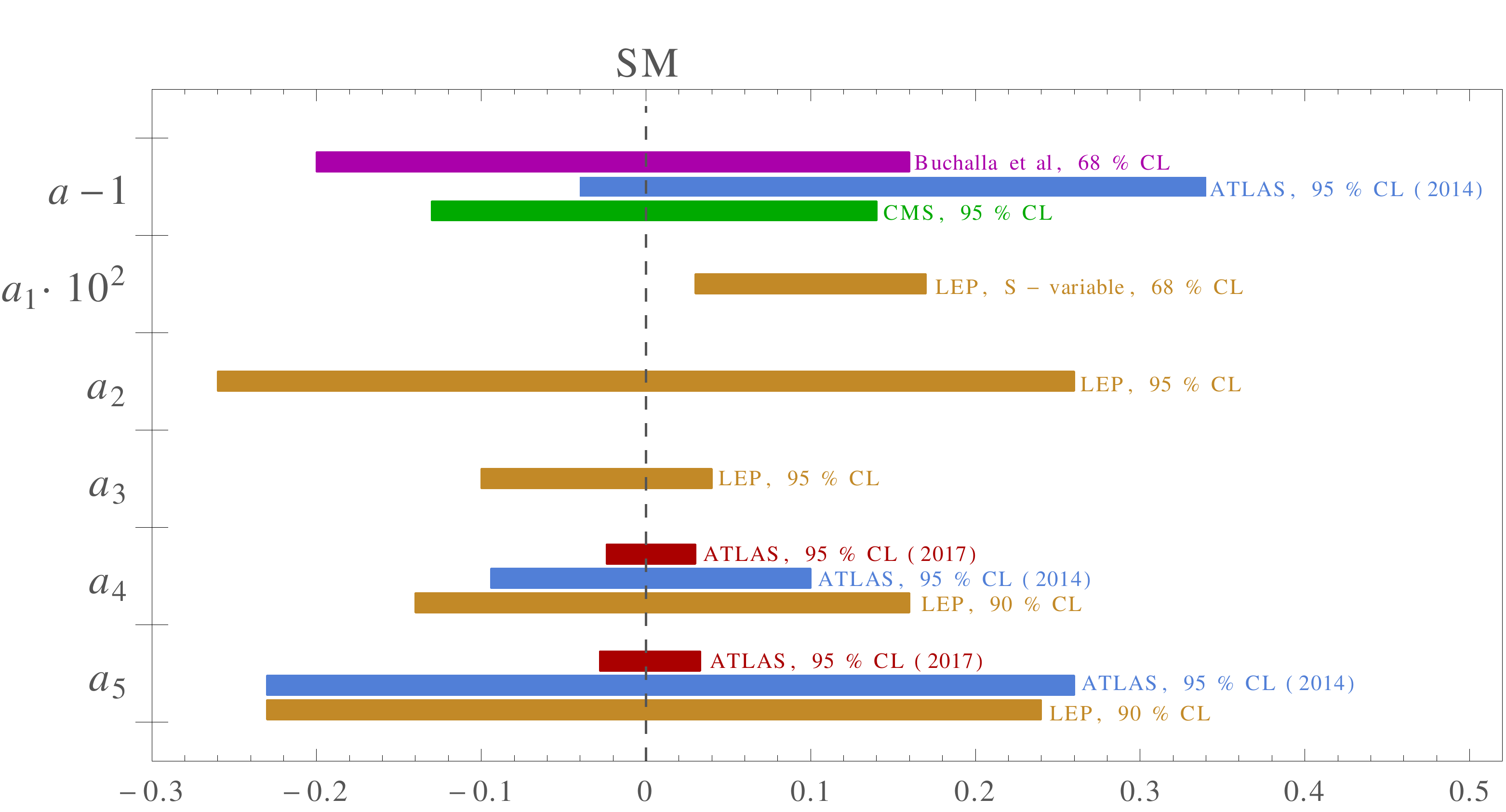}
\caption{Present experimental constraints on the EChL coefficients.
They are extracted from Refs.~\cite{Falkowski:2013dza, Brivio:2013pma, Khachatryan:2014jba, Aad:2014zda, ATLAS:2014yka, Fabbrichesi:2015hsa, Buchalla:2015qju,deFlorian:2016spz,Aaboud:2016uuk}.}
\label{fig:constraints}
\end{center}
\end{figure}

With these building blocks one then constructs the EChL up to a given order in the chiral
expansion. We require this Lagrangian to be $CP$ invariant, Lorentz invariant, $SU(2)_L \times U(1)_Y$ gauge invariant and custodial preserving. For the present work we include the terms with
chiral dimension up to  $\mO(p^4)$, therefore, the EChL can be generically written as:
\be
\mL_{\rm EChL} = \mL_2 + \mL_4 +\mL_{\rm GF} +\mL_{\rm FP}\, ,
\ee
where  $\mL_2$ refers to the terms with chiral dimension 2, i.e $\mO(p^2)$,  $\mL_4$ refers
to the terms with chiral dimension 4, i.e $\mO(p^4)$, and
$\mL_{\rm GF}$ and $\mL_{\rm FP}$ are the gauge-fixing (GF) and the corresponding
non-abelian Fadeev-Popov (FP) terms. The relevant terms for the description of EW gauge boson
scattering amplitudes  are\footnote{Our notation is taken from \cite{Herrero:1993nc,Herrero:1994iu}
and compares: 1) with~\cite{Longhitano:1980iz} as, $a_1=(g/g') \alpha_1$, $a_2=(g/g') \alpha_2$,
$a_3= -\alpha_3$, $a_4= \alpha_4$, $a_5= \alpha_5$; 2) with~\cite{Gasser:1983yg} as, $\ell_1=4 a_5$, $\ell_2= 4 a_4$, $\ell_5= a_1$, $\ell_6= 2(a_2 -a_3)$; and with~\cite{Gasser:1984gg} as, $L_1=a_5$, $L_2=a_4$, $L_9=a_3-a_2$, $L_{10}=a_1$.}:
\begin{align}
\mL_2 =&    -\Frac{1}{2 g^2} {\rm Tr}\Big(\hat{W}_{\mu\nu}\hat{W}^{\mu\nu}\Big) -\Frac{1}{2 g'^{2}}
{\rm Tr} \Big(\hat{B}_{\mu\nu} \hat{B}^{\mu\nu}\Big)\nn\\
& +\Frac{v^2}{4}\left[%
  1 + 2a \frac{H}{v} + b \frac{H^2}{v^2}\right] {\rm Tr} \Big(D^\mu U^\dagger D_\mu U \Big)
 + \Frac{1}{2} \partial^\mu H \, \partial_\mu H + \dots\, ,
\label{eq.L2}\\
{\mL}_{4} =& %
 ~ a_1 {\rm Tr}\Big( U \hat{B}_{\mu\nu} U^\dagger \hat{W}^{\mu\nu}\Big)
  + i a_2  {\rm Tr}\Big ( U \hat{B}_{\mu\nu} U^\dagger [{\cal V}^\mu, {\cal V}^\nu ]\Big)
  - i a_3 {\rm Tr}\Big (\hat{W}_{\mu\nu}[{\cal V}^\mu, {\cal V}^\nu]\Big) \nn \\
+&
  ~ a_4 \Big[{\rm Tr}({\cal V}_\mu {\cal V}_\nu) \Big]  \Big[{\rm Tr}({\cal V}^\mu {\cal V}^\nu)\Big] 
  + a_5  \Big[{\rm Tr}({\cal V}_\mu {\cal V}^\mu)\Big]  \Big[{\rm Tr}({\cal V}_\nu {\cal V}^\nu)\Big] 
\nn \\
  -&
  ~ c_{W} \frac{H}{v} {\rm Tr}\Big(\hat{W}_{\mu\nu} \hat{W}^{\mu\nu}\Big)
  - c_B \frac{H}{v}\, {\rm Tr} \Big(\hat{B}_{\mu\nu} \hat{B}^{\mu\nu} \Big)\, +\dots\label{eq.L4}
\end{align}
 Regarding the present experimental constraints on the previous EW chiral coefficients, we have
summarized in  \figref{fig:constraints} the most recent available
set from the literature~\cite{Falkowski:2013dza, Brivio:2013pma, Khachatryan:2014jba, Aad:2014zda, ATLAS:2014yka, Fabbrichesi:2015hsa, Buchalla:2015qju,deFlorian:2016spz,Aaboud:2016uuk}.  
  From the previous set of constraints we can see that the most constrained EW chiral couplings at present
are $a_1$, from its relation with the oblique $S$ parameter, and  $a_3$ where the most important
constraints come from its relation with the anomalous triple gauge couplings. Also $a_2$ is constrained, although more mildly,  by triple gauge couplings. On the other hand, the chiral couplings  $a_4$ and $a_5$ are constrained mainly by the studies of the anomalous quartic gauge couplings at the LHC and LEP~\cite{ Brivio:2013pma, Aad:2014zda, Fabbrichesi:2015hsa,Aaboud:2016uuk}.
In addition, $a$ is constrained to be close to the SM value ($a_{\rm SM}=1$)
up to $\mathcal{O}(10\%)$ deviations, the coefficient $b$ is unknown
so far, see however~\cite{Delgado:2014dxa}. Regarding $c_W$ and $c_B$, the best constraint comes from the related coefficient appearing in the photonic
$\frac{e^2}{16 \pi^2}c_{\gamma\gamma}\frac{H}{v}F_{\mu\nu}F^{\mu\nu}$ Lagrangian term. It has been
experimentally constrained to $ c_{\gamma\gamma}=-0.24\pm 0.37$~\cite{Buchalla:2015qju}.
A recent summary of constraints and some phenomenological issues of $\mL_{\rm EChL}$ for LHC physics
can be found in~\cite{deFlorian:2016spz}.

\section{Selection of scenarios with vector resonances in $WZ$ scattering} 
\label{scenarios}
In this section we present the specific EChL scenarios that will be explored in our forthcomming study at the LHC,  having dynamical vector resonances $V$ emerging in  $WZ$ scattering. First we show the results of the cross-sections for $WZ \to WZ$ from the EChL, which are compared with the SM predictions. Then we unitarize these EChL results, and finally, within these unitarized results, we select the scenarios with emergent vector resonances $V$.    

Even though all the EW chiral coefficients in the previously introduced EChL will
enter in the description of the subprocesses of our interest, i.e. the scattering of EW gauge bosons,
not all of them are equally relevant for all channels. As stated in the introduction, here we will be mostly
interested in studying the deviations with respect to the SM predictions for the specific scattering
process $W_L Z_L \to W_L Z_L$, since it provides one of the cleanest windows to look for charged vector
resonances at the LHC. On the other hand, we know by means of the ET~\cite{Cornwall:1974km, Vayonakis:1976vz, Lee:1977eg, Gounaris:1986cr}, which applies to renormalizable gauges and is valid also for the
EChL~\cite{Dobado:1993dg,Dobado:1994vr,Dobado:1997fv,He:1993qa}, that the
scattering amplitude for this subprocess $W_L Z_L \to W_L Z_L$ can be approximated, at large
energies compared to the gauge boson masses, by the scattering amplitude of the corresponding
would-be Goldstone bosons,
\begin{equation}
A(W_L Z_L \to W_L Z_L) \simeq A(wz \to wz)\,.
\label{ET}
\end{equation}
Since the relevant EW chiral coefficients in the amplitude $A(wz \to wz)$ (i.e., those that remain
even switching off the gauge interactions, $g=g'=0$),  are just $a$, $b$, $a_4$ and $a_5$, we conclude
that for our purpose of describing the most relevant departures  from the SM in $A(W_L Z_L \to W_L Z_L)$
it will be sufficient to work with just this subset of EChL parameters.

As we have said, in the present work we deal with massive gauge bosons in the external legs of the VBS amplitudes and not with their corresponding Goldstone bosons. 
The various contributing terms from the EChL to the EW gauge boson scattering amplitude of our interest  are the following:
\begin{eqnarray}
A(W_L Z_L \to W_L Z_L)^{{\rm EChL}} &=& A^{(0)}(W_L Z_L \to W_L Z_L)  + A^{(1)}(W_L Z_L \to W_L Z_L)\,,
\label{WLZL}
\end{eqnarray}
where the leading order (LO), ${\cal O}(p^2)$, and next to leading order contributions (NLO), ${\cal O}(p^4)$, are
denoted as  $A^{(0)}$ and $A^{(1)}$ respectively, and are given by:
 \begin{eqnarray}
 A^{(0)}(W_L Z_L \to W_L Z_L)&=&  A^{{\rm EChL}^{(2)}_{\rm tree}}\,,    \nn\\
 A^{(1)}(W_L Z_L \to W_L Z_L)&=&  A^{{\rm EChL}^{(4)}_{\rm tree}}+ A^{{\rm EChL}^{(2)}_{\rm loop}}\,.
\label{LOandNLO}
\end{eqnarray}

For completeness, we have also collected in the appendices the necessary Feynman rules, Feynman diagrams
and resulting scattering amplitudes, for the simplest case of a tree level computation,
i.e.,
\begin{equation}
A(W_L Z_L \to W_L Z_L)^{{\rm EChL}^{(2+4)}_{\rm tree}}=A^{{\rm EChL}^{(2)}_{\rm tree}}+ A^{{\rm EChL}^{(4)}_{\rm tree}}\,.
\label{EChL-tree}
\end{equation}
The analytical result is given in terms of the three EChL parameters, $a$, $a_4$ and $a_5$ involved,
and has been found with the help of FeynArts~\cite{Hahn:2000kx} and FormCalc~\cite{Hahn:1998yk}. We have also included in the appendices
the corresponding results for the SM amplitude at the tree level, to illustrate clearly the differences
with respect to the EChL results. It should be noticed that the $b$ parameter does not enter in $WZ$ scattering at the tree level, and 
it just enters in $A^{{\rm EChL}^{(2)}_{\rm loop}}$. It should also be noticed that, to our knowledge,
a full one-loop EChL computation is not available in the literature for this process,
 i.e., the full analytical result
of $A^{{\rm EChL}^{(2)}_{\rm loop}}$ is unknown. However, we will use an approximation
to estimate the size of this one-loop contribution,
following~\cite{Espriu:2012ih,Espriu:2013fia,Espriu:2014jya}. Concretely, the real part of the loop diagrams is computed 
using the ET (but keeping $m_H\neq 0$) and the imaginary part of the loops is   calculated exactly through the tree-level result by
making use of the optical theorem. In the following, we will refer to this NLO computation, ${\rm EChL}^{(2+4)}_{\rm loop}$,  as quasi exact one-loop EChL result.

\begin{figure}[t!]
\begin{center}
\includegraphics[width=.6\textwidth]{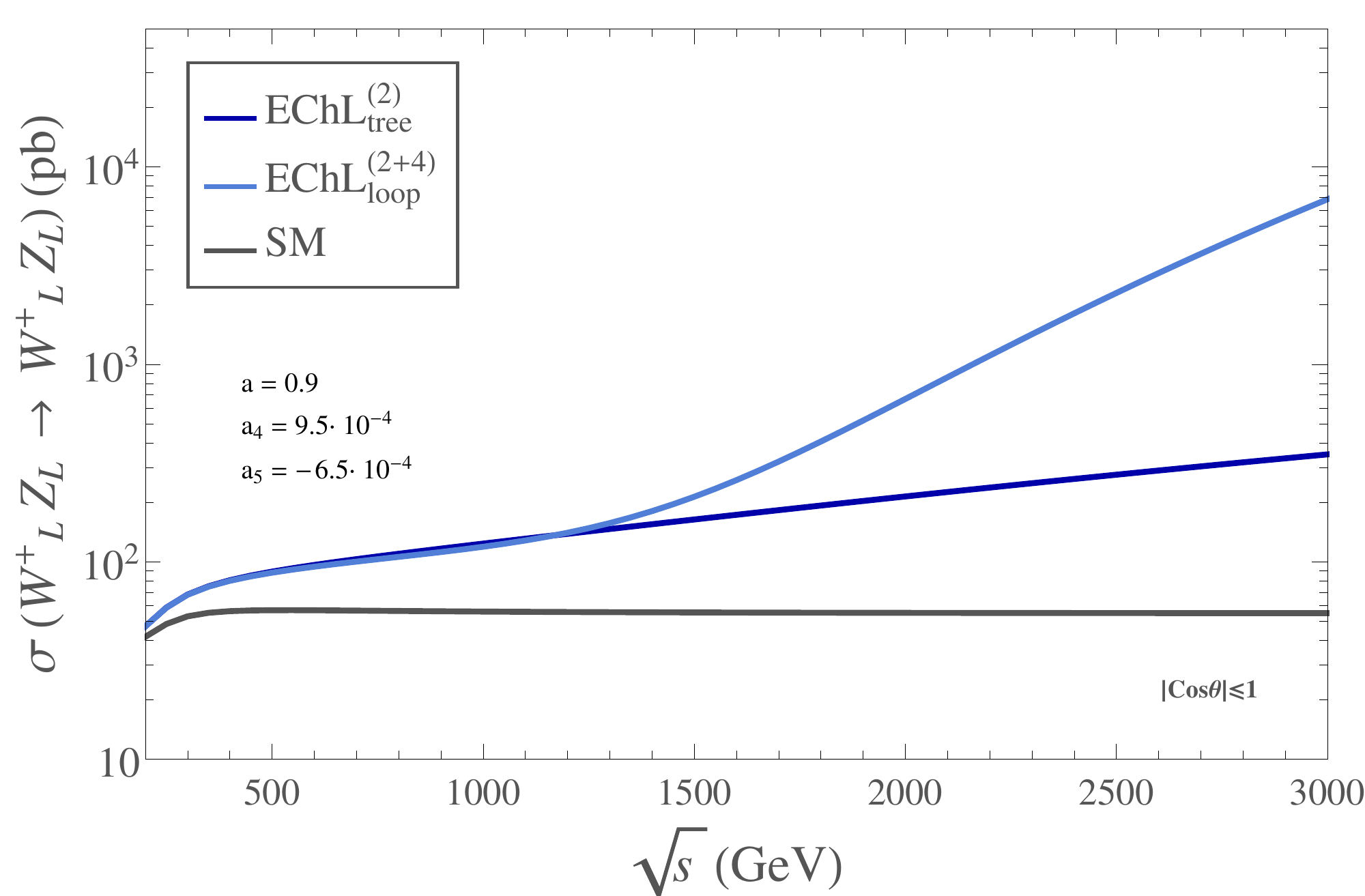}
\caption{Predictions of the cross section $\sigma(W_LZ_L \to W_LZ_L)$ as a function of the center of mass
energy $\sqrt{s}$ from the EChL. The predictions at leading order, ${\rm EChL}^{(2)}_{\rm tree}$,
and next to leading order, ${\rm EChL}^{(2+4)}_{\rm loop}$,
are displayed separately. The EChL coefficients
are set here to $a=0.9$, $b=a^2$, $a_4=9.5 \times 10^{-4}$ and $a_5=-6.5 \times 10^{-4}$. Here the integration is done in the whole $|\cos \theta| \leq 1$ interval of the centre of mass scattering angle $\theta$.
The prediction of the SM cross
section is also included, for comparison. All predictions have been obtained using FormCalc and our private Mathematica code and checked with MadGraph5.} 
\label{fig:xsecEChLLLLL}
\end{center}
\end{figure}

We have chosen one example to illustrate numerically and graphically
the energy behavior of the EChL cross section and the comparison with the SM prediction.
This is displayed in \figref{fig:xsecEChLLLLL}, where the chiral parameters have been
set to $a=0.9$, $b=a^2$, $a_4=9.5 \times 10^{-4}$ and $a_5=-6.5 \times 10^{-4}$.
As we can see in \figref{fig:xsecEChLLLLL} the predictions from the EChL grow with energy,
and they depart clearly from the SM prediction which for $|\cos\theta|\leq 1$  is nearly flat with energy in the explored 
interval of $\sqrt{s}  \in (500,3000)$~GeV. This growth is more pronounced as larger the values
of $|a_4|$ and/or $|a_5|$ are, and it leads to amplitudes that cross over the unitarity bound   at
some energy $\sqrt{s}$, whose particular value obviously depends on the assumed $(a,a_4,a_5)$ parameters.
We have checked that by using input $(a,a_4,a_5)$ parameters in the allowed region by the experimental
constraints in \figref{fig:constraints}, this crossing, which is defined in terms of the $IJ$ partial waves as $|a_{IJ}|=1$, may indeed occur at the TeV energies explored by the LHC,
even for as small values as
$|a_{4,5}|\sim 10^{-3}$. For instance, in the example of \figref{fig:xsecEChLLLLL} this crossing takes place first for the $|a_{00}|$ partial wave, and it happens at around 2 TeV. Larger values of $a_{4,5}$ would lead to the unitarity violation happening at even lower energies.

At this stage, it is also interesting to comment on the goodness of our assumption of neglecting other loop contributions in our computation of $WZ$ scattering. In particular, as we have said, we are ignoring in this work the contributions from fermions. Since the fermions would only contribute via loops to this $WZ$ scattering process, and since the dominant contributions would come from the third generation-quark loops, we have performed an estimate of the size of these loop contributions to be sure that they are indeed negligible. For this estimate we have assumed that all the fermion interactions are the same as in the SM and we have used the analytical results of  \cite{Dawson:1990cp} which are provided for the SM within the ET. Our numerical estimate of the heavy fermion loops indicates that for the high energies of our interest here, say between 1 and 3 TeV, the contributions from the top loops to $\sigma(w z \to wz)$ decrease with   $\sqrt{s}$, in contrast to the contributions from the EChL loops which increase with energy, and they are indeed very small, between $10^{-1}$\,pb and $10^{-2}$\,pb. These are more than three orders of magnitude below the prediction of $\sigma(W_LZ_L \to W_LZ_L)$ from the EChL (specifically, from our quasi exact prediction   ${\rm EChL}^{(2+4)}_{\rm loop}$ in \figref{fig:xsecEChLLLLL}). Therefore we conclude that our assumption in this work of ignoring the fermion loops is well justified.   
 
\begin{figure}[t!]
\begin{center}
\includegraphics[width=0.49\textwidth]{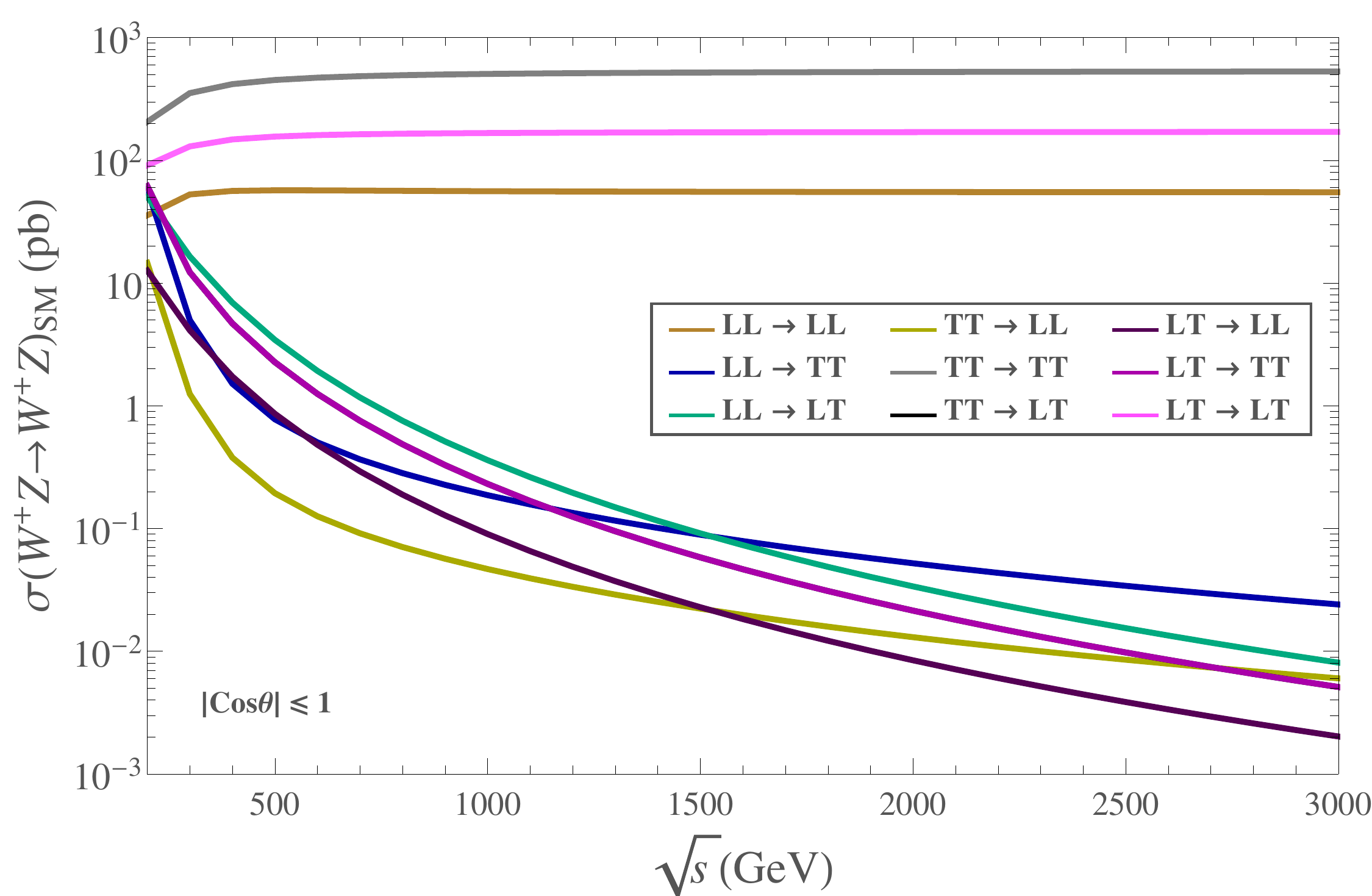}
\includegraphics[width=0.49\textwidth]{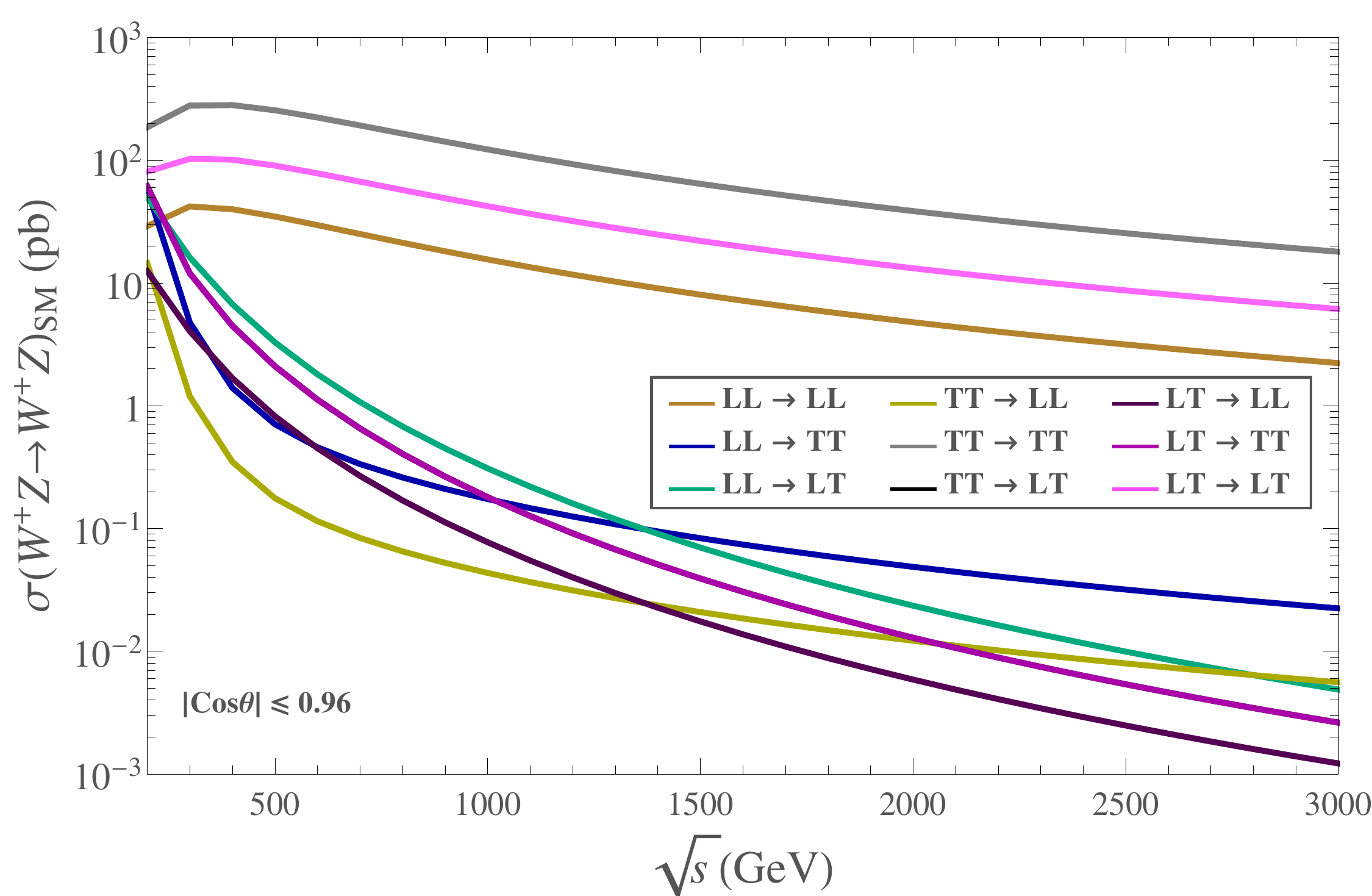}
\caption{Predictions of the SM cross section as a function of the center of mass energy,
$\sqrt{s}$, of the process $W Z \to W Z$ for different polarizations of the initial $W_AZ_B$ ($AB=LL,TT,LT$)
and final $W_CZ_D$ ($CD=LL,TT,LT$) bosons. 
We display the different polarization cross sections integrated in two choices of the center of mass scattering angle, $|\cos\theta|\leq 1$ (left panel) and $|\cos\theta|\leq 0.96$ (right panel), corresponding the latter to $|\eta_{W,Z}|<2$. All predictions have been obtained with FormCalc and checked with MadGraph5.}
\label{fig:xsWZWZSM}
\end{center}
\end{figure}

The above commented deviations of the EChL predictions with respect to the SM ones in the scattering
of longitudinally polarized gauge bosons, are by themselves an interesting result and suggest that they could lead
to signals above the SM background given by an enhancement in events with $W_LZ_L$ in the final state.
However, the polarization of the final gauge bosons is not expected
to be measured at the LHC, and therefore the realistic SM background will come from the full unpolarized SM
cross section. The relevance of the various polarization channels in the SM prediction is shown in
\figref{fig:xsWZWZSM}. We display the different polarization cross sections integrated in two choices of the center of mass scattering angle, $|\cos\theta|\leq 1$ and $|\cos\theta|\leq 0.96$. We have checked that we get the same results with FormCalc and MadGraph5.
It is clear that the channel $W_TZ_T \to W_TZ_T$ (gray lines in \figref{fig:xsWZWZSM}) is the dominant one, then go  $W_LZ_T \to W_LZ_T$ and  $W_TZ_L \to W_TZ_L$ (pink lines) which we denote together here and along this work as $LT \to LT$,  and next $W_LZ_L \to W_LZ_L$ (orange lines). For instance, in the energy interval
$\sqrt{s} \in (1000,3000)$~GeV, the size of  $\sigma(W_LZ_L \to W_LZ_L)$ is approximately one order of magnitude smaller than that
of the $\sigma(W_TZ_T \to W_TZ_T)$. Therefore, in order to extract clear signals at the LHC from departures in
the $W_LZ_L \to W_LZ_L$ channel we will have to produce cross-sections emerging above this irreducible SM background.
It is one of our main motivations here to consider dynamically generated resonances as leading emergent signals from
the EChL in $WZ \to WZ$ scattering, instead of considering just smooth enhancements over the SM background.

Finally, the previously mentioned violation of unitarity of the EChL scattering amplitudes leads to our
major concern in this work: the need of an unitarization method in order to provide realistic predictions at the
LHC. We choose here one of the most used unitarization methods for the partial waves, the IAM,  
which has the advantage over other methods of being able to
generate dynamically the vector resonances that we are interested in. In terms of fixed isospin $I$ and angular momentum $J$, and following a similar notation  
as in \eqref{WLZL}, for the LO $a_{IJ}^{(0)}$ and NLO $a_{IJ}^{(1)}$ contributions, the IAM partial waves are given by (for a review, see for instance Ref.~\cite{Pelaez:2015qba}): 
\begin{equation}
   a^{\rm IAM}_{IJ} = \frac{\big(a_{IJ}^{(0)}\big)^2}
 {a_{IJ}^{(0)}-a_{IJ}^{(1)}} \, .
\label{IAMdef}
\end{equation}

Other unitarization procedures such as N/D and the improved K matrix (IK) were also studied and compared with the IAM in the present context in detail in Ref.~\cite{Delgado:2015kxa}. In this reference the IAM, N/D and the IK unitarization methods are implemented in a particular way compatible with the electroweak chiral expansion.   All of these three methods turn out to be acceptable, since they produce partial waves which are: IR and UV finite, renormalization scale $\mu$ independent, elastically unitary, have the proper analytical structure (they feature a right and a left cut) and they reproduce the expected low energy results of the EChL up to the one-loop level. Thus the three methods can provide an UV completion of the low-energy chiral amplitudes. Moreover, for some region of the chiral couplings parameter space, they can have a pole in the second Riemann sheet with similar properties. These poles have a natural interpretation as dynamically generated resonances with the quantum numbers of the corresponding channel\footnote{The simplest and better known case, where this machinery is known to work very well,
is provided by $\pi\pi$ scattering. There, unitarization of the $IJ=11$ partial wave provides the position
and properties of the $\rho$ meson when the measured values of the low-energy chiral couplings in the chiral Lagrangian
are used. Note that these couplings are measured at energies well below $m_\rho$. Likewise determining
the corresponding anomalous coefficients in VBS at the LHC would give valuable information
on resonances to be found at higher values of $s$.}.  By comparison of the three methods for different values of the chiral couplings it is possible to realize that all of them normally produce the same qualitative results and, in many cases, the agreement is also quantitative up to high energies. This is particularly  true  for the $I=J=0$ channel. However, as it is explained in detail in
Ref.~\cite{Delgado:2015kxa}, the N/D and the IK methods cannot be applied to the $I=J=1$ channel considered in this work in the particular case of $b=a^2$, since it leads to contributions from the left and right cuts which cannot be separated in a $\mu$-invariant way, as required by these two methods. 
Therefore, in the following we will use only the IAM method.  
Contrary to the perturbative expansion of the EChL amplitudes, the IAM amplitudes
 fulfill all the analyticity and elastic unitarity requirements. In addition, $a^{\rm IAM}_{IJ}$ may or may not exhibit 
a pole as discussed above. If present, it can be interpreted as a dynamically generated resonance. In that case we use here the usual convention for the position of the pole
in terms of the mass, $M_R$,  and width, $\Gamma_R$, of the corresponding resonance $R$:
$s_{\rm pole}=(M_R-\frac{i}{2}\Gamma_R)^2$. 
Finally,  it is worth mentioning that the IAM is actually derived from the re-summation of bubbles in the s-channel and therefore accounts for re-scattering effects. The dynamical generation of resonances can be understood from the inclusion of this infinite chain of diagrams.
Concretely, in the present case of $WZ \to WZ$ scattering, such re-summation of infinite bubbles in the s-channel means in practice to consider the sequential chain of diagrams with $W$ and $Z$ in the internal bubbles, {\it i.e.}, $WZ \to WZ  \to \dots \to WZ \to WZ$. The charged vector resonance $V^\pm$ is then understood as emerging from this chain. 

The solution to the position of the pole in the case of $a^{\rm IAM}_{11}$ is very simple if the ET is used, and gives  simple predictions for the mass and the width of the dinamically generated vector resonances in terms of the EChL parameters, $a$, $b$, $a_4$ and $a_5$,  given by~\cite{Delgado:2013loa,Delgado:2013hxa}: 
\begin{align}
(M_V^2)_{\rm ET} &=\dfrac{1152 \pi^2v^2 (1-a^2)}{8(1-a^2)^2-75(a^2-b)^2+4608\pi^2(a_4(\mu)-2a_5(\mu))}\,,\\
(\Gamma_V)_{\rm ET}&=\dfrac{(1-a^2)}{96\pi v^2}M_V^3\,\left[1+\dfrac{(a^2-b)^2}{32\pi^2 v^2 (1-a^2)}M_V^2\right]^{-1},
\label{MVGVET}
\end{align}
with $a_4(\mu)$ and $a_5(\mu)$ the scale dependent parameters whose running  equations for arbitrary $a$ and $b$ can be found in \cite{Espriu:2012ih,Espriu:2013fia,Espriu:2014jya,Delgado:2013loa,Delgado:2013hxa}. These solutions apply to narrow resonances, i.e., for $\Gamma_V \ll M_V$, which is indeed our case. It should be noticed that, as it is well known, the case with $a=1$ cannot be treated in the IAM within the ET framework. This will not be the case in our quasi-exact predictions, as we will see in the following.

\begin{figure}[t!]
\begin{center}
\includegraphics[width=.49\textwidth]{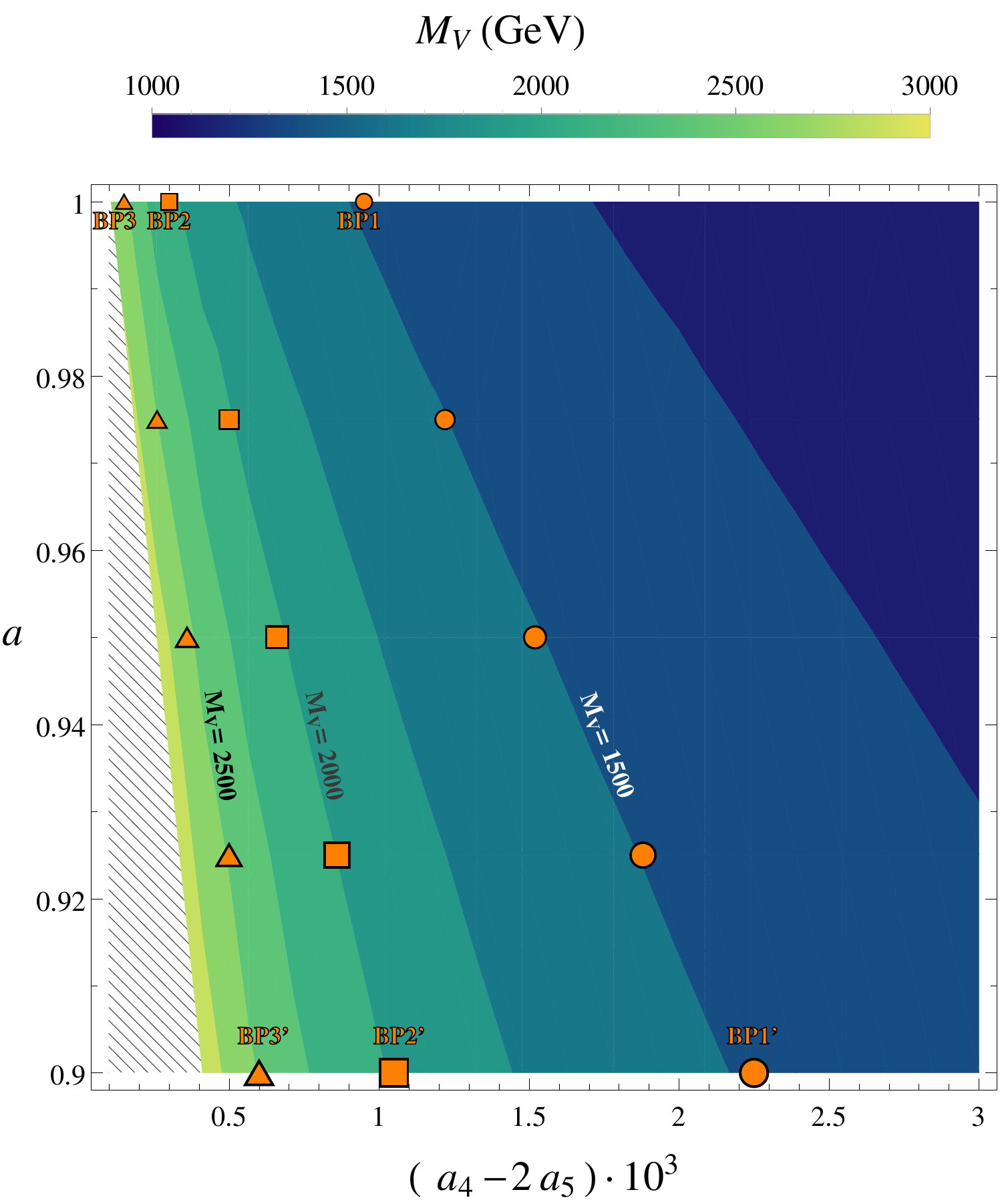}
\includegraphics[width=.49\textwidth]{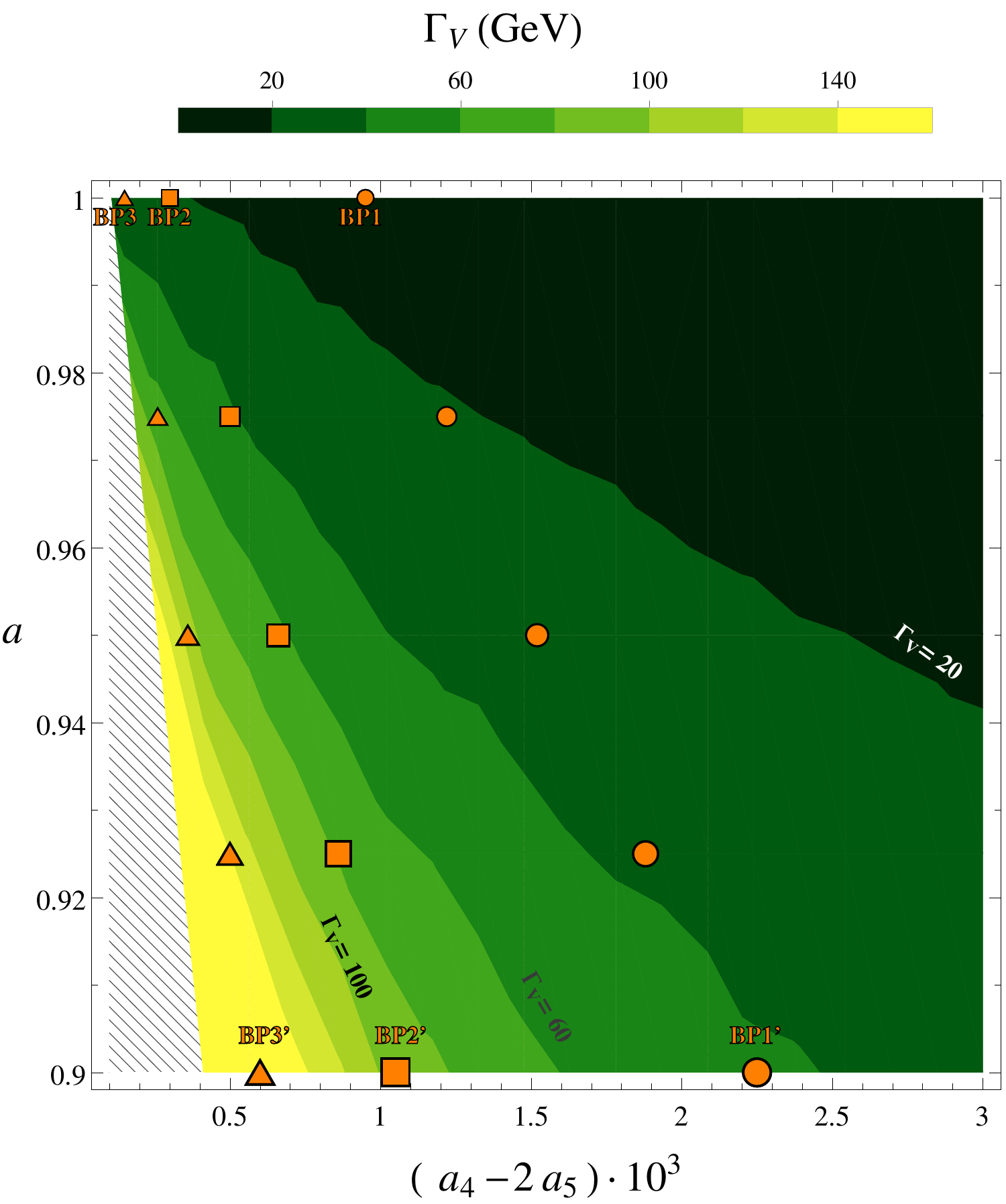}\\
\caption{Predictions for masses (left panel) and widths (right panel) of vector resonances as a function of $a$ and the combination $(a_4-2a_5)$ in the EChL+IAM. Our fifteen selected scenarios lay approximately over the contour lines of fixed $M_V$, 1500 GeV (circles), 2000 GeV (squares), and 2500 GeV (triangles), and have values for $a$ fixed, respectively,  to 0.9 (biggest symbols, corresponding to BP1', BP2' and BP3'), 0.925, 0.95, 0.975 and 1 (smallest symbols, corresponding to BP1, BP2, and BP3). All studied cases with vector resonances are such that no corresponding scalar or tensor resonances appear. The stripped area denotes the region with resonances heavier than 3000 GeV.}
\label{fig:contourMW}
\end{center}
\end{figure}

The solution to the position of the $a^{\rm IAM}_{11}$ pole in the quasi-exact case
with $m_{W,Z} \neq 0$ is more involved~\cite{Espriu:2012ih,Espriu:2013fia,Espriu:2014jya}, but it basically shares the main qualitative features of the previous ET results. First, the main contribution from the parameters $a_4$ and $a_5$ appears also in the particular combination $(a_4 - 2 a_5)$ which is $\mu$-scale independent if $b=a^2$. We have checked explicitly that other contributions from $a_4$ and $a_5$ not going as $(a_4 - 2 a_5)$ vanish in the isospin limit where $m_W=m_Z$. Second, the main dependence with $a$ also comes in the combination $(1-a^2)$, and the main dependence with $b$ also comes in the combination $(a^2-b)^2$. All these generic features can also be seen in our numerical results, displayed in \figref{fig:contourMW}, which we have generated with the FORTRAN code that implements
the quasi-exact EChL+IAM framework, borrowed from the authors in Refs.~\cite{Espriu:2012ih,Espriu:2013fia,Espriu:2014jya}.

The plots in \figref{fig:contourMW} show the contour lines of fixed $M_V$ and $\Gamma_V$ in the $\left[(a_4-2a_5), a\right]$ EChL parameter space plane. Here we have explored values of these parameters in the intervals that are allowed by present constraints, specifically, $a \in (0.9,1)$ and 
$(a_4-2 a_5) \in {\cal O}(10^{-4}, 10^{-3})$. The particular contour lines with $M_V=1500,\, 2000,\, 2500$ GeV are highlighted since they will be chosen as our reference mass values in our next study at the LHC. This figure assumes $b=a^2$, but we have checked explicitly that other choices for the $b$ parameter with $b\neq a^2$ do not change appreciably these results. In fact, the contour lines of $M_V$ and $\Gamma_V$ in the  $\left[(a_4-2a_5), b\right]$ plane with $a$ fixed in the interval $a \in (0.9,1)$ (not included here),  do not show any appreciable dependence with $b$ if this parameter is varied in the interval
$b \in (0.8,1)$. The distortions due to $b\neq a^2$ are clearly subleading in comparison to the leading effects from $(1-a^2)$ and $(a_4-2a_5)$, as explicitly shown in the ET formulas of \eqref{MVGVET}, and will be neglected from now on. The main reason of this secondary role of $b$, versus $a$, $a_4$ and $a_5$ is because, as we have previously said, in the $a_{11}$ amplitude $b$ enters only via loops, whereas $a$, $a_4$ and $a_5$ enter already at the tree level. Therefore our selection of scenarios will be done in terms of $a$, $a_4$ and $a_5$, and $b$ will be fixed to $b=a^2$, for simplicity. This choice of $b=a^2$ is also motivated in several theoretical models~\cite{Coleman:1985,Halyo:1991pc,Goldberger:2008zz}. Our final results will not change appreciably for other choices of $b$.

\begin{table}[t!]
\begin{center}
\vspace{.2cm}
\begin{tabular}{ |c|c|c|c|c|c|c| }
\hline
\rowcolor{gray! 50}
\rule{0pt}{1ex}
{\footnotesize {\bf BP}} & {\footnotesize {\bf $M_V ({\rm GeV})$}}  & {\footnotesize  {\bf $\Gamma_V ({\rm GeV)}$}}  & {\footnotesize {\bf $g_V(M_V^2)$}}  & {\footnotesize {\bf $a$}}  & {\footnotesize {\bf $a_4 \cdot 10^{4}$}}  & {\footnotesize {\bf $a_5\cdot 10^{4}$}}
\\[5pt] \hline
\rule{0pt}{1ex}
BP1  & $\quad 1476 \quad $ & $\quad 14 \quad $ & $ \quad 0.033  \quad $ & $ \quad 1 \quad $ & $ \quad 3.5 \quad $ & $ \quad -3 \quad $
\\[5pt] \hline
\rule{0pt}{1ex}
BP2  & $\quad 2039 \quad $ & $\quad 21 \quad $ & $ \quad 0.018  \quad $ & $ \quad 1 \quad $ & $ \quad 1 \quad $ & $ \quad -1 \quad $
\\[5pt] \hline
\rule{0pt}{1ex}
BP3  & $\quad 2472 \quad $ & $\quad 27 \quad $ & $ \quad 0.013  \quad $ & $ \quad 1 \quad $ & $ \quad 0.5 \quad $ & $ \quad -0.5 \quad $
\\[5pt] \hline
\rule{0pt}{1ex}
BP1' & $\quad 1479 \quad $ & $\quad 42 \quad $ & $ \quad 0.058  \quad $ & $ \quad 0.9 \quad $ & $ \quad 9.5 \quad $ & $ \quad -6.5 \quad $
\\[5pt] \hline
\rule{0pt}{1ex}
BP2'  & $\quad 1980 \quad $ & $\quad 97 \quad $ & $ \quad 0.042  \quad $ & $ \quad 0.9 \quad $ & $ \quad 5.5 \quad $ & $ \quad -2.5\quad $
\\[5pt] \hline
\rule{0pt}{1ex}
BP3'  & $\quad 2480 \quad $ & $\quad 183 \quad $ & $ \quad 0.033  \quad $ & $ \quad 0.9 \quad $ & $ \quad 4\quad $ & $ \quad -1 \quad $
\\[5pt] \hline
\end{tabular}
\caption{\small Selected benchmark points (BP) of dynamically generated vector resonances.
The mass, $M_V$, width, $\Gamma_V$, coupling to gauge bosons, $g_V(M_V)$,
and relevant chiral parameters, $a$, $a_4$ and $a_5$ are given for each of them.
$b$ is fixed to $b=a^2$. This table is generated using the FORTRAN code that implements
the EChL+IAM framework, borrowed from the authors in Refs.~\cite{Espriu:2012ih,Espriu:2013fia,Espriu:2014jya}. 
The effective coupling $g_V(M_V^2)$ is defined in section~\ref{sec-model}.    }
\label{tablaBMP}
\end{center}
\end{table}

In  \tabref{tablaBMP} we present a number of selected benchmark points (BP); namely, some specific sets of
values for the relevant parameters
$a, a_4$ and $a_5$ that yield to dynamically generated vector resonances emerging in the $IJ=11$ channel with masses around
the values  1.5, 2 and  2.5 TeV and not to resonances in the $IJ=00$ (isoscalar) and $IJ=20$ (isotensor) channels, which we do not consider in this work. These particular mass values for the vector resonances,
belonging to the interval (1000, 3000)~GeV have been chosen on purpose as illustrative examples
of the a priori expected reachable masses at the LHC. 
In the following sections we will use these benchmark points to predict the visibility of vector resonances that may exist in
the $IJ=11$ channel, and therefore resonate in the process $WZ \to WZ$ at the LHC. For the $IJ=00$ channel there are recent alternative studies of the IAM scalar resonances and their production at the LHC, see for instance~\cite{BuarqueFranzosi:2017prc}. 

The selected points in \tabref{tablaBMP} are also included in our previous contour plots in Fig.~\ref{fig:contourMW}. They are placed at the upper and lower horizontal axes in these plots, and are chosen on purpose at the two boundary values
of the $a$ parameter: 1) $a=1$ for BP1, BP2 and BP3 and 2) $a=0.9$ for  
BP1', BP2' and BP3'. These will be our main reference scenarios to which we will devote most of our LHC analysis. However, in order to provide a complementary study of the sensitivity to the $a$ parameter we have also defined a family of additional scenarios belonging to these contour lines of fixed $M_V=1500$, $2000$ and $2500$ GeV, respectively,   but with different values of $a$ in the interval $(0.9,1)$. These BP points are specified by circles, squares and triangles in Fig.~\ref{fig:contourMW} and will also be discussed in the final section.

\section{Dealing with IAM vector resonances in $\boldsymbol{WZ}$ scattering}
\label{sec-model}
In order to study how the vector resonances that are predicted in the IAM could be seen at the LHC with a MonteCarlo analysis, we need first to
establish a diagrammatic procedure for $WZ \to WZ$ scattering to implement the basic ingredients of these IAM
resonances in a Lagrangian framework. The use of MonteCarlo event generators like MadGraph requires the
model ingredients to be implemented in a Lagrangian language, which means in our case that we have to specify the
interactions of the emergent vector resonances with the gauge bosons (and Goldstone bosons). Thus, instead
of implementing the $A(W_L Z_L \to W_L Z_L)$ scattering amplitude in terms of the predicted IAM
partial waves, we simulate this scattering amplitude with a simple model that contains the
basic ingredients of the emergent vector resonances. Namely, the mass, the width and the proper couplings to
the gauge bosons $W$ and $Z$.  The simplest Lagrangian to include these vector resonances, $V$, that shares the
chiral and gauge symmetries of the EChL is provided in Refs.~\cite{Ecker:1989yg,DAmbrosio:2006xmn,Pich:2015kwa,Pich:2016lew}.
In the Proca 4-vector formalism,  the corresponding $P$-even Lagrangian is given by:
\begin{equation}
\mL_V = -\Frac{1}{4}{\rm Tr}( {\hat V}_{\mu\nu} {\hat V}^{\mu\nu}) +
 \Frac{1}{2} M_V^2 {\rm Tr}( {\hat V}_\mu {\hat V}^\mu )
 \, + \,\Frac{f_V}{2\sqrt{2}} {\rm Tr}( {\hat V}_{\mu\nu} f_+^{\mu\nu})
 \, + \,
\Frac{i g_V}{2\sqrt{2} } {\rm Tr}( {\hat V}_{\mu\nu} \, [u^\mu,u^\nu ] )\, ,
\label{Proca}
\end{equation}
which includes the isotriplet vector resonances, $V^{\pm}$ and $V^0$, via the ${\hat V}_\mu$ fields and the a priori
free parameters: mass $M_V$, and couplings $f_V$ and $g_V$. 
The basic definitions in \eqref{Proca}  are~\cite{Pich:2012dv,Pich:2013fea,Pich:2012jv}:
\begin{align}
{\hat V}_\mu &=\Frac{ \tau^a {V}_\mu^a}{\sqrt{2} } \,=\,
\left(\begin{array}{cc}
\Frac{V^0_\mu }{\sqrt{2}} & V^+_\mu \\ V^-_\mu  &-\Frac{V_\mu^0}{\sqrt{2}}
\end{array}
\right)\, ,
\\
{\hat V}_{\mu\nu} &= \nabla_\mu {\hat V}_\nu -\nabla_\nu {\hat V}_\mu\, , 
\\
u_\mu  &=    \,
i\, u\, \Big(D_\mu U\Big)^\dagger u \,\,, {\rm with}\,\, u^2=U\\
f_+^{\mu\nu} & =
\, -\, \left(
 u^\dagger \hat{W}^{\mu\nu}  u +   u \hat{B}^{\mu\nu} u^\dagger
\right)\, ,
\\
\nabla_\mu \mX \,& =\, \partial_\mu \mX \, +\, [\Gamma_\mu , \mX ] \,\,, {\rm with}\,\, \Gamma_\mu =
\Frac{1}{2} \Big(\Gamma_\mu^{L} +\Gamma_\mu^{R}\Big)\, , 
\\
\Gamma_\mu^{L} &= u^\dagger \left(\partial_\mu + i\,\frac{g}{2} \vec{\tau}\vec{W}_\mu
\right) u^{\phantom{\dagger}} 
\, , \quad\;
\Gamma_\mu^{R} = u^{\phantom{\dagger}}  \left(\partial_\mu + i\, \Frac{g'}{2} \tau^3 B_\mu\right)u^\dagger 
\, .
\end{align}

In the unitary gauge (convenient for tree-level collider analyses)
we have $u=U=\mathbb{I}$, and one finds a simpler result. In particular, after rotating to the mass eigenstate basis,
where the unphysical mixing terms between the $V$'s and the gauge bosons (introduced by $f_V \neq 0$)
are removed, and after bringing the kinetic and mass terms into
the canonical form, we find:
\begin{align}
\mL_V &=    -\Frac{1}{4} \Big( 2V^+_{\mu\nu}  V^{-\mu\nu}+V^0_{\mu\nu}‚ V^{0\mu\nu}\Big) +
 \Frac{1}{2} M_V^2 \Big( 2V^+_{\mu} V^{-\mu}+V^{0}_{\mu} V^{0\mu} \Big)
\nn\\
&-\,\Frac{i f_V }{v^2}
\bigg[m_W^2 V^0_\nu  (W^+_\mu W^{-\,\mu\nu} -W^-_\mu W^{+\,\mu\nu} )
+ m_W m_Z V^+_\nu (W^-_\mu Z^{\mu\nu} -Z_\mu W^{-\, \mu\nu})
\nn\\
&\hspace{1.2cm}+ m_W m_Z V^-_\nu (Z_\mu W^{+\, \mu\nu}- W^+_\mu Z^{\mu\nu} )
\bigg]
\nn\\
&
+ \Frac{i 2g_V }{v^2} \bigg[ m_W^2 V^{0\,\, \mu\nu} W_\mu^+ W_\nu^-
+  m_W \, m_Z\,  V^{+\,\, \mu\nu} W_\mu^- Z_\nu
+  m_W \, m_Z\,  V^{-\,\, \mu\nu} Z_\mu W_\nu^+   \bigg]\, ,
\label{LVugauge}
\end{align}
where we have used the short-hand notation
$V^a_{\mu\nu}= \partial_\mu V^a_\nu - \partial_\nu V^a_\mu$ (for $a=\pm ,0$),
$W^a_{\mu\nu}= \partial_\mu W^a_\nu - \partial_\nu W^a_\mu$  (for $a=\pm$), and
$Z_{\mu\nu}= \partial_\mu Z_\nu - \partial_\nu Z_\mu$.

It should be noticed that in the previous Lagrangian of \eqref{LVugauge} there are not interaction terms between the
vector resonances and two neutral gauge bosons, $VZZ$, (as there are not either  $Vzz$ interactions in
\eqref{Proca} of $V$ with two neutral Goldstones $z$) and this explains why the vector resonances cannot
emerge in the s-channel of $WW \to ZZ$ nor
$ZZ \to ZZ$\footnote{Notice that scalar resonances could resonate in these channels, but we do not considered them here.}. This is a clear consequence of exact custodial invariance and it also confirms that $W^{\pm}Z \to W^{\pm}Z$ are the proper channels to look for emergent
signals from the charged vector resonances $V^{\pm}$. The relevant set of Feynman rules extracted from the
above Lagrangian in \eqref{LVugauge} is collected in the appendices, for completeness.

Since we are mostly interested here in the deviations with respect to the SM predictions in the case of the
longitudinal modes,  we will mainly focus on their scattering amplitudes.
Therefore, from now on we will simplify our study by setting $f_V=0$. This is well justified since
this $f_V$ predominantly affects the couplings of the resonances to transverse gauge bosons and,
in consequence, $g_V$ is the most relevant coupling to the longitudinal modes.
Some additional comments on the behavior of the scattering amplitudes for the other modes will be made at the end of this section. 

Our aim here is to use the Lagrangian $\mL_V$ in \eqref{LVugauge} as a practical tool to
mimic the main features of
the vector resonances found with the IAM. Specifically, we wish to introduce all these features by means of a tree level computation of 
$A(WZ \to WZ)$ with
$\mL_{\rm model}=\mathcal{L}_2 +\mathcal{L}_V$.
This leads us to the issue of relating $g_V$, $M_V$ and  $\Gamma_V$ to the properties of the
IAM vector resonances found from $a_{11}^{\rm IAM}$. On one hand, the mass and the width are obviously related to
the position of the pole, $s_{\rm pole}=(M_V-\frac{i}{2}\Gamma_V)^2$,  of $a_{11}^{\rm IAM}(s)$. On the other hand,
the coupling $g_V$ should also be related to the properties of $a_{11}^{\rm IAM}(s)$ in the resonant region.
For instance, one could extract a value of $g_V$ by identifying the residues of $a_{11}^{\rm model}(s)$ and
$a_{11}^{\rm IAM}(s)$ at $s_{\rm pole}$. If for simplicity we had used the ET version of the relevant amplitudes,
this would have led to the simple relation $g_V^2=2(a_4-2a_5)$. Alternatively,
one could follow the approach of
Refs.~\cite{Pich:2015kwa,Pich:2016lew}
where close to the resonance mass shell, they find $\mL_{\rm model}$
to be equivalent to a more general Lagrangian\footnote{
The Lagrangian in Refs.~\cite{Pich:2015kwa,Pich:2016lew} considers the antisymmetric tensor representation
for the spin--1 resonances, which is fully equivalent to the Proca four-vector representation
provided appropriate non-resonant operators are added to the Lagrangian.}
in which the on-shell vector coupling $g_V$ is related to the $\mathcal O(p^4)$ low-energy chiral parameters
in the form $a_4=-a_5=g_V^2/4$.

\begin{figure}[t!]
\begin{center}
\includegraphics[width=.6\textwidth]{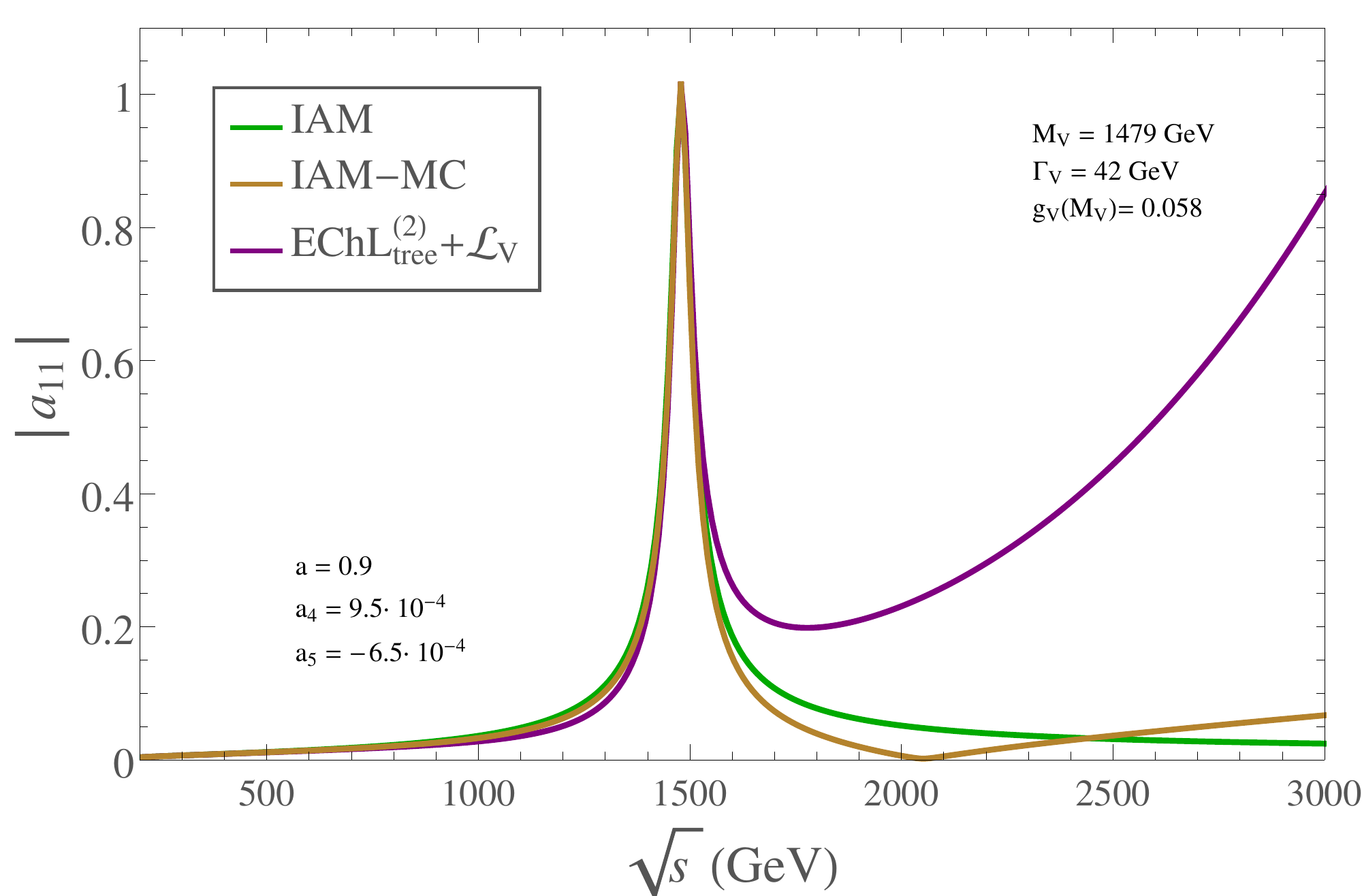}\\
\caption{Prediction of the $|a_{11}|$ partial wave as a function of the center of mass energy $\sqrt{s}$ in the three models explained in the text: IAM (green), IAM-MC (orange) and $\mL_2+\mL_V$ with constant $g_V$ (purple). The values of the parameters are those of BP1' in \tabref{tablaBMP}.}
\label{fig:procagvconst}
\end{center}
\end{figure}

However, this Lagrangian $\mL_2+\mL_V$ leads to problems if a  constant $g_V$ is assumed.
Even though it gives a reasonable estimate of the partial wave at $s\sim M_V^2$,
it does not work satisfactorily away from the resonance region. Indeed, it yields to a bad high energy
behavior for $s>M_V^2$: the subsequent partial wave $a_{11}(s)$ grows too fast with energy and crosses
the unitary bound at energies of a few TeV. This unwanted violation of unitarity happens, indeed, for any choice of the
constant $g_V$ in the Lagrangian $\mL_2+\mL_V$.
We depict this failure in \figref{fig:procagvconst} for one particular example with $a=0.9$, $a_4=9.5\times 10^{-4}$
and $a_5=-6.5\times 10^{-4}$ that produces a IAM vector pole at $M_V=1479$ GeV and
$\Gamma_V=42$ GeV, and where we have assumed a constant value of $g_V=0.058$.  In this case we have found that the crossing over the
unitarity bound occurs at around 3 TeV. From this study, we conclude then that the $a_{11}(s)$ resulting
from $\mL_2+\mL_V$ with constant $g_V$ does not simulate correctly the behaviour of $a_{11}^{\rm IAM}$, which is
by construction unitary and therefore we will not take $g_V$ as a constant coupling.

We will define in the following the specific model that we choose to mimic with a chiral Lagrangian the IAM amplitude, which is referred in \figref{fig:procagvconst} as IAM-MC. 
This will obviously lead us to consider again $\mL_2+\mL_V$ but with a momentum dependent $g_V$.  This will be done in the next subsection.

\subsection{Our model: IAM-MC}
We work with the Lagrangian  $\mL_2+\mL_V$, first introduced in the EW interaction basis in \eqrefs{eq.L2} and (\ref{Proca}),
 to mimic the IAM amplitude of $WZ$ scattering but with an energy dependent coupling $g_V(s)$ (remember that we are setting $f_V=0$ in all our numerical estimates), which leads to unitary results in the way that will be described in this subsection. Firstly, our $A(W_L Z_L \to W_L Z_L)$ amplitudes have by construction the resonant behavior of the IAM amplitudes at $s_{\rm pole}=(M_{V}-\frac{i}{2}\Gamma_{V })^2$, as commented above. Secondly, it is illustrative to notice that
the effective coupling $g_V(s)$ is in fact related to a form factor, as can be seen for instance using a current algebra language. Concretely, the matrix element of a vector current between two longitudinal $W$ bosons
and the vacuum is described by an energy dependent form factor  $G_V(s)$ given by \cite{Arnan:2015csa}:
\be
\langle W_L^i(k_1)W_L^j(k_2)|J^k_\mu|0\rangle=(k_1-k_2)_\mu G_V(s)\epsilon^{ijk},
\label{genformf}
\ee
\begin{figure}[t!]
\begin{center}
\includegraphics[width=.49\textwidth]{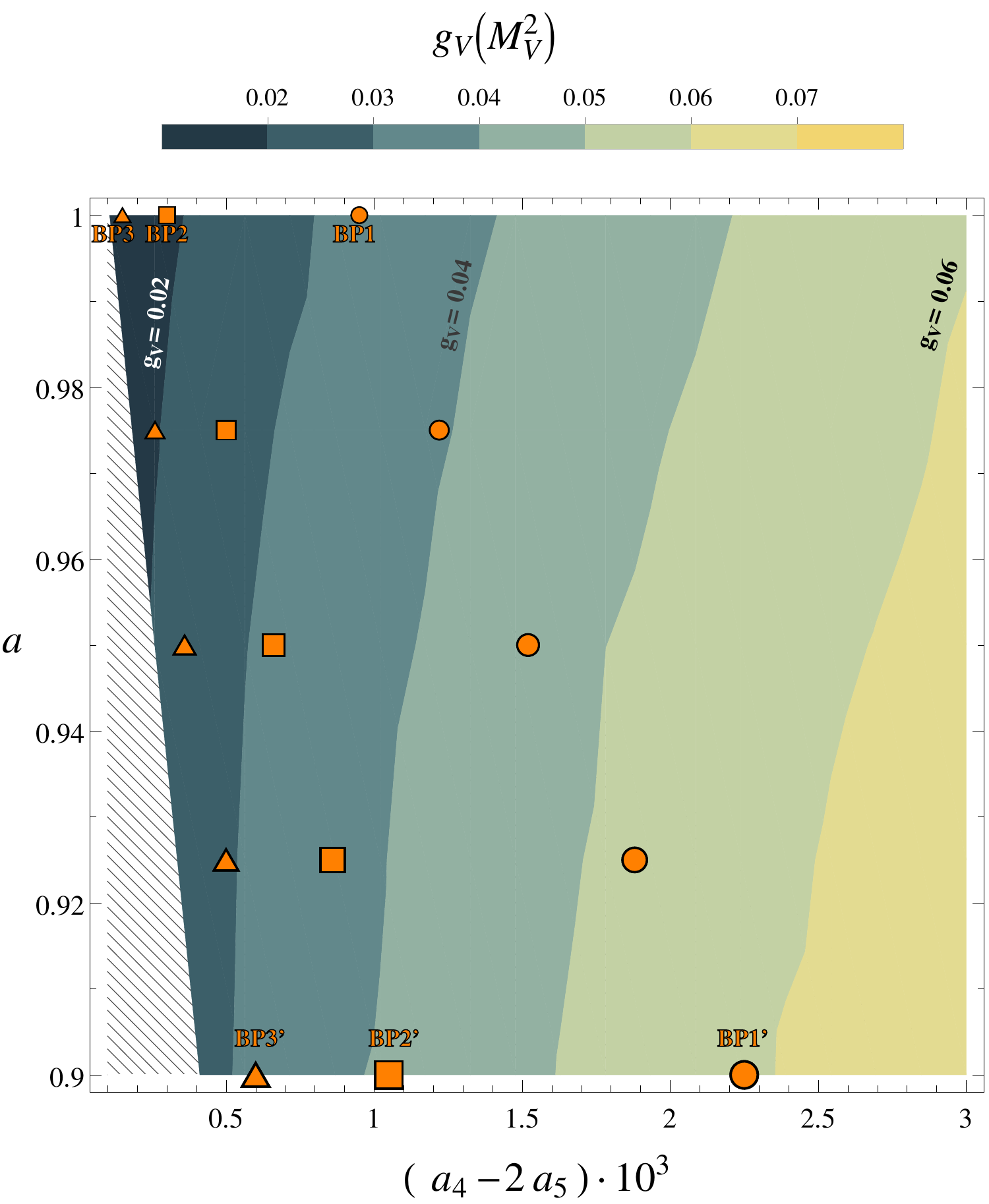}
\caption{Predictions of $g_V(M_V^2)$ as a function of $a$ and $(a_4-2a_5)$   computed from Eq.(\ref{a11MV}), as discussed in the text. The benchmark points specified with geometric symbols correspond respectively to those in \figref{fig:contourMW}.}
\label{fig:contourgv}
\end{center}
\end{figure}
where $J^k_\mu$ is the interpolating vector current with isospin index $k$ that creates a resonance $V$. This form factor $G_V(s)$ can be easily related to $g_V(s)$ at $s=M_V^2$ by $G_V(M_V^2)=\sqrt{2}M_V^2g_V(M_V^2)/v^2$. In practice, $g_V(M_V^2)$ is determined by the matching procedure described next.

In order to build our resonant
$A(W_L Z_L \to W_L Z_L)$ amplitudes we use the following prescription. First, we impose the matching at the partial waves level. Concretely, it is performed by identifying the tree level predictions
from $\mL_2+\mL_V$ with the predictions from the IAM at $M_V$, i.e:
\begin{equation}
\Big|a_{11}^{{\rm EChL}_{\rm tree}^{(2)}+\mL_V}(s=M_V^2)\Big|=\Big|a_{11}^{\rm IAM}(s=M_V^2)\Big|\,,
\label{a11MV}
\end{equation}
where $a_{11}^{{\rm EChL}_{\rm tree}^{(2)}+\mL_V}$ is the partial wave amplitude computed from $\mL_2+\mL_V$.

Solving (numerically) this  \eqref{a11MV} for the given values of $(a,a_4,a_5)$
and the corresponding values of $(M_V,\Gamma_V)$ leads to the wanted solution for $g_V=g_V(M_V^2)$.
For instance, in the previous example of $a=0.9$, $a_4=9.5\times 10^{-4}$ and $a_5=-6.5\times 10^{-4}$ (our benchmark point BP1' in \tabref{tablaBMP}) with
corresponding $M_V=1479$ GeV and $\Gamma_V=42$ GeV, we found $g_V(M_V^2)=0.058$. For the other selected benchmark
points the corresponding values found for $g_V(M_V^2)$ are collected in  \tabref{tablaBMP} and in \figref{fig:contourgv}.  Interestingly, these numerical results in \figref{fig:contourgv} for $g_V(M_V^2)$ show a clear correlation with the previously predicted $M_V$ and $\Gamma_V$ values in \figref{fig:contourMW}, which fulfill approximately:  $\Gamma_V \simeq M_V^5 g_V^2/(48 \pi v^4)$, as naively expected from the Proca Lagrangian for $f_V=0$. 
 
One may notice at this point that the computation of the IAM partial waves has been done with electroweak gauge bosons in the external legs and not with Goldstone bosons. The ET has only been used to compute the real part of the loops involved, as explained before in the previous section.

Away from the resonance we consider an energy dependence in $g_V(s)$ with the following requirements:
\begin{itemize}
\item[i)] Below the resonance, at low energies, one should find compatibility with the result from ${\rm EChL}^{(2+4)}_{\rm loop}$, which implies that the predictions from $\mL_V$ should match those from $\mL_4$ at these energies.
This is what happens indeed to $a_{11}^{\rm IAM}$ below the resonance, by construction.
\item[ii)] Above the resonance, at large energies, we require the cross section not to grow faster than the Froissart bound~\cite{PhysRev.123.1053}, which can be written as:
\begin{equation}
\sigma(s)  \le \sigma_0 \log^2\bigg(\frac{s}{s_0}\bigg)\,,\label{froisbound}
\end{equation}
with $\sigma_0$ and $s_0$ being energy independent quantities.
Notice that when using this bound we are implicitly assuming that there are no other resonances (in addition to $V$) emerging in the spectrum, at least until very high energies.
\end{itemize}
We have found that these requirements above are well approximated by setting the following simple function:
\begin{align}
g_V^2(s)&=g_V^2(M_V^2) \frac{M_V^2}{s} \,\,\, {\rm for} \,\, s< M_V^2 \nn\,, \\
g_V^2(s)&=g_V^2(M_V^2) \frac{M_V^4}{s^2} \,\,\, {\rm for} \,\, s> M_V^2\,.
\label{gvenergy}
\end{align}
This $g_V(s)$ coupling should be used when $V$ is propagating in the $s$-channel. In the other channels where the resonance could also propagate, $t$ and/or $u$ channels, the coupling should be the same described in \eqref{gvenergy} in terms of the corresponding $t$ or $u$ variables to be fully crossing symmetric. Nevertheless, we have checked that a completely crossing symmetric energy-dependent coupling, given by $g_V^2(z)=\theta(M_V^2-z)g_V^2(M_V^2)\frac{M_V^2}{z}+\theta(z-M_V^2)g_V^2(M_V^2)\frac{M_V^4}{z^2}$, leads to a moderate violation of the Froissart bound in \eqref{froisbound} at energies in the TeV range. To avoid this violation of unitarity, we propose the following expression for the coupling in terms of the $t$ and $u$ variables:
\begin{align}
g_V^2(z)&=g_V^2(M_V^2) \frac{M_V^2}{z} \,\,\, {\rm for} \,\, s< M_V^2 \nn\,, \\
g_V^2(z)&=g_V^2(M_V^2) \frac{M_V^4}{z^2} \,\,\, {\rm for} \,\, s> M_V^2\,,
\label{gvenergytu}
\end{align}
with $z=t,u$ corresponding to the $t,u$ channels, respectively, in which the resonance is propagating.

The accuracy of the result with this choice of energy dependent coupling in comparison with the previous constant coupling can be seen in \figref{fig:procagvconst}. It is clear from this figure that the result for $a_{11}$ using this energy dependent coupling simulates much better the IAM result than that with a constant $g_V$, and it also provides a good low and high energy behaviors.  It is worth commenting that we have tried other choices for the dependence with energy of this $g_V(s)$ coupling, but none of these alternative tries have passed all the above required conditions.
We have also checked explicitly that our hypothesis in \eqrefs{gvenergy}-(\ref{gvenergytu}) leads to a high-energy behavior of the cross section that is always below and close to the saturation of this Froissart bound.

The above described method, which will be called from now on IAM-MC (named after IAM for MonteCarlo), is the one we choose
to simulate the IAM with a Lagrangian formalism. We find that it is the most appropriate one for the forthcoming MonteCarlo
analysis with MadGraph5 of LHC generated events.

In summary, we follow the subsequent steps to get $A(W_LZ_L \to W_LZ_L)_{\rm IAM-MC}$ for each of the
given $(a,a_4,a_5)$ input values:
\begin{itemize}
\item[1)] Compute the amplitude from the tree level diagrams with the Feynman rules from $\mL_2+\mL_V$.
This gives a result in terms of $a,M_V,g_V$ and $\Gamma_V$.
\item[2)] For the given values of $(a,a_4,a_5)$, then set $M_V$ and $\Gamma_V$ to the corresponding values found from the
poles of $a_{11}^{\rm IAM}$.
\item[3)] Extract the value of $g_V(M_V^2)$ by solving numerically \eqref{a11MV}.
\item[4)] Substitute $g_V$ by $g_V(s)$ in the $s$-channel and by $g_V(u)$ in the $u$-channel (for the process of study, $WZ\to WZ$, the charged vector resonance only propagates in these two channels) and use \eqrefs{gvenergy} and (\ref{gvenergytu}).
\item[5)] Above the resonance we assume that the deviations with respect to the SM come dominantly from $\mL_V$, which means in practice that the proper Lagrangian for the computation
of the IAM simulated amplitude is
$\mL_{\rm SM}+\mL_V$ rather than $\mL_2+\mL_V$. This is obviously equivalent to use $\mL_2+\mL_V$ with $a=1$ at energies above the resonance.
\end{itemize}

\begin{figure}[t!]
\begin{center}
\includegraphics[width=.49\textwidth]{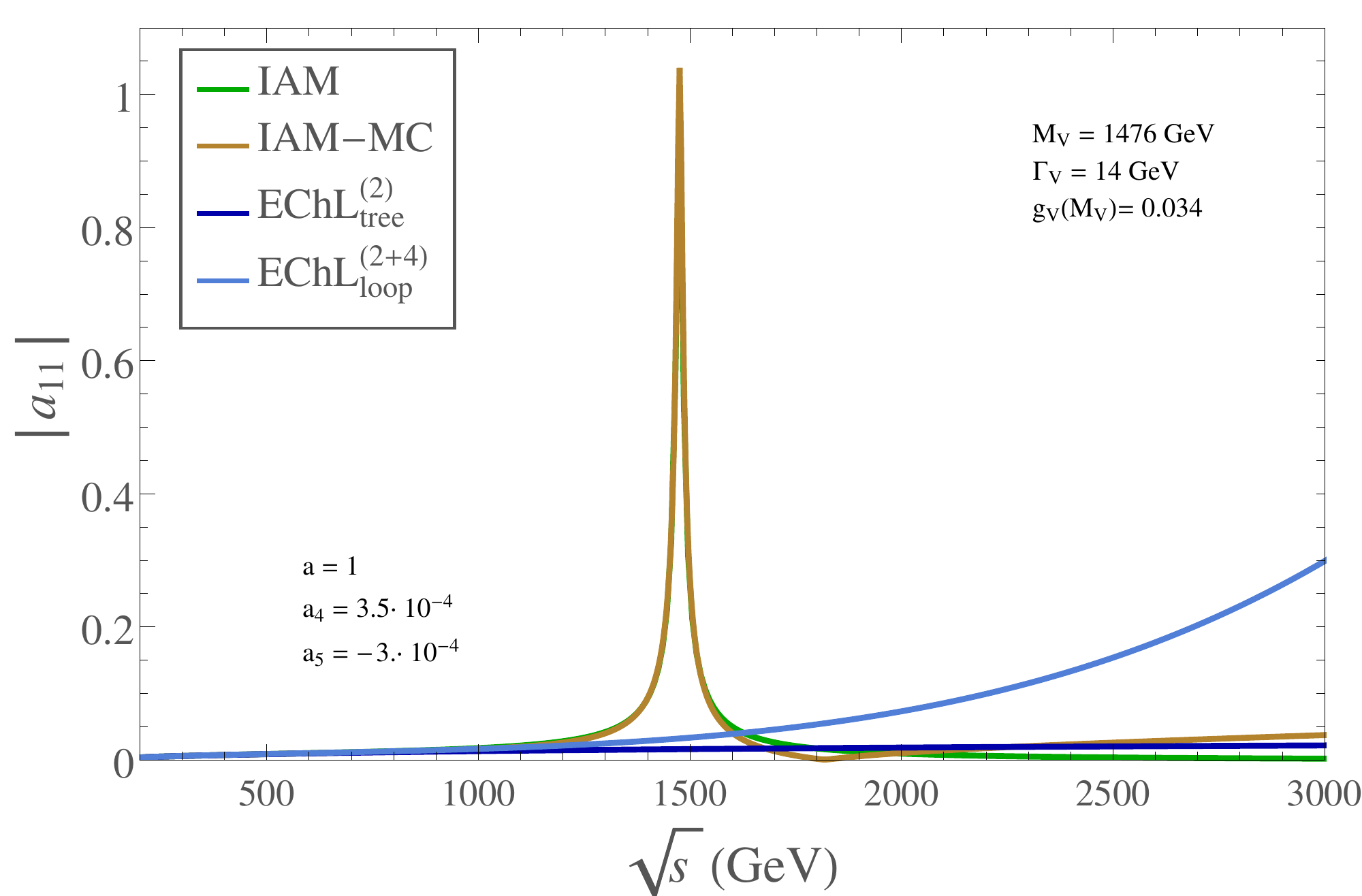}
\includegraphics[width=.49\textwidth]{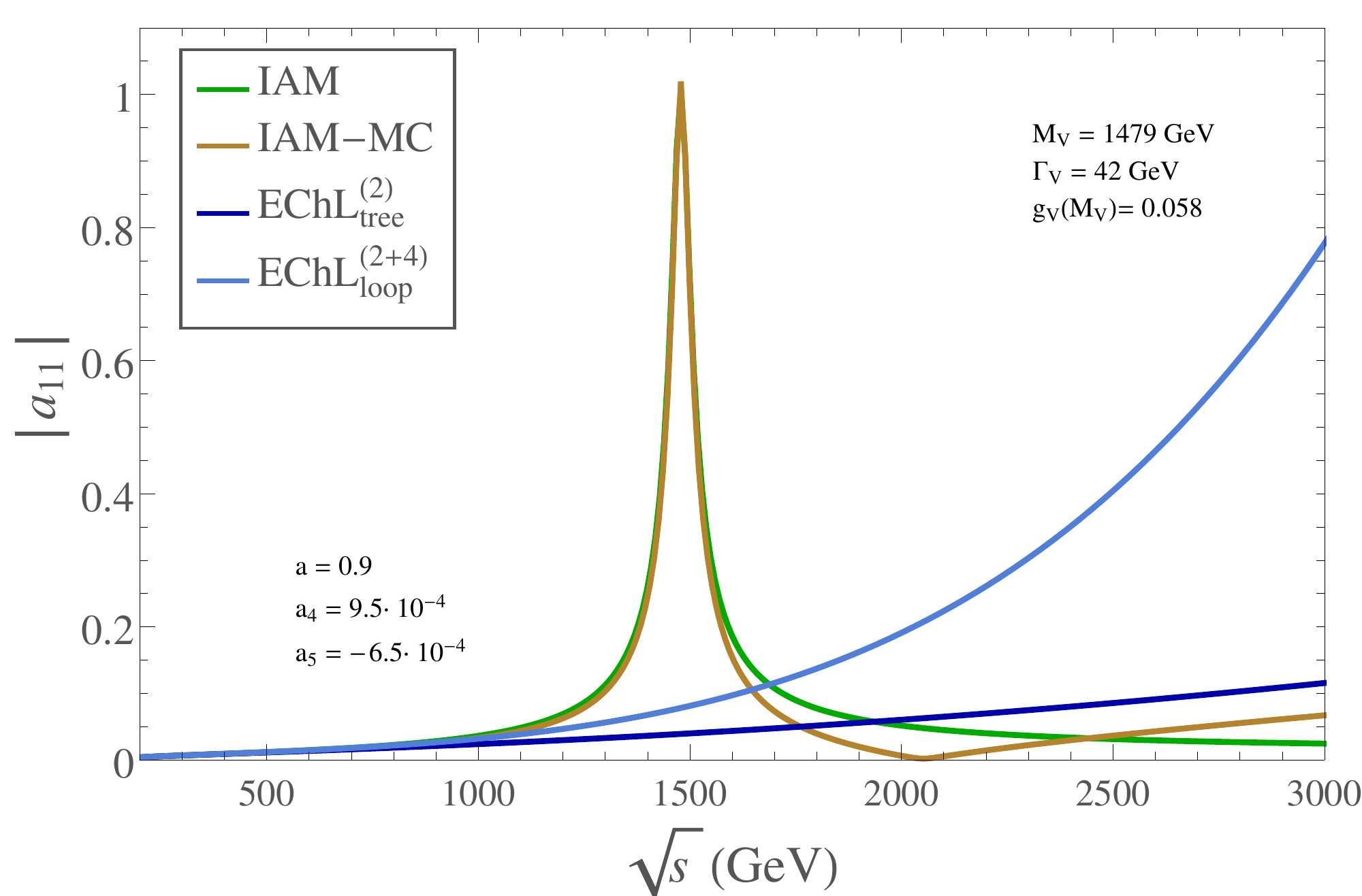}\\
\includegraphics[width=.49\textwidth]{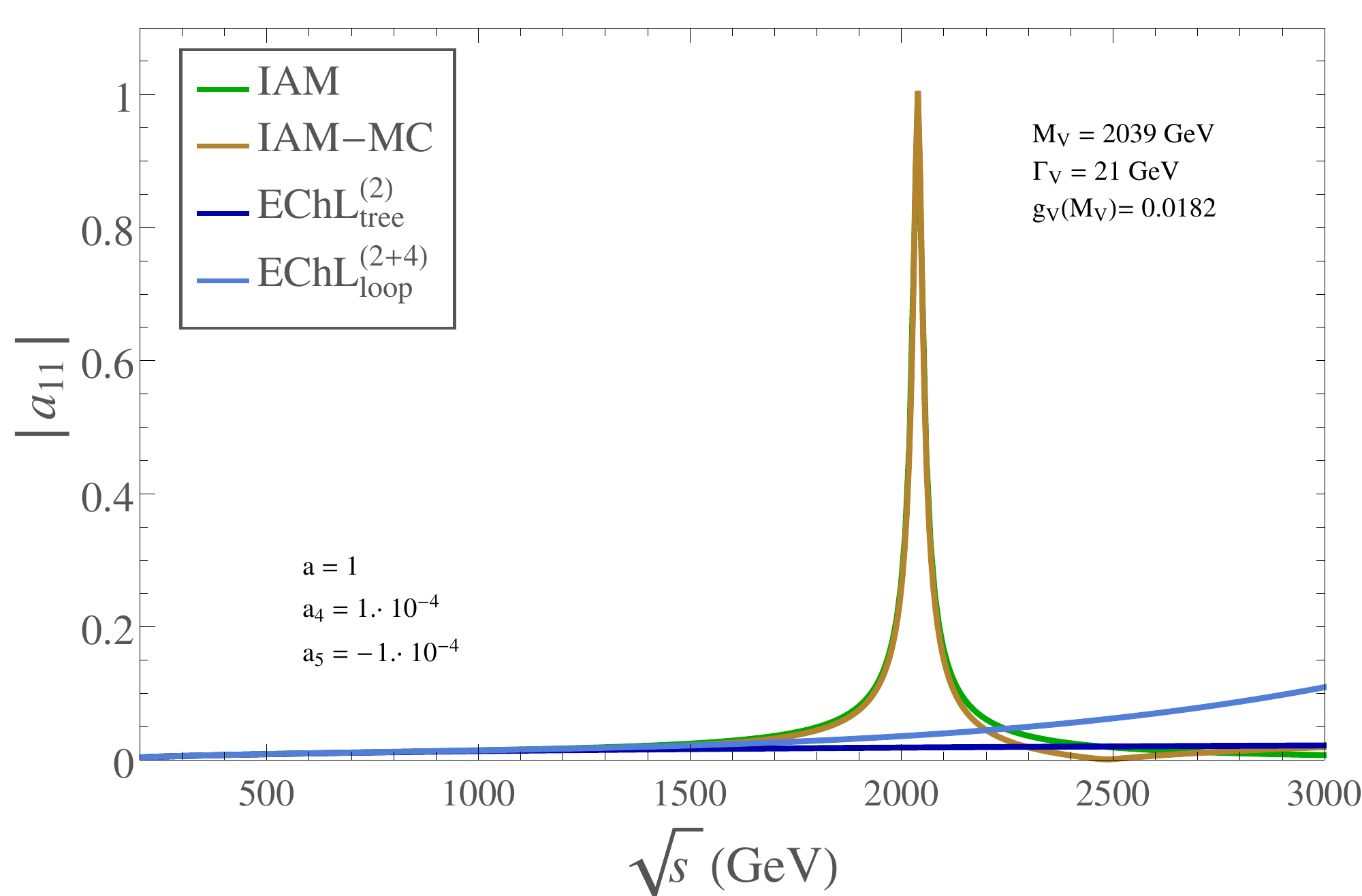}
\includegraphics[width=.49\textwidth]{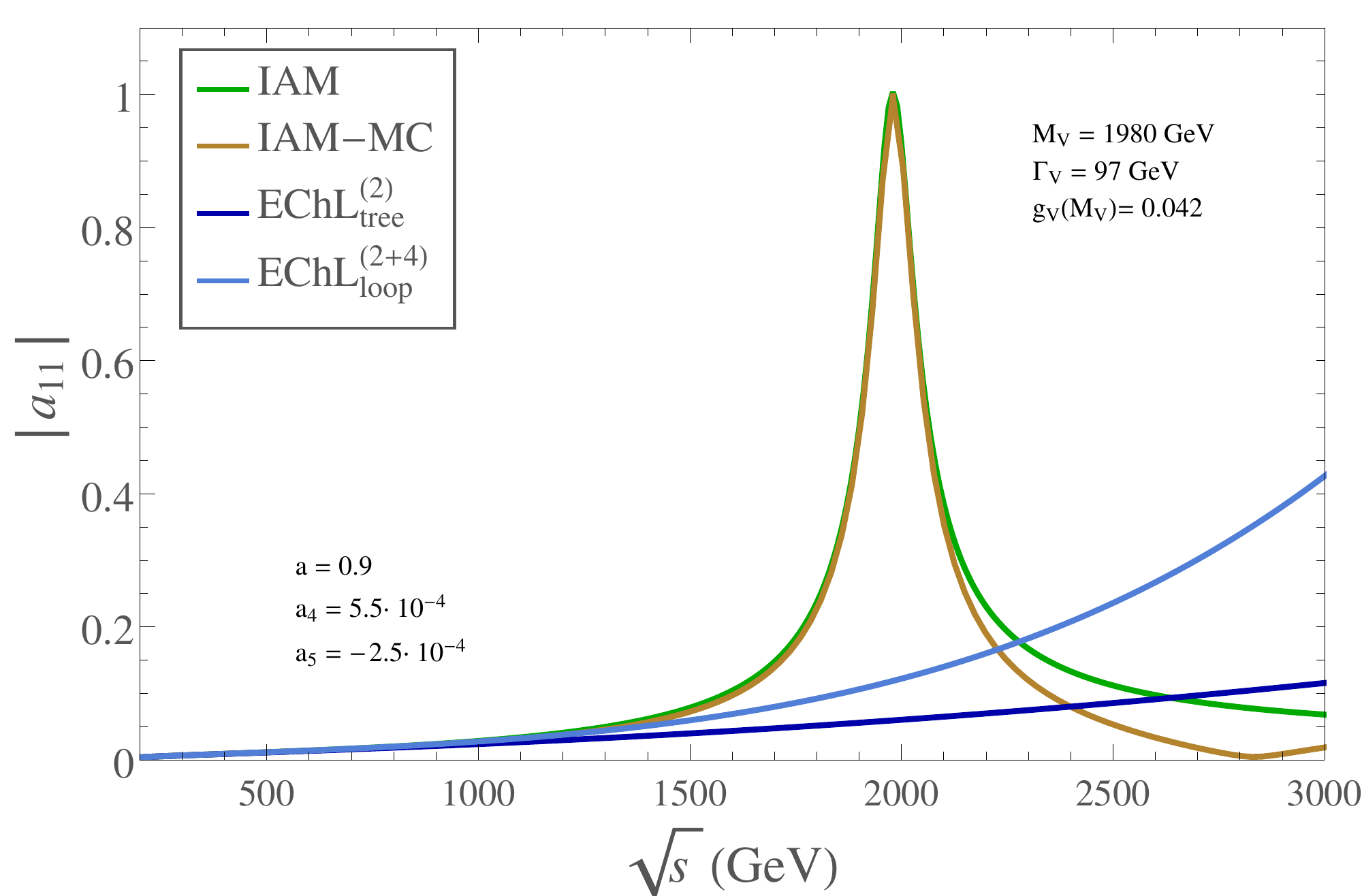}\\
\includegraphics[width=.49\textwidth]{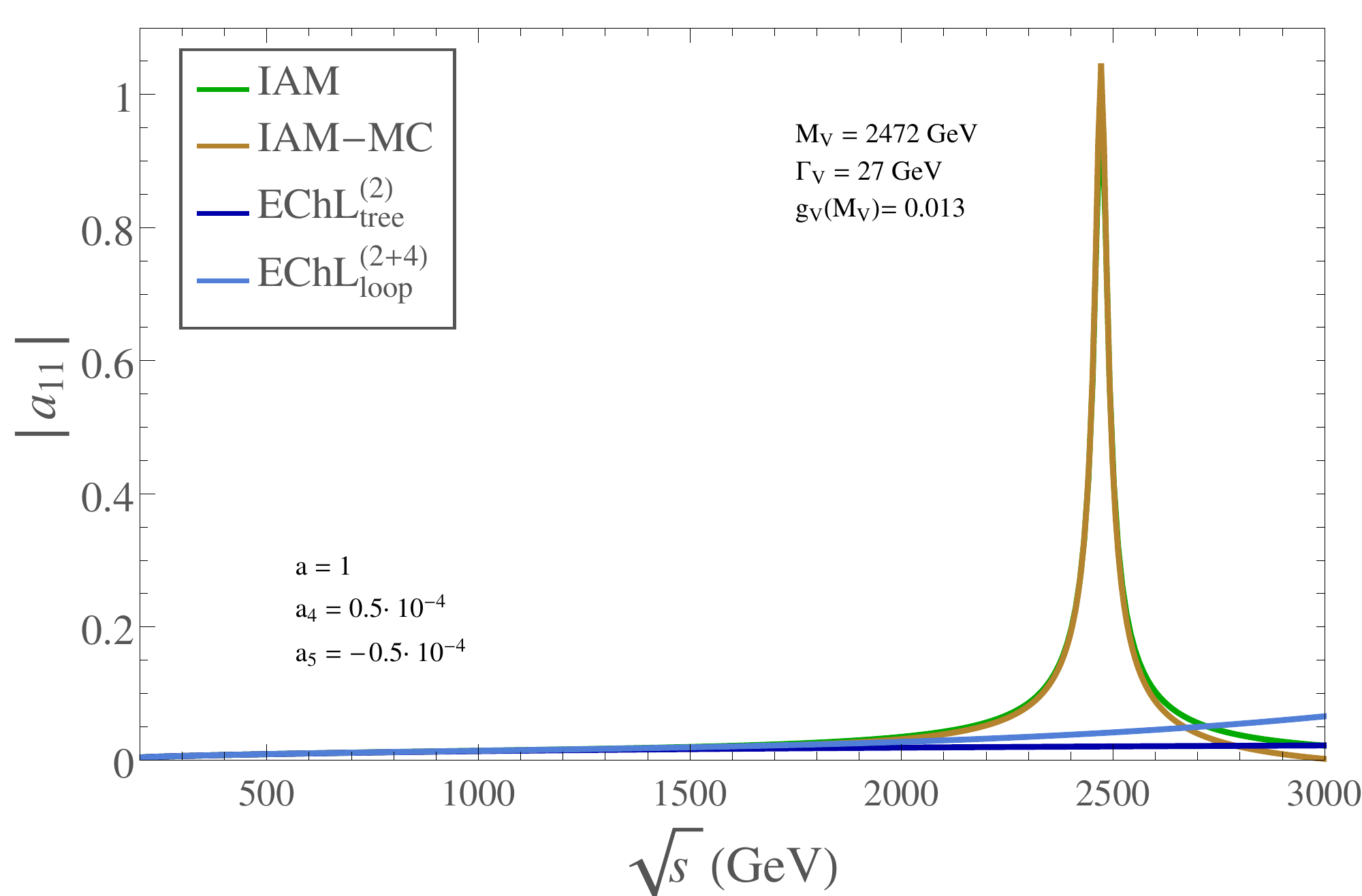}
\includegraphics[width=.49\textwidth]{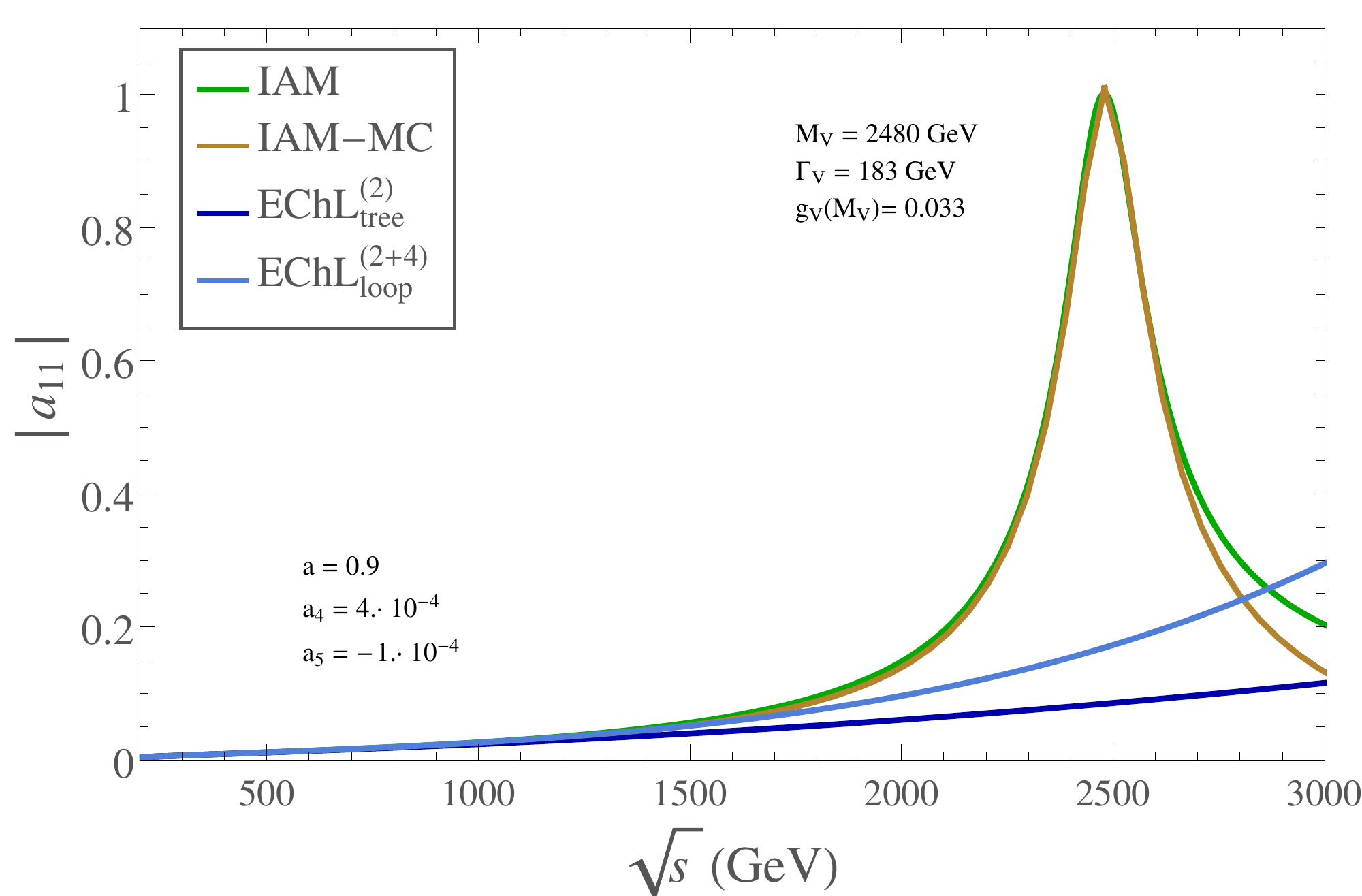}\\
\caption{Predictions of the $|a_{11}|$ partial waves as a function of the center of mass energy $\sqrt{s}$ for all the selected benchmark points in \tabref{tablaBMP}. Different lines correspond to the different models considered in the text:  EChL unitarized with the IAM (green), our IAM-MC model (orange),  non-unitarized EChL up to $\mO(p^2)$ (dark blue) and  non-unitarized EChL up to $\mO(p^4)$ including loop contributions (light blue).}
\label{fig:pw1}
\end{center}
\end{figure}

The detailed description and the analytical results of this computation are collected in the appendices. We emphasize again that these analytical results of the $WZ$ scattering amplitudes do not make use of the ET and they are obtained by a tree level diagrammatic computation with massive external $W$ and $Z$ gauge bosons. For completeness and comparison we have also included in the appendices the predictions for the three cases of our interest, the IAM-MC, the SM, and the EChL, as well as the corresponding Feynman rules.

As for the numerical results, we present in \figref{fig:pw1} our predictions of the partial waves $a_{11}^{\rm IAM-MC}$ for all the selected benchmark points of \tabref{tablaBMP}. We have also included in these plots the corresponding predictions from the IAM and from the EChL, at both LO and NLO, for comparison. In these plots we clearly see the accuracy of our IAM-MC model in simulating the behavior of the IAM amplitudes. This happens not only at the close region surrounding the resonance, where it is clearly very good, but also below and above the resonance, inside the displayed energy interval of $\sqrt{s} \in (200,3000)$ GeV.

 \begin{figure}[t!]
\begin{center}
\includegraphics[width=.49\textwidth]{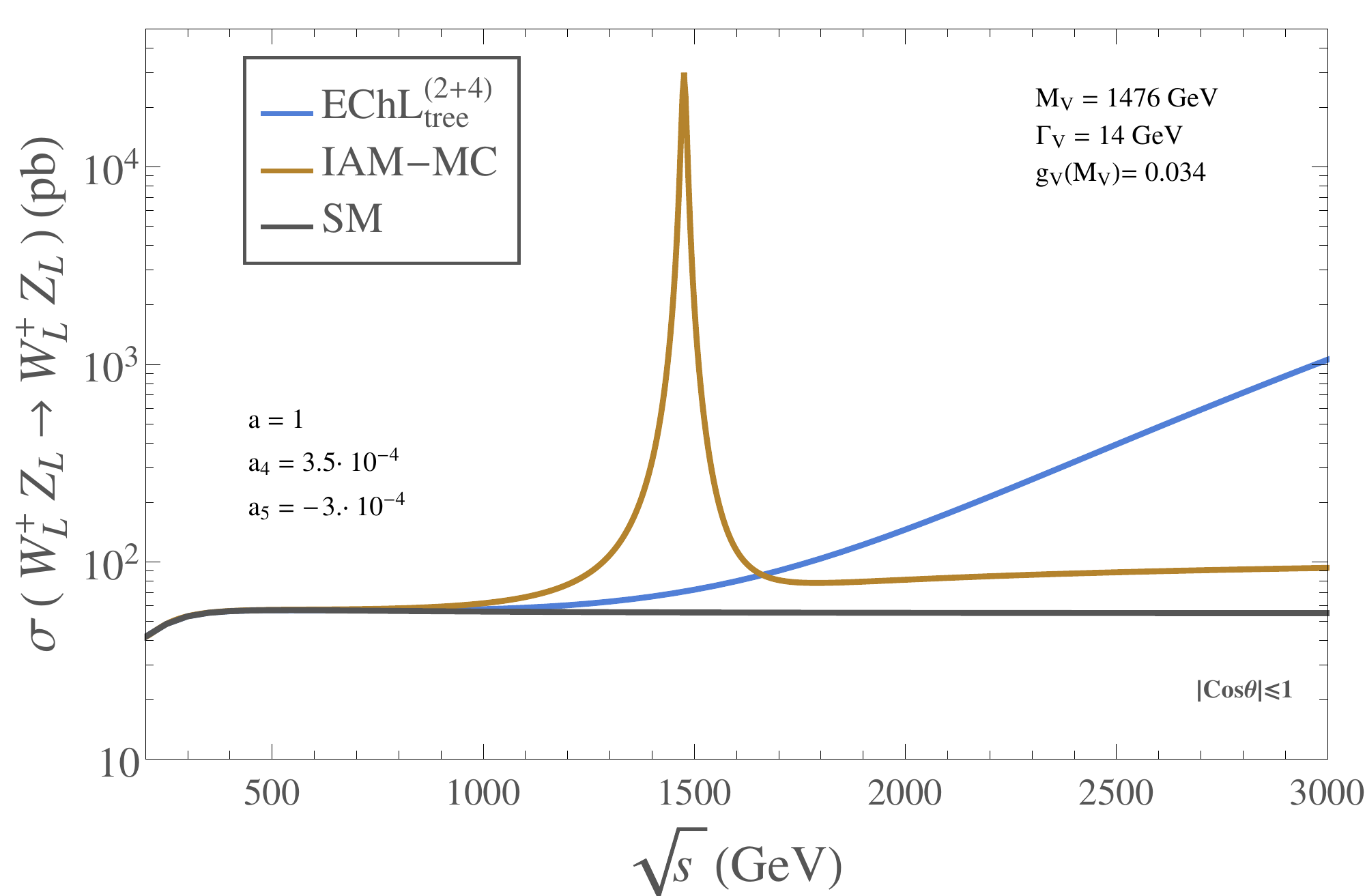}
\includegraphics[width=.49\textwidth]{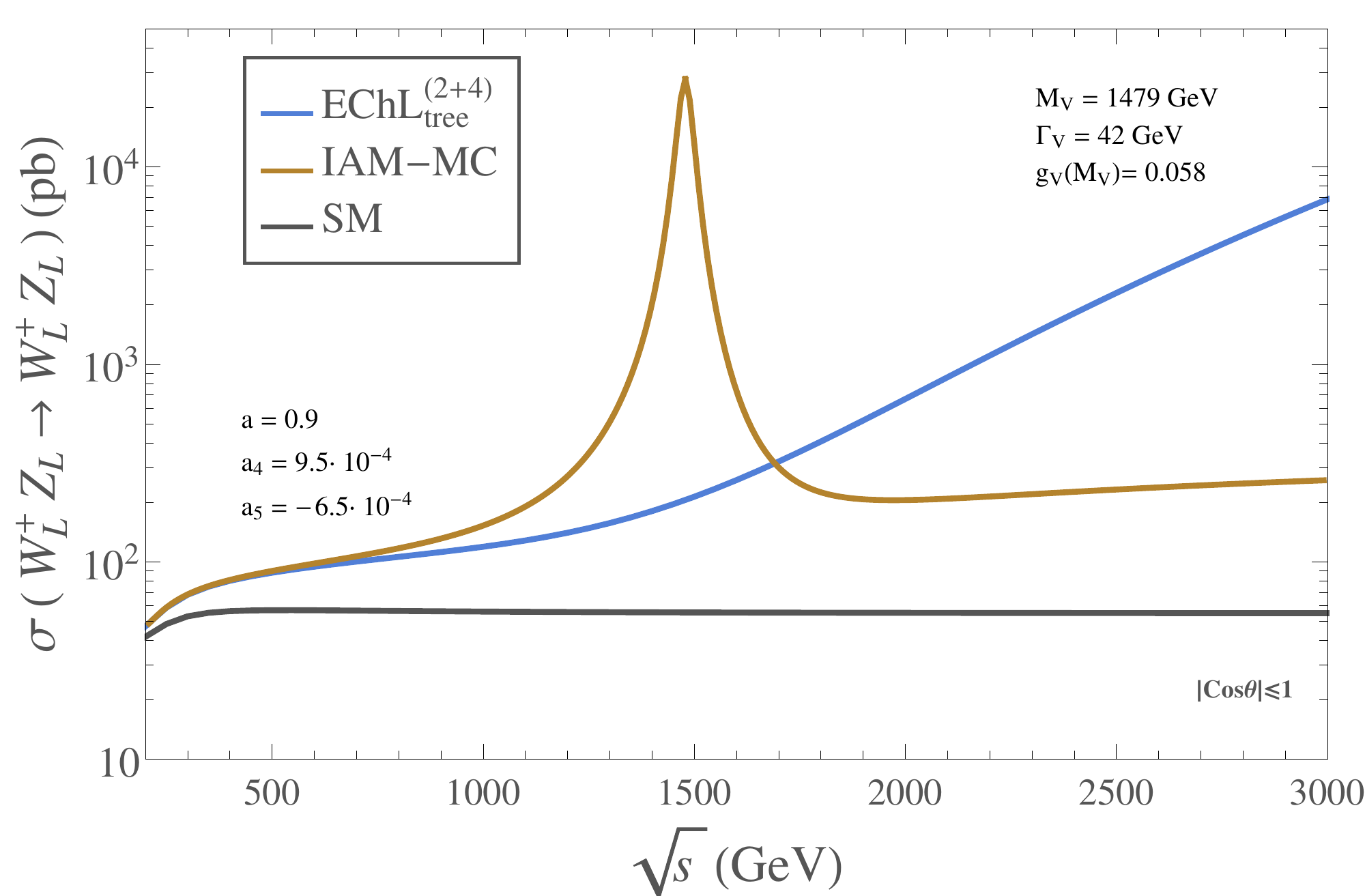}\\
\includegraphics[width=.49\textwidth]{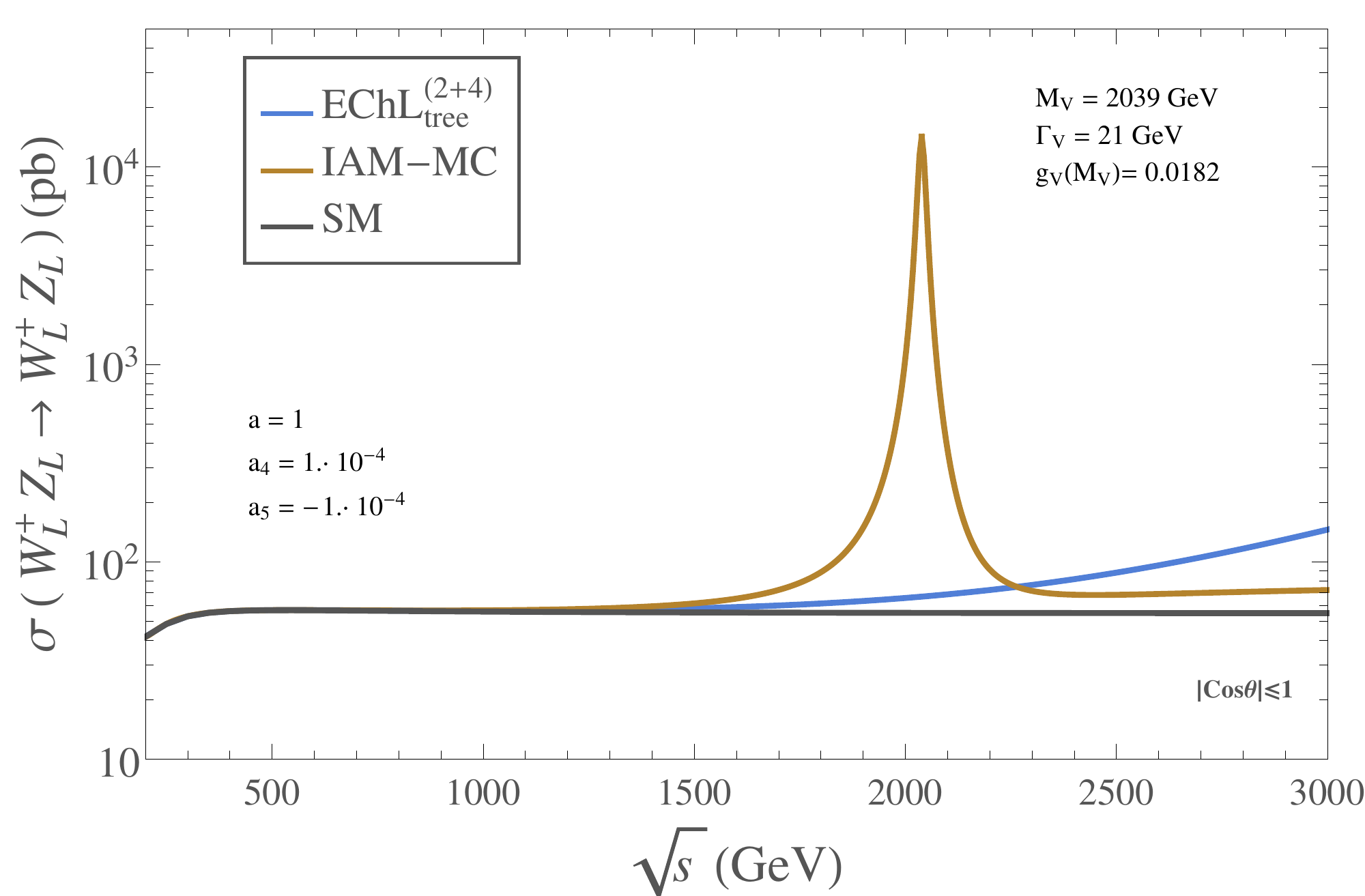}
\includegraphics[width=.49\textwidth]{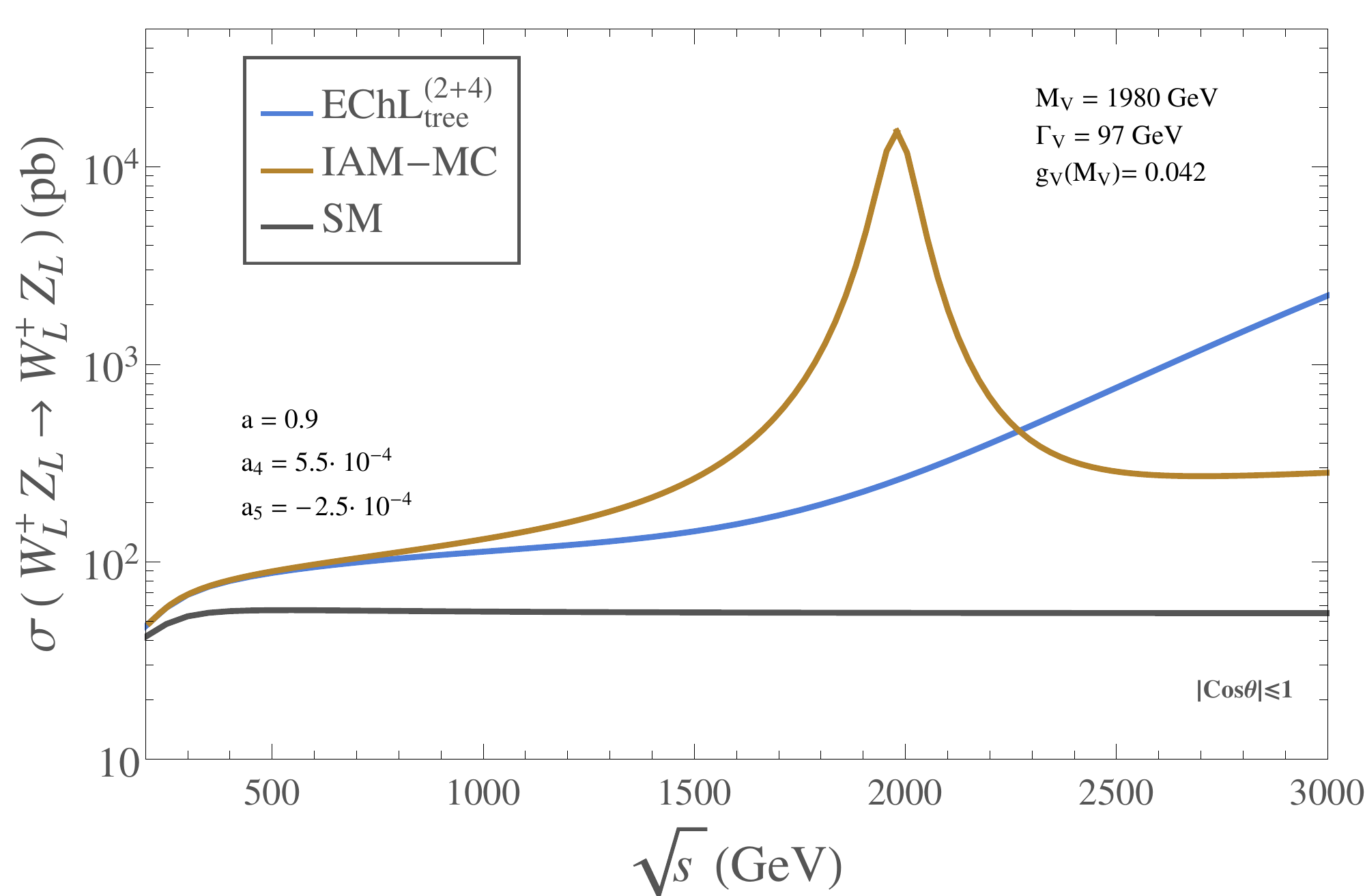}\\
\includegraphics[width=.49\textwidth]{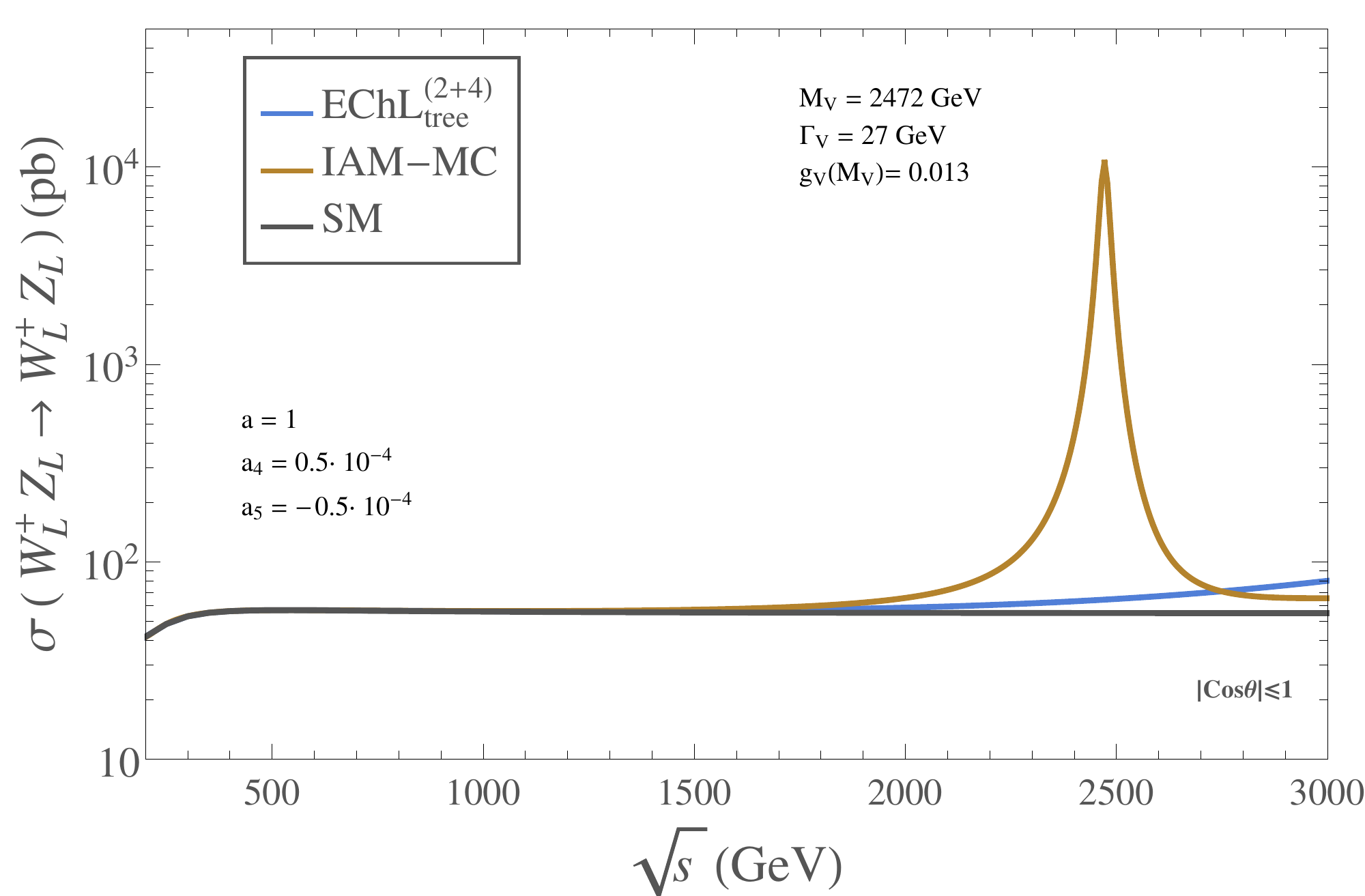}
\includegraphics[width=.49\textwidth]{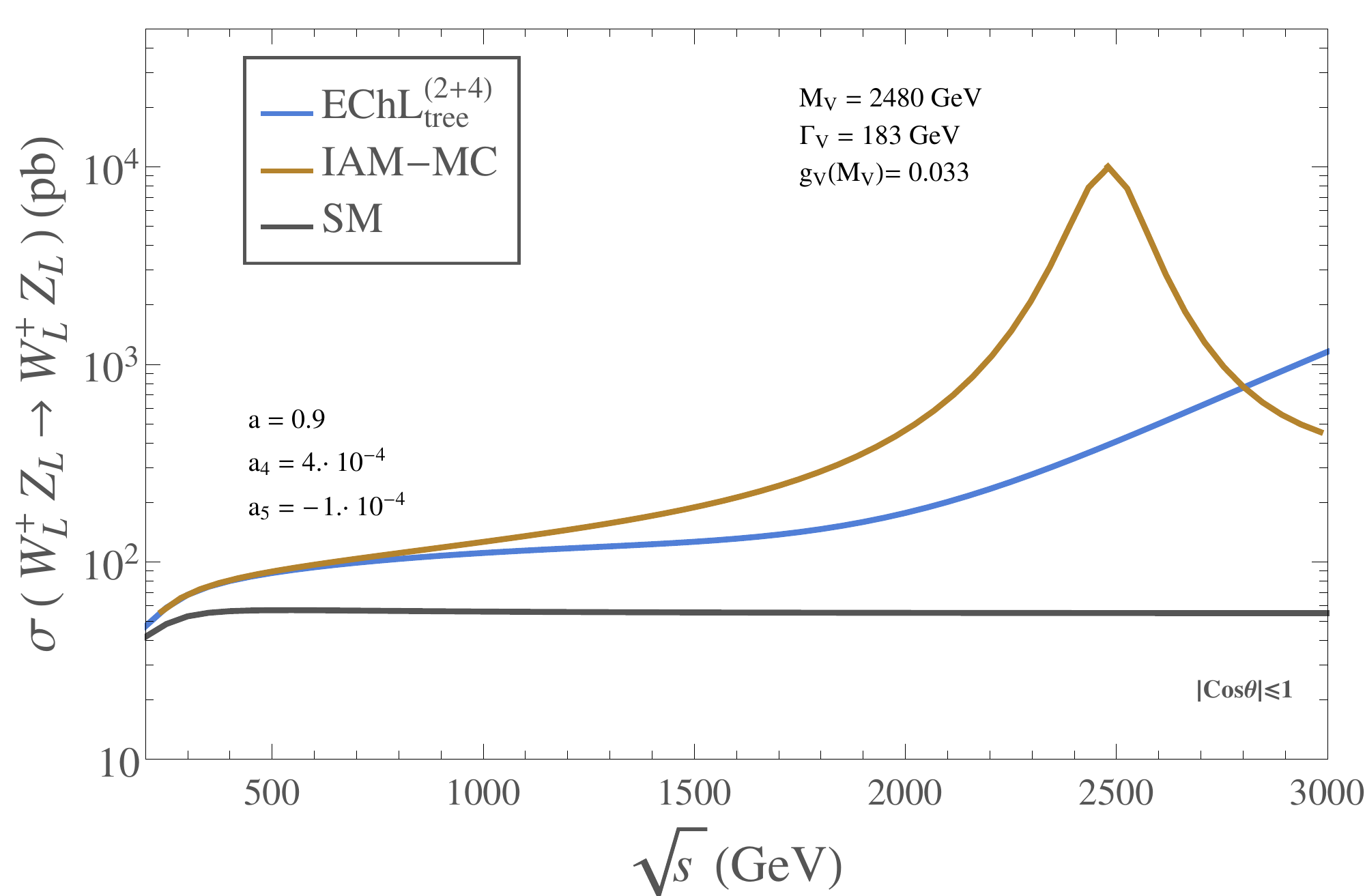}\\
\caption{Predictions of the cross section $\sigma(W^+_LZ_L\to W^+_LZ_L)$ as a function of the center of mass energy $\sqrt{s}$ for all the selected benchmark points in \tabref{tablaBMP} integrated over the whole center of mass scattering angle, $|\cos\theta|\leq1$.
Different lines correspond to the different models considered in the text:
SM (black), our IAM-MC model (orange) and non-unitarized  EChL up to $\mO(p^4)$ (blue).}
\label{fig:xsec1}
\end{center}
\end{figure}

For the numerical computation that is relevant for the forthcoming study of the LHC events we will not use the decomposition in  partial waves, but the complete amplitude instead. This is an important point, since a description of
$\sigma(W_LZ_L \to W_LZ_L)$ in terms of only the lowest partial waves would not give a realistic result for energies away from the resonant region, which we have checked explicitly. Therefore, before starting the analysis of the LHC events, it is convenient to learn first about the predictions of the cross section at the $WZ \to WZ$ subprocess level. Thus, we present in \figref{fig:xsec1} our numerical results for $\sigma(W_LZ_L \to W_LZ_L)$  within our IAM-MC framework and for the same benchmark points of \tabref{tablaBMP}. In these plots we have also included the predictions from the SM and from the EChL for comparison. What we learn from these figures is immediate: the vector resonances do emerge clearly in the scattering of the longitudinal modes, well above the SM background. We also see that the predictions from the IAM-MC match those from the EChL at low energies, as expected. The main features of the resonances, i.e., the mass, the width and the coupling are obviously manifested in each profile of the resonant IAM-MC lines.  It is also worth mentioning our explicit test that all these cross sections in \figref{fig:xsec1} respect the Froissart unitary bound in \eqref{froisbound}.

So far we have been discussing about the predictions of the scattering amplitudes for the longitudinal gauge boson modes. However, for a realistic study with applications to LHC physics, as we will do in the next section, we must explore also the behavior of the scattering of the transverse modes. In fact, the transverse $W_T$ and $Z_T$ gauge bosons are dominantly radiated from the initial quarks at the LHC, as compared to the longitudinal ones and, consequently, they will be relevant and have to be taken into account in the full computation. Of course we will make our predictions at the LHC taking into account all the polarization channels as it must be.

\begin{figure}[t!]
\begin{center}
\includegraphics[width=.49\textwidth]{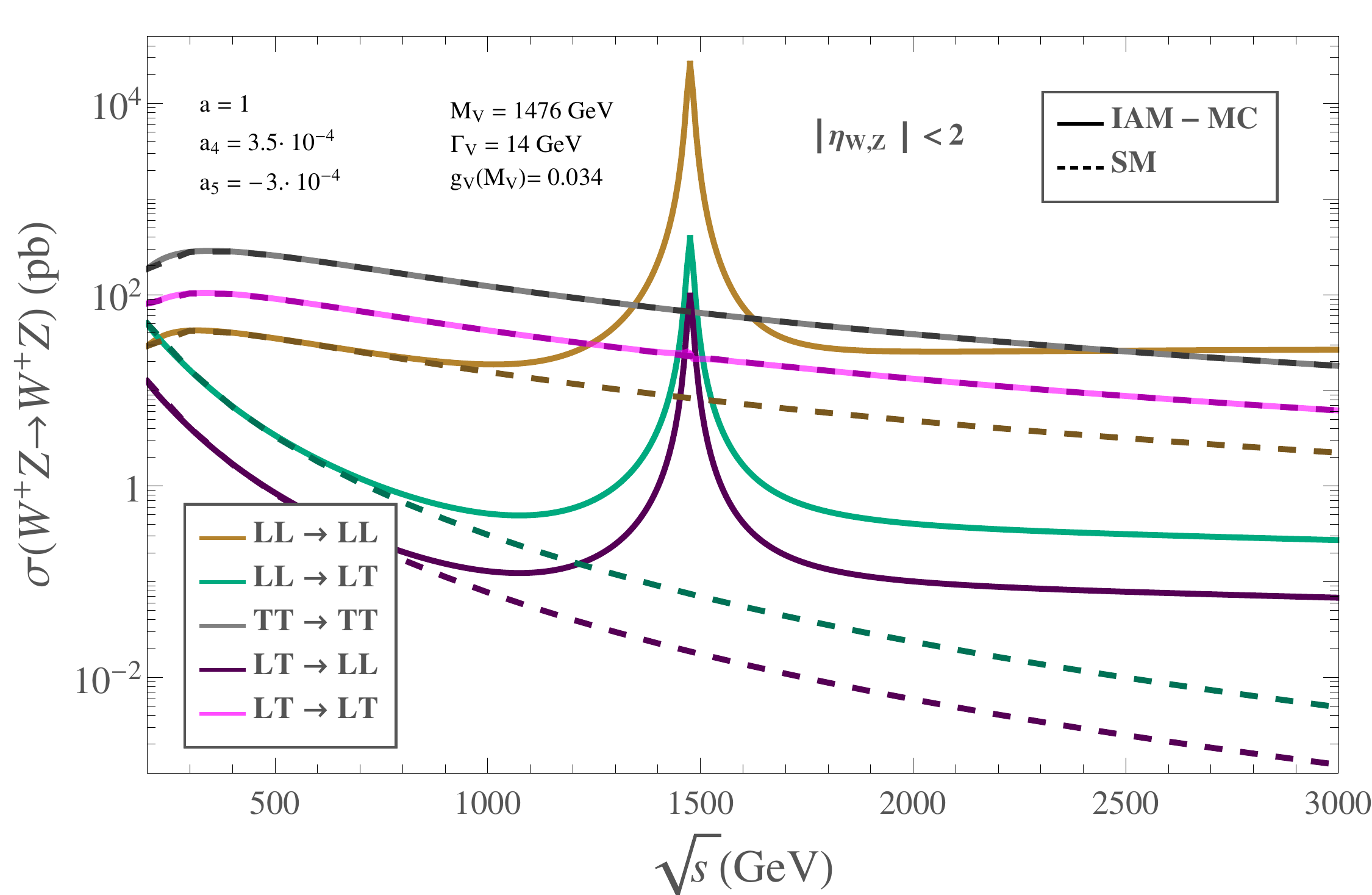}
\includegraphics[width=.49\textwidth]{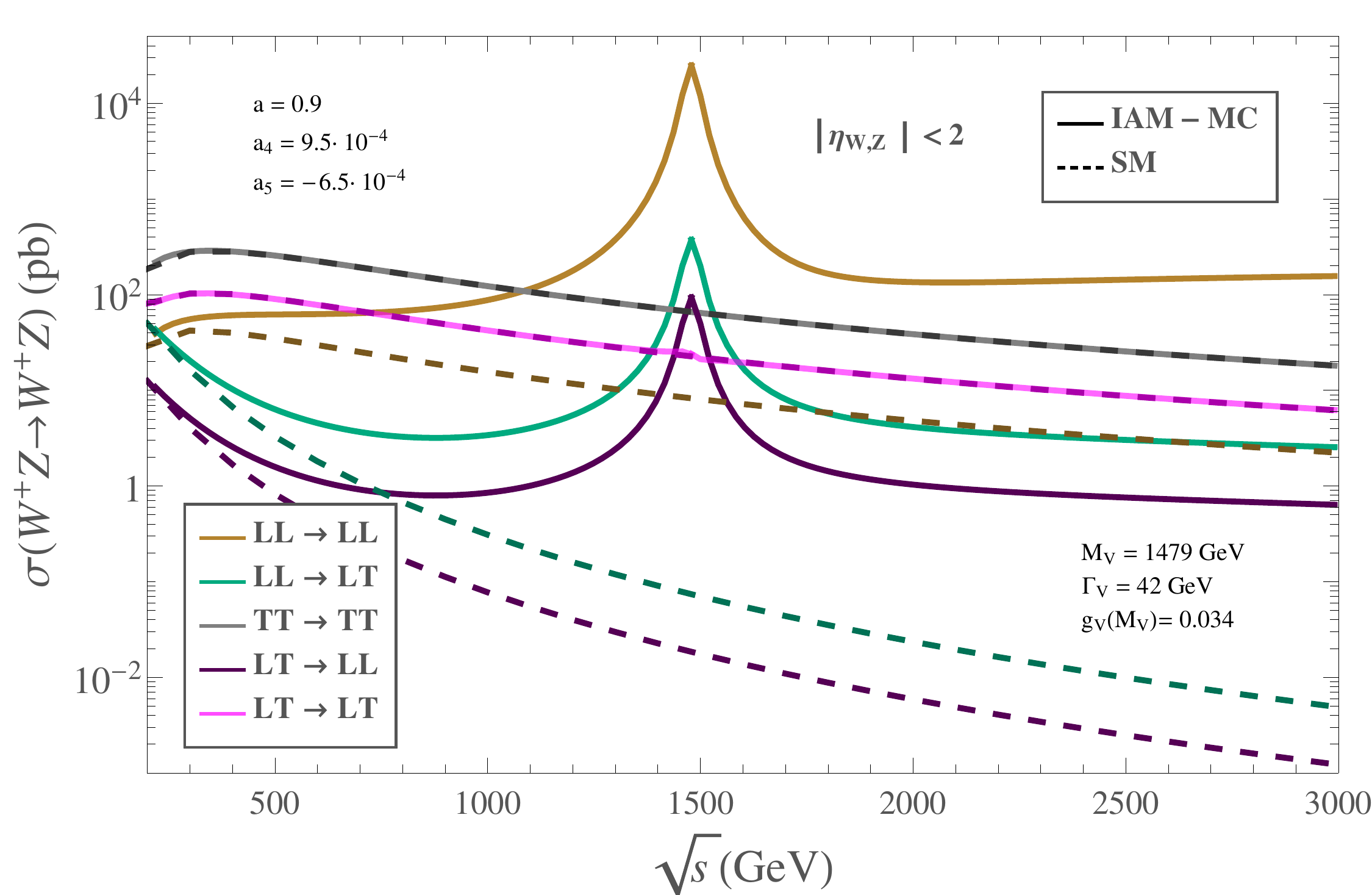}\\
\caption{Cross section $\sigma(W^+Z\to W^+ Z)$ as a function of the center of mass energy $\sqrt{s}$ for  the most relevant polarization channels and for the two selected benchmark points, BP1 (left panel) and BP1' (right panel). Results were obtained imposing a cut on the center of mass scattering angle that corresponds to $|\eta_{W,Z}|<2$. This cut will be used as a detector acceptance cut in the LHC process. Solid lines are the predictions from our IAM-MC model and dashed lines are the predictions from the SM.}
\label{fig:IAM-MC-polarization}
\end{center}
\end{figure}

To compute the various amplitudes $A(W_A Z_B \to W_C W_D)$ with all the polarization possibilities for $A,B,C,D$ being either $L$ or $T$, we proceed as described above for the case of the longitudinal modes. We use the same analytical results for the amplitudes given in the appendices in terms of the generic polarization vectors and substitute there the proper polarization vectors according to the corresponding $L$ or $T$ cases. The numerical results of the cross sections $\sigma(W_A Z_B \to W_C W_D)$ for the most relevant polarizations channels are presented in \figref{fig:IAM-MC-polarization} for the two
benchmark points BP1 and BP1' that we have chosen as illustrative examples. We have also included the corresponding predictions of the cross sections in the SM for comparison. All these results have been computed with FeynArts and FormCalc, and have been checked with MadGraph5.

Regarding this  \figref{fig:IAM-MC-polarization}, one can confirm that at the subprocess level, $WZ \to WZ$,  the scattering of longitudinal modes in our IAM-MC model clearly dominates over the other polarization channels in the region surrounding the resonance. This is  in contrast with the SM case, where the $TT\to TT$ channel dominates by far in the whole energy region studied .  This feature of the IAM-MC was indeed expected since, as already said, the coupling $g_V$ affects mainly to the longitudinal modes. Secondly, the predictions of the resonant peaks in the IAM-MC are clearly above the SM background in all the polarization channels that resonate. Thirdly, we also learn that the $LL\to LL$ channel is not the only one that resonates. In fact, also the $LL\to LT$, $LT\to LL$ and $LT\to LT$ channels  manifest a resonant behavior (barely appreciated in the figure in the $LT\to LT$ case) in the IAM-MC, although with  much lower cross sections at the peak than the dominant $LL\to LL$ channel. In these examples the hierarchy found in the IAM-MC predictions at the peak is the following:
\be
\sigma(LL \to LL) \gg \sigma(LL \to LT) > \sigma(LT \to LL)> \sigma(TT \to TT)
> \sigma(LT \to LT) ,
\ee
where $\sigma(AB\to CD)$ is short-hand notation for $\sigma(W_A Z_B\to W_C Z_D)$, and where $LT$ corresponds to $W_LZ_T+W_TZ_L$. Also from \figref{fig:IAM-MC-polarization} one can see that $\sigma(LL\to LT)$ is approximately two orders of magnitude smaller than  $\sigma(LL\to LL)$. Therefore, we conclude that the main features found previously for the $\sigma(W_LZ_L\to W_LZ_L)_{\rm IAM-MC}$, in the region close to the resonance, should emerge in the total cross section, $\sigma(WZ \to WZ)_{\rm{IAM-MC}}$, given the fact that this channel is by far the domminant one. This will be confirmed in the next section.
We would like to mention that all the plots presented in this section have been done with FormCalc and checked with MadGraph5.

\section{Production and sensitivity to vector resonances in  $\boldsymbol{pp\to WZjj}$ events at the LHC}
\label{LHC}

\begin{figure}[t!]
\begin{center}
\includegraphics[width=0.7\textwidth]{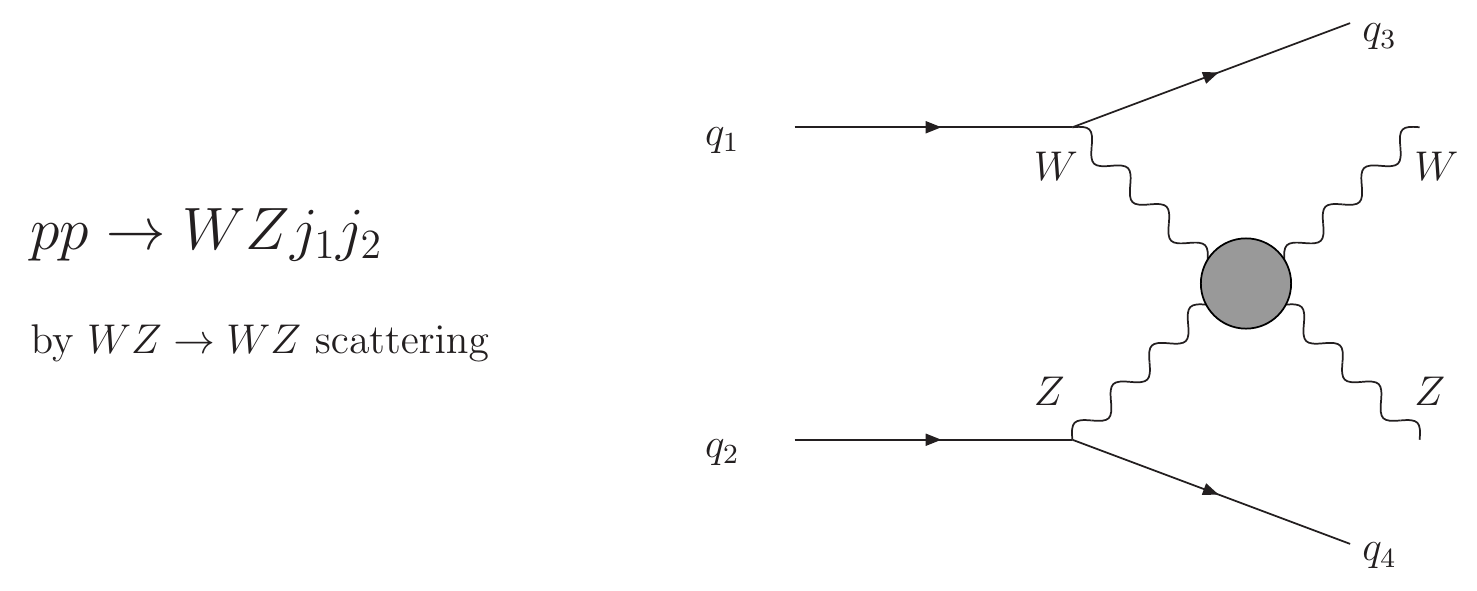}
\caption{Graphical representation of the $pp \to WZjj$ process at the LHC, at the parton level,  by means of $WZ \to WZ$ scattering. The initial $W$ and $Z$ gauge bosons are radiated from the constituents quarks of the protons and are generically virtual particles which re-scatter to produce the final $W$ and $Z$.}
\label{fig:diagppWZjj}
\end{center}
\end{figure}

The process that we wish to explore here is $pp\to WZjj$ at the LHC via the 
 VBS subprocess $WZ \to WZ$, as generically depicted in  \figref{fig:diagppWZjj}. Concretely, we select the process
 with $W^+$ instead of $W^-$ since the former is  more copiously produced from
the initial protons.
However, these type of events containing two gauge bosons $W^+$ and $Z$ and two jets in the final state can happen at the LHC in many different ways, not only by means of VBS.  Therefore, in order to be able to select efficiently these VBS mediated processes, one has to perform the proper optimal cuts in the kinematical variables of the outgoing particles of the collision. These cuts should favor the VBS configuration versus other competing processes. Thus, we are going first to specify our selection of these VBS cuts in terms of the kinematical variables of the two final jets and the final $W^+$ and $Z$ gauge bosons.

There are many studies in the literature searching for these optimal VBS cuts (see, for instance, Refs.~\cite{Haywood:1999qg, Doroba:2012pd, Szleper:2014xxa, Fabbrichesi:2015hsa, Aad:2016ett}) and where different kinematical variables like transverse momenta, pseudorapidities, and invariant masses of the final particles have been considered. The common feature explored by all these studies is the generic topology showed in these type of VBS mediated events, which  have two opposite-sided large pseudorapidity jets together with two gauge bosons, $W^+$ and $Z$ in our case, within the acceptance of the LHC detectors. This is in contrast to pure QCD events which produce mainly jets in the low pseudorapidity region.

For the present work, we have first selected the cuts in the pseudorapidities of the final jets, $j_1,j_2$, and of the final  $W^+,Z$ gauge bosons by giving the following basic VBS cuts: $|\eta_{j_1,j_2}|<5\,,~\eta_{j_1} \cdot \eta_{j_2} < 0\,,~p_T^{j_1,j_2}>20 ~{\rm GeV} \,,~|\eta_{W,Z}| < 2, $ of Ref.~\cite{Doroba:2012pd}. For all the results and plots presented in this section we use MadGraph5, and set the LHC energy to 14 TeV.  For the parton distribution functions we set the option
NNPDF2.3~\cite{Ball:2013hta}. The results from our IAM-MC model, which has been described in the previous section, are generated by means of a specific UFO file that contains the model and the  needed four point function
$\Gamma_{WZWZ}^{\rm IAM-MC}$ of the blob represented in  \figref{fig:diagppWZjj},  whose analytical result is also collected in the appendices in terms of the IAM-MC model parameters,
see \eqrefs{fpfUFOMJ}-(\ref{fpfUFOparts}). This four point function has obviously momentum dependence and is treated by MadGraph5  as an effective four point vertex  which is then used  by the MonteCarlo to generate the signal events that we are interested in. With the simplifications assumed in this work, the IAM-MC parameters contained in the UFO file are basically the chiral coefficient $a$ and the vector resonance parameters $M_V$, $\Gamma_V$ and $g_V(M_V)$, which are fixed from the given  input values of $a$, $a_4$ and $a_5$ accordingly to our previous discussion. Concretely, we use the selected points in \figref{fig:contourMW} to make our predictions with MadGraph5 of the signal events at the LHC from the IAM-MC model.

\subsection{Study of the most relevant backgrounds}

Regarding the background events from the SM we also generate them with MadGraph5. We only consider here the main irreducible $WZjj$ backgrounds  since we are assuming that the final $W$ and $Z$ gauge bosons can be reasonably identified and disentangled from pure QCD (${\cal O}(\alpha_S^n)$) events leading to fake `$WZjj$' configurations. For the same reason, we do not consider either the potential backgrounds from top quarks production and decays. This will be totally justified in the final part of this study where we will focus on the leptonic decays of the final $W$ and $Z$ leading to a very clear signal with three leptons, two jets and missing energy in the final state and with very distinct kinematics.
We therefore focus here on the two main irreducible SM backgrounds:
\begin{itemize}
\item[1)] The pure SM-EW background, from parton level amplitudes $A(q_1 q_2  \to q_3 q_4 WZ)$  of order
${\cal O}(\alpha^2)$.
\item[2)] The mixed SM-QCDEW background, from parton level amplitudes   $A(q_1 q_2  \to q_3 q_4 WZ)$  of order
${\cal O}(\alpha \alpha_S)$.
\end{itemize}

\begin{figure}[t!]
\begin{center}
\includegraphics[width=.49\textwidth]{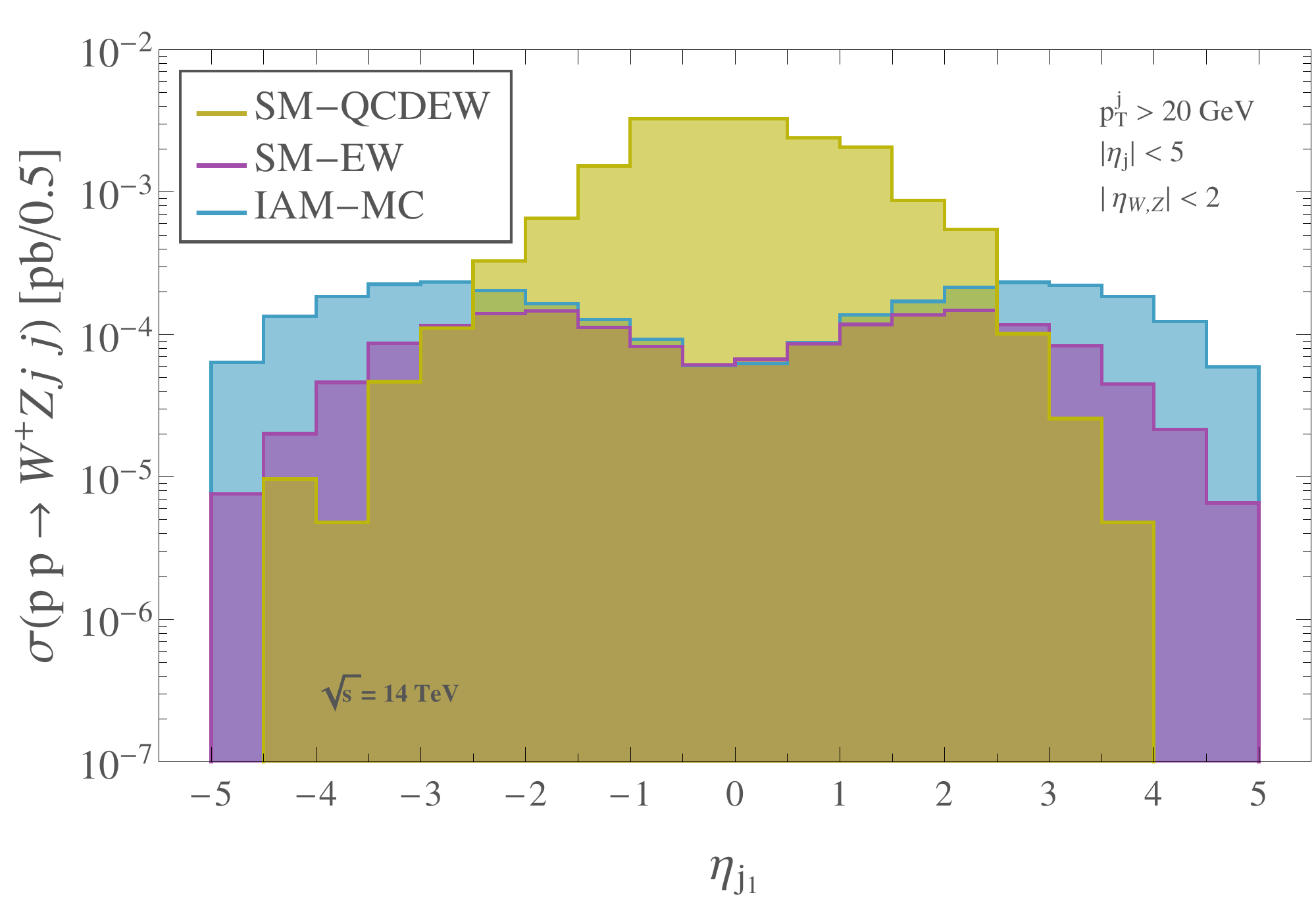}
\includegraphics[width=.49\textwidth]{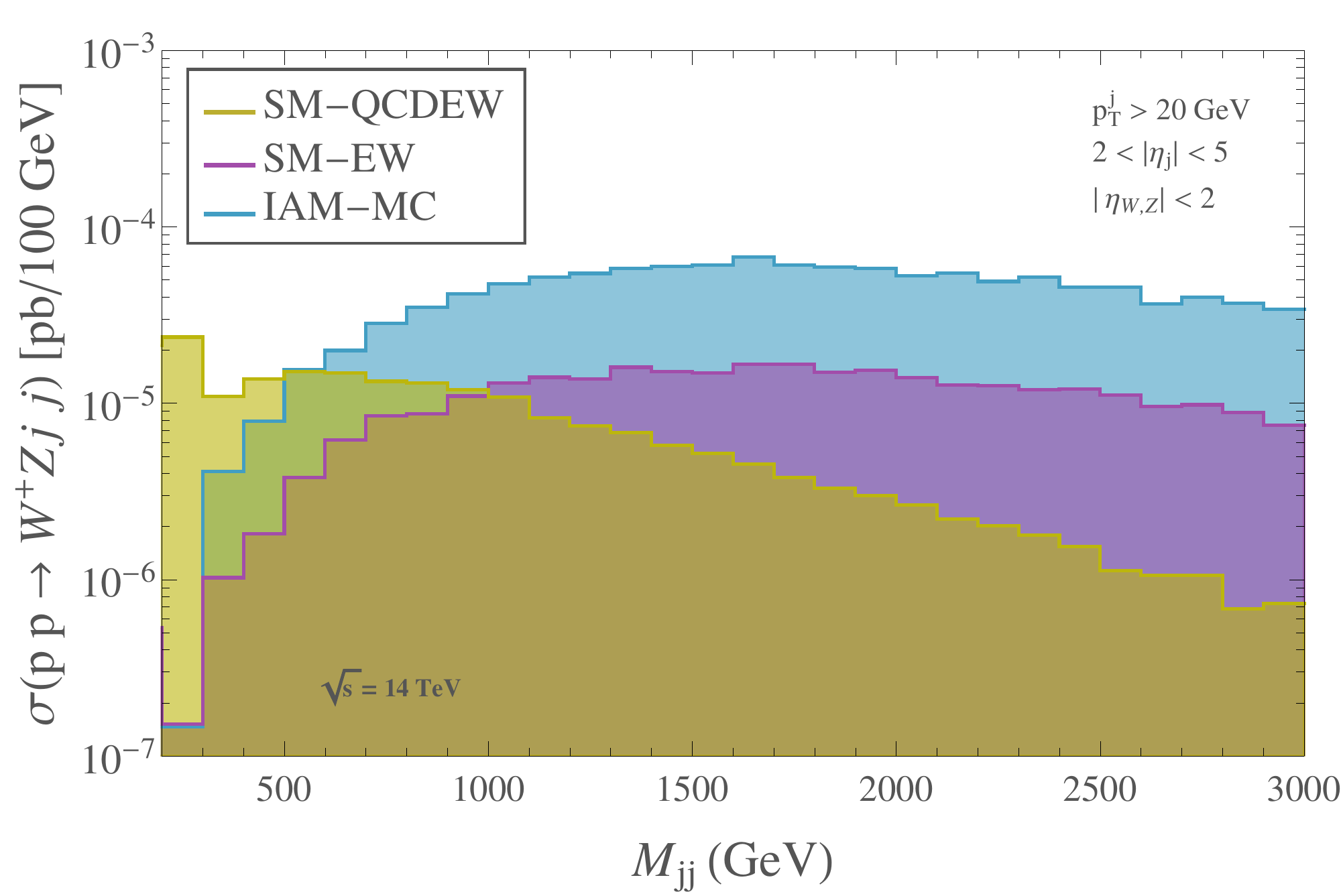}\\
\caption{$\sigma(pp \to W^+Zjj)$ distributions with the pseudorapidity of the outgoing jet $\eta_{j_1}$ (left panel) and with the invariant mass of the final jet pair $M_{jj}$ (right panel).
The predictions for the IAM-MC signal for the selected BP1' scenario (blue) and the two main SM backgrounds, SM-QCDEW (yellow) and SM-EW (purple), are shown separately.}
\label{fig:etajmjj}
\end{center}
\end{figure}

 We show our predictions of the IAM-MC signal for the selected BP1' scenario together with those of the two main irreducible SM-EW and SM-QCDEW backgrounds in   \figref{fig:etajmjj}, for the simple VBS cuts specified in the figure. The selected distributions for this signal versus background comparison are the final jet pseudorapidity, $\eta_{j_1}$ (with $j_1$ being the most energetic jet),  and the invariant mass of the two final jets, $M_{jj}$.
As we can clearly see in this figure, the signal is mainly produced in the interval $2<|\eta_{j_1}|<5$ and with a rather large jet invariant mass of $M_{jj} > 500$ GeV, whereas the SM-QCDEW background is mainly centrally produced, with $|\eta_{j_1}|  <2$ and at lower invariant masses $M_{jj} <500$ GeV. Therefore, this suggests our more refined selection of cuts for discriminating the IAM-MC signal from the SM-QCDEW background given by the following optimal VBS cuts:
 \begin{eqnarray}
&2<|\eta_{j_1,j_2}|<5\,,  \nn \\
&\eta_{j_1} \cdot \eta_{j_2} < 0, \nn \\
&p_T^{j_1,j_2}>20 ~{\rm GeV} \,, \nn \\
&M_{jj}>500\,\,{\rm GeV}\,, \nn \\
&|\eta_{W,Z}| < 2\,.
\label{optimal-cuts}
\end{eqnarray}
\begin{figure}[t!]
\begin{center}
\includegraphics[width=.49\textwidth]{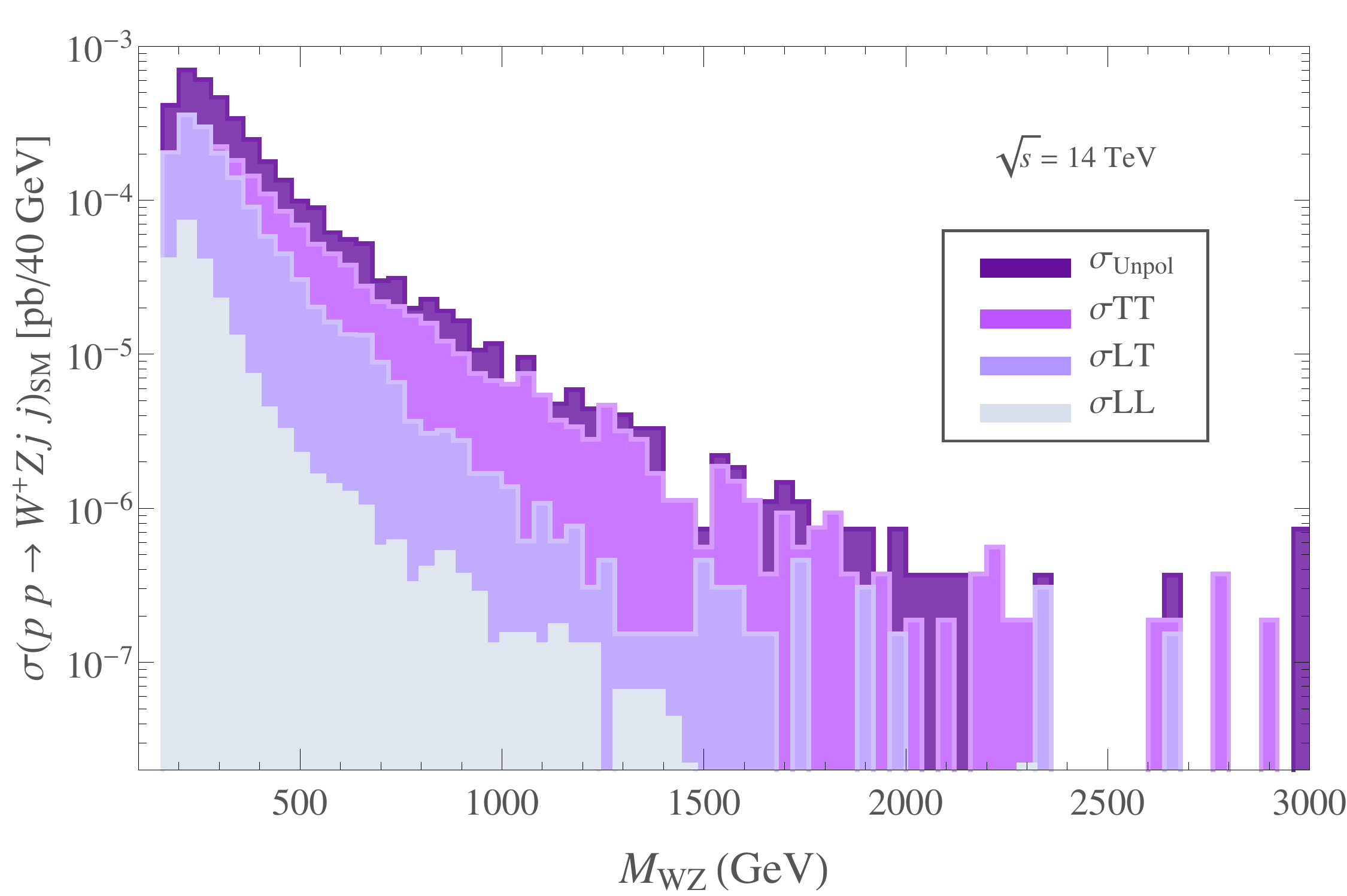}
\includegraphics[width=.49\textwidth]{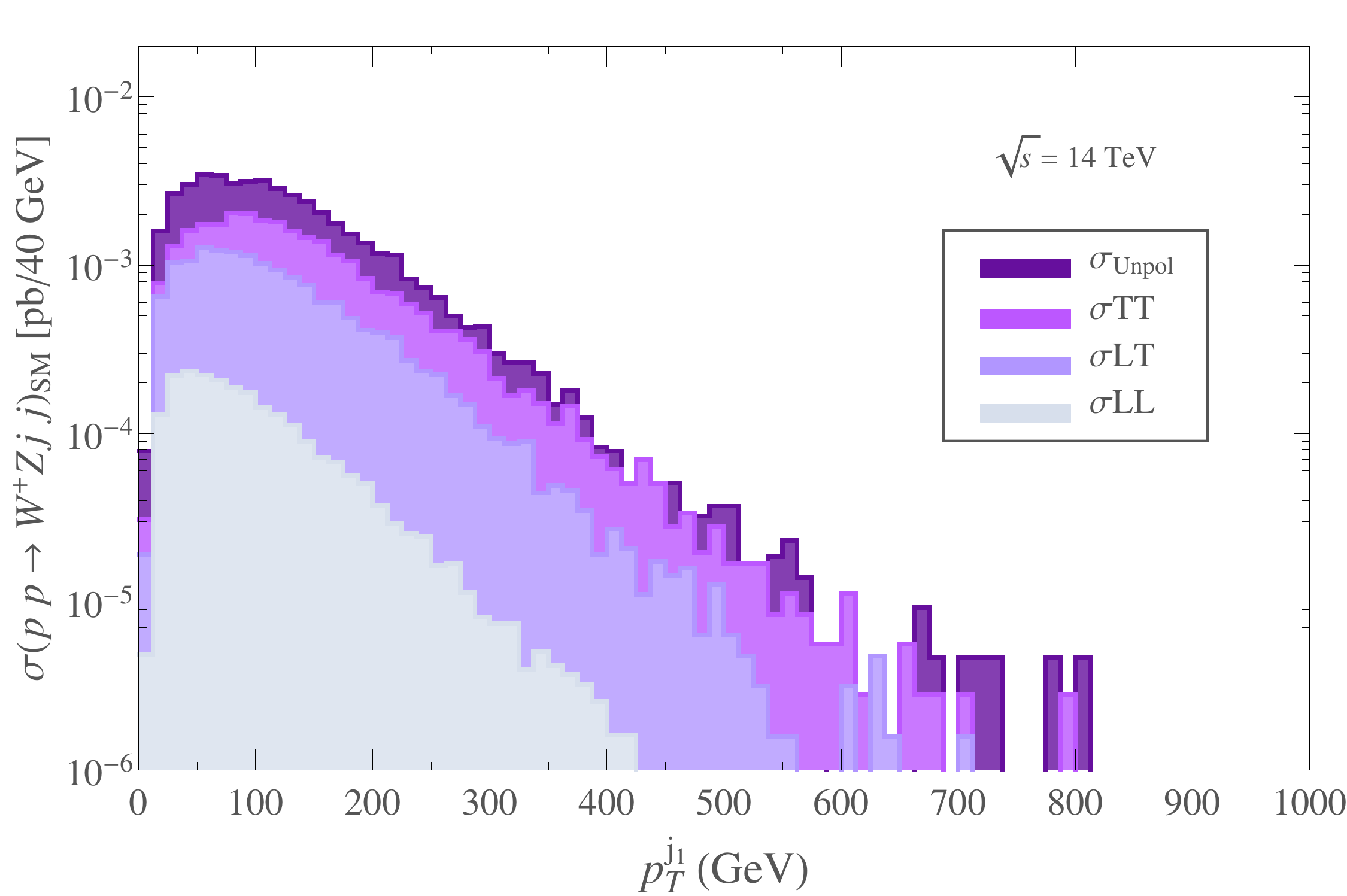}\\
\caption{$\sigma(pp \to W^+Zjj)$ distributions of the SM-EW background with the invariant mass of the $WZ$ pair, $M_{WZ}$ (left panel) and with the transverse momentum of the most energetic jet, $p_T^{j_1}$ (right panel). The imposed cuts are $|\eta_{j_1,j_2}|<5\,,~\eta_{j_1} \cdot \eta_{j_2} < 0\, {\rm and}~|\eta_{W,Z}| < 2$. 
The predictions for the various polarizations $\sigma_{AB}$ of the final $W_AZ_B$ pair as well as the total unpolarized, $\sigma_{\rm Unpol}$, result are displayed separately, for comparison.
Starting from the upper to the lower lines they correspond respectively to: $\sigma_{\rm Unpol}$, $\sigma_{\rm TT}$, $\sigma_{\rm LT}$ and $\sigma_{\rm LL}$.}
\label{fig:SM-EW}
\end{center}
\end{figure}
Regarding the SM-EW background, as we can see in   \figref{fig:etajmjj}, it has very similar kinematics with respect to our IAM-MC signal in these two jet variables $\eta_{j_1}$ and  $M_{jj}$. This was expected, since, after applying the basic VBS cuts, both receive dominant contributions from the VBS kind of configurations. In order to disentangle our signal from this SM-EW background one has to rely on additional discriminants. As suggested by our previous analysis in section~\ref{sec-model},  the most powerful of these discriminants would be a devoted study of the final gauge boson polarizations, since the IAM-MC signal produces mainly $W_LZ_Ljj$ events whereas the SM-EW background produces mainly $W_TZ_Tjj$ events.
This latter case can be clearly seen in our results in \figref{fig:SM-EW},  where we show the separated predictions of the SM-EW backgrounds for the various polarizations of the final gauge bosons, $W_LZ_Ljj$,  $W_LZ_Tjj$+$W_TZ_Ljj$ and $W_TZ_Tjj$. Both distributions, the one in the invariant mass of the $WZ$ pair, $M_{WZ}$, and the one in the transverse momentum of the most energetic final jet, $p_T^{j_1}$, show the clear dominance of the $W_TZ_Tjj$ type of events in this SM-EW background.  This was expected, since as shown in \figref{fig:xsWZWZSM}, the polarizations are practically preserved in the SM, and these background $W_TZ_Tjj$ events are basically mediated by $W_T Z_T \to W_T Z_T$, which is the dominant VBS SM channel. We also see in \figref{fig:SM-EW} that the $p_T^{j_1}$ distribution of these  SM-EW background events peaks towards lower values in  $p_T^{j_1}$  in the  $W_LZ_Ljj$ events than in the $W_TZ_Tjj$ events.
This can be understood by the fact that longitudinally polarized vector bosons tend to be emitted at a smaller angle with respect to the beam, and hence smaller transverse momentum, with respect to the incoming quark direction than the transversely polarized ones. As a consequence, the final quark (and thus the final jet) accompanying a longitudinal gauge boson is more forward than the one accompanying a transverse $W$ or $Z$. This translates into  different $p_T^j$ distributions. Whereas the ones coming from events with transverse gauge bosons tend to peak closer to the EW boson mass, the ones with longitudinally polarized $W$ or $Z$ peak normally around half of the EW boson mass.

These features are very interesting regarding future prospects of polarization studies. As we have argued, being able to disentangle the polarization of the gauge bosons in the final state will be enormously helpful to discriminate signal versus background in these scenarios. Indeed, a more detailed study of the relevant kinematical variables to perform this kind of discrimination deserves some future development, although there are already some analysis in this direction, see for instance
Ref.~\cite{Fabbrichesi:2015hsa}. However, as sophisticated techniques to distinguish among the polarizations of the final $W$ and $Z$ are not yet well stablished, we are not going to use a polarization analysis as a discriminant in this work. We prefer to leave this issue for a forthcoming work. Thus, we will rely in the following in the most obvious and simple way to discriminate the IAM-MC signal and the SM backgrounds, which is by looking for resonant peaks in the $M_{WZ}$ invariant mass distributions of the unpolarized cross sections.

\subsection{Results for the resonant signal events}
 \begin{figure}[t!]
\begin{center}
\includegraphics[width=.49\textwidth]{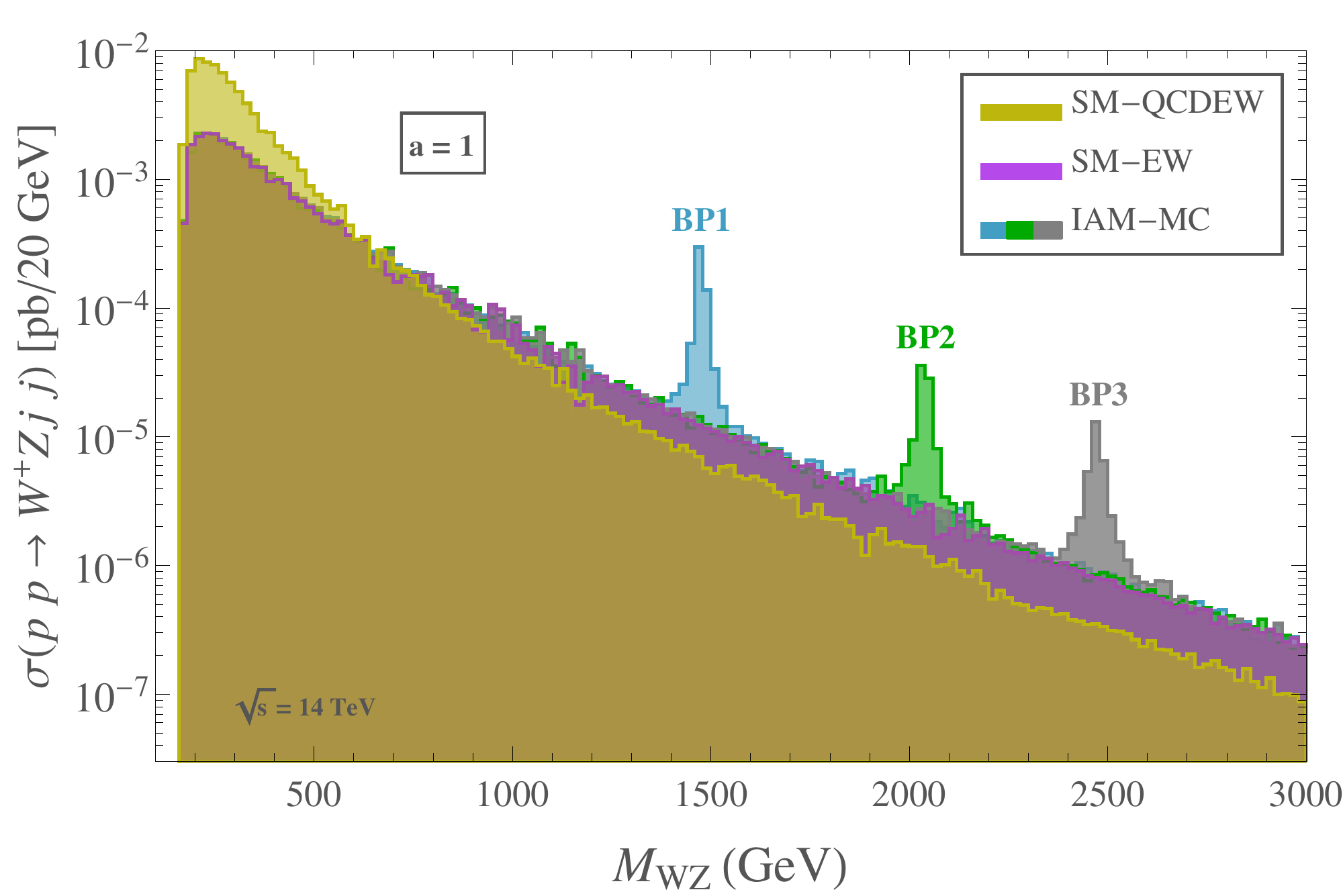}
\includegraphics[width=.49\textwidth]{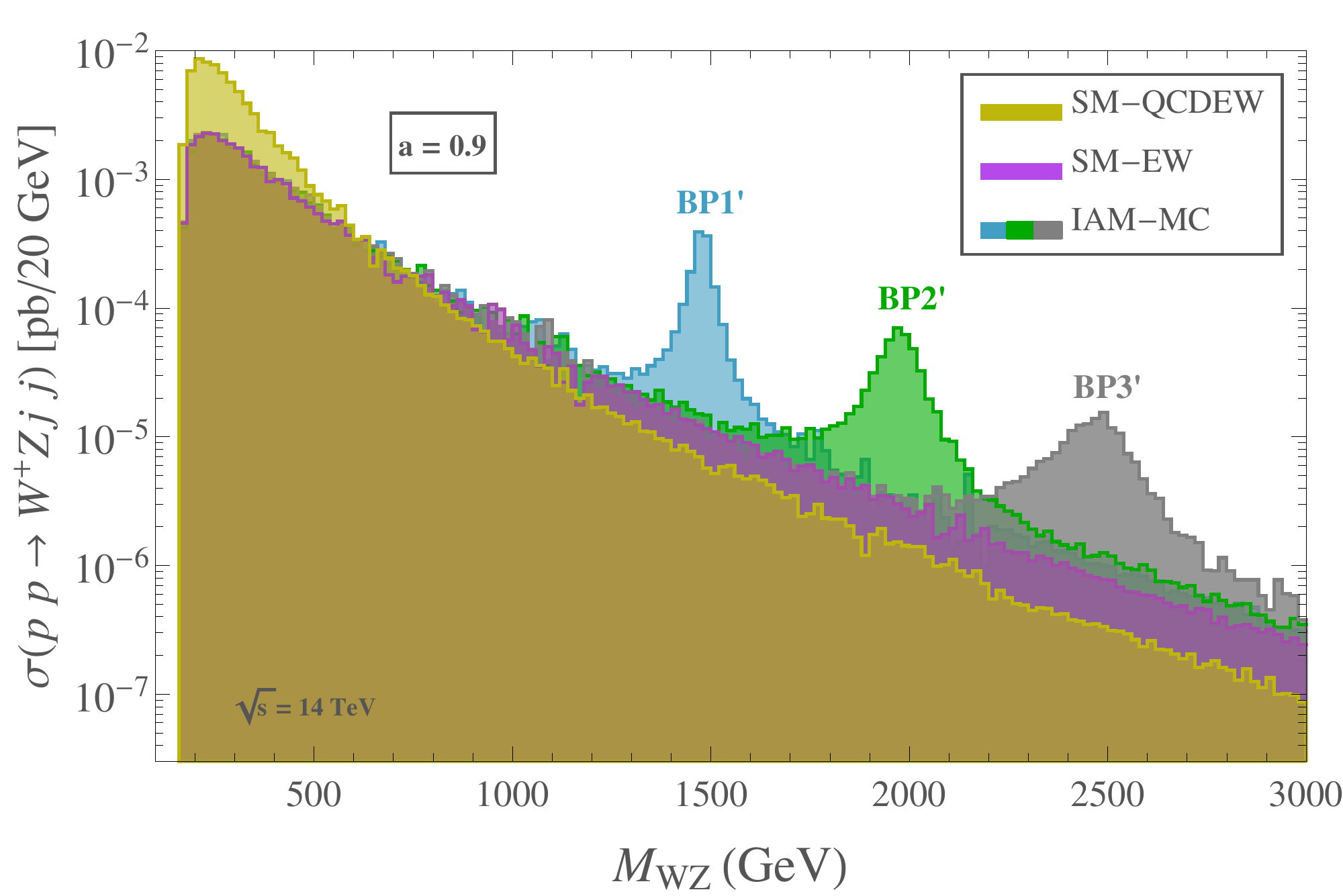}\\
\caption{Predictions of the $\sigma(pp \to W^+Zjj)$ distributions with the invariant mass of the  $WZ$ pair, $M_{WZ}$, for the benchmark points of the IAM-MC model BP1 (blue), BP2 (green), BP3 (gray) in the left panel and BP1' (blue), BP2' (green), BP3'  (gray) in the right panel, and of the two main SM backgrounds, SM-QCDEW (yellow) and SM-EW (purple). The cuts in \eqref{optimal-cuts} have been applied.}
\label{fig:WZdistributions}
\end{center}
\end{figure}

In this subsection we present the main results of our IAM-MC resonant signal events together and compared with the relevant backgrounds explored previously. Our predictions of the above mentioned $M_{WZ}$ distributions for the IAM-MC signal and of the two main SM backgrounds, SM-QCDEW and SM-EW, are displayed in \figref{fig:WZdistributions}.   We have summarized in these plots the results for all the selected benchmark points in  \tabref{tablaBMP}, after applying the optimal cuts in \eqref{optimal-cuts}. We see in these figures that the resonant peaks, coming mainly from the interaction of longitudinally polarized gauge bosons, clearly emerge above the SM backgrounds (dominated by the transverse modes) in all these distributions and in all the studied BP scenarios. In order to quantify the statistical significance of these emergent peaks, we define $\significance_{WZ}$ in terms of the predicted events in our IAM-MC model,
${\rm N}(pp\to W^+Zjj)^{\footnotesize \rm IAM-MC}$, and the background events, ${\rm N}(pp\to W^+Zjj)^{\footnotesize \rm SM}$,  as follows:
\be
\significance_{WZ} =\frac{S_{WZ}}{\sqrt{B_{WZ}}}\,,
\label{SS-WZ}
\ee
with,
\begin{align}
S_{WZ}&={\rm N}(pp\to W^+Zjj)^{\footnotesize \rm IAM-MC}- {\rm N}(pp\to W^+Zjj)^{\footnotesize \rm SM}\,, \nn \\
B_{WZ}&={\rm N}(pp\to W^+Zjj)^{\footnotesize \rm SM}\,.
\end{align}
Here the event rates are summed over the interval in $M_{WZ}$
surrounding the corresponding resonance mass. 
In the SM predictions we have summed the purely EW contribution and the QCDEW contributions.
We display in  \tabref{TablasigmasWZ-paper} the results for these $\significance_{WZ}$ of the  $pp\to W^+Z j j$ events, for different LHC luminosities: $\mL=300 ~{\rm fb}^{-1}$, $\mL=1000 ~{\rm fb}^{-1}$ and $\mL=3000 ~{\rm fb}^{-1}$, that are expected for the forthcoming runs~\cite{Barachetti:2120851}. 
We have included the results of two intervals for comparison. First, the events are summed in $M_{WZ}$  over the corresponding narrow
 $(M_V-0.5\,\Gamma_V, M_V+0.5\,\Gamma_V)$ interval.
 Second, they are summed over the wider interval around the resonances of $(M_V-2\,\Gamma_V, M_V+2\,\Gamma_V)$. The results differ a bit in the two chosen intervals, as expected, but the conclusions are basically the same: we find very high statistical significances for all the studied BP scenarios in this case of $pp\to W^+Z j j$ events.

 \begin{table}[t!]
\begin{center}
\begin{tabular}{cc|c|c|c|c|c|c|}
\hhline{~~|-|-|-|-|-|-|}
& & \cellcolor{gray! 50}BP1 & \cellcolor{gray! 50}BP2 &\cellcolor{gray! 50} BP3 & \cellcolor{gray! 50}BP1' & \cellcolor{gray! 50}BP2' & \cellcolor{gray! 50}BP3' \\
\hhline{~|-|-|-|-|-|-|-|}
\multicolumn{1}{c|}{\multirow{ 3}{*}{ \begin{sideways}$\mL=300\,{\rm fb}^{-1}$\end{sideways} }}&\cellcolor{gray! 15} ${\rm N}^{\rm IAM-MC}_{WZ}$ & 89\,(147)& 19\,(25) &4\,(9) & 226\,(412) & 71\,(151) &33\,(59)\\ [0.5ex]
\hhline{~|-|-|-|-|-|-|-|}
\multicolumn{1}{c|}{} & \cellcolor{gray! 15}${\rm N}^{\rm SM}_{WZ}$ & 6\,(17)& 2\,(4) &0.3\,(2) & 11\,(45) & 5\,(27) &3\,(14)\\[0.5ex]
\hhline{~|-|-|-|-|-|-|-|}
\multicolumn{1}{c|}{} &\cellcolor{gray! 15} $\significance_{WZ}$ & 34.8\,(31.1)& 10.8\,(9.7) &6\,(5.4) & 64.9\,(54.4) & 28.9\,(23.8) &16.1\,(12)\\[0.5ex]
\hhline{~|-|-|-|-|-|-|-|}\\[-4ex] \hhline{~|-|-|-|-|-|-|-|}
\multicolumn{1}{c|}{\multirow{ 3}{*}{  \begin{sideways} $\mL=1000\,{\rm fb}^{-1}$\end{sideways}}} &\cellcolor{gray! 15}${\rm N}^{\rm IAM-MC}_{WZ}$ & 298\,(488)& 64\,(82) &13\,(30) & 752\,(1374) & 237\,(504) &110\,(196) \\[0.5ex]
\hhline{~|-|-|-|-|-|-|-|}
\multicolumn{1}{c|}{}&\cellcolor{gray! 15} ${\rm N}^{\rm SM}_{WZ}$ & 19\,(57)& 8\,(15) &1\,(6) & 36\,(151) & 17\,(90) &11\,(46)\\[0.5ex]
\hhline{~|-|-|-|-|-|-|-|}
\multicolumn{1}{c|}{}& \cellcolor{gray! 15}$\significance_{WZ}$ & 63.5\,(56.8)& 19.8\,(17.7) & 11\,(9.9) & 118.5\,(99.4) & 52.7\,(43.5) & 29.3\,(22)\\[0.5ex]
\hhline{~|-|-|-|-|-|-|-|}\\[-4ex] \hhline{~|-|-|-|-|-|-|-|}
\multicolumn{1}{c|}{\multirow{ 3}{*}{   \begin{sideways}$\mL=3000\,{\rm fb}^{-1}$\end{sideways}}} &\cellcolor{gray! 15} ${\rm N}^{\rm IAM-MC}_{WZ}$ & 893\,(1465)& 193\,(246) &39\,(89) & 2255\,(4122) &710\,(1511) &331\,(589)\\ [0.5ex]
\hhline{~|-|-|-|-|-|-|-|}
\multicolumn{1}{c|}{}& \cellcolor{gray! 15}${\rm N}^{\rm SM}_{WZ}$ & 58\,(172)& 24\,(44) &3\,(17) & 109\,(454) & 52\,(271) &34\,(139)\\[0.5ex]
\hhline{~|-|-|-|-|-|-|-|}
\multicolumn{1}{c|}{}& \cellcolor{gray! 15}$\significance_{WZ}$ & 110\,(98.5)& 34.3\,(30.6) &19\,(17.1) & 205.3\,(172.2) & 91.3\,(75.3) &50.8\,(38.1)\\[0.5ex]
\hhline{~|-|-|-|-|-|-|-|}
\end{tabular}
\caption{Predicted number of $pp\to W^+Z j j$ events of the IAM-MC,  ${\rm N}^{\rm IAM-MC}_{WZ}$, for the selected BP scenarios in \tabref{tablaBMP} and of the SM background (EW+QCDEW), ${\rm N}^{\rm SM}_{WZ}$,  at 14 TeV,  for different LHC luminosities: $\mL=300 ~{\rm fb}^{-1}$, $\mL=1000 ~{\rm fb}^{-1}$ and $\mL=3000 ~{\rm fb}^{-1}$. We also present the corresponding statistical significances, $\significance_{WZ}$, calculated according to \eqref{SS-WZ}. These numbers have been computed summing events in the bins contained in the interval of  $\pm 0.5\,\Gamma_V ~(\pm 2\,\Gamma_V$) around each resonance mass, $M_V$. The cuts in \eqref{optimal-cuts} have been applied.}
 \label{TablasigmasWZ-paper}
\end{center}
\end{table}

 
\begin{figure}[t!]
\begin{center}
\includegraphics[width=.49\textwidth]{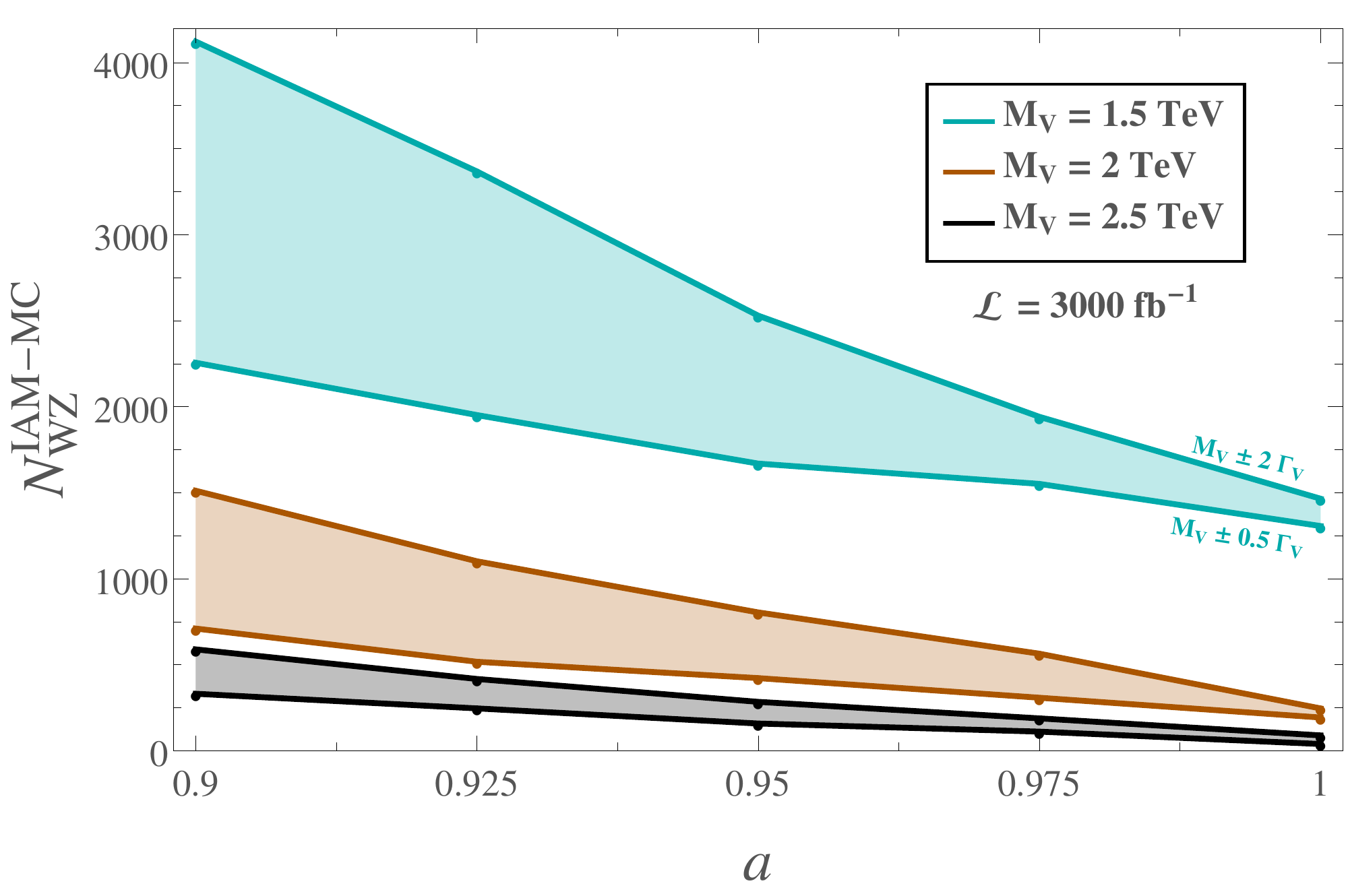}
\includegraphics[width=.48\textwidth]{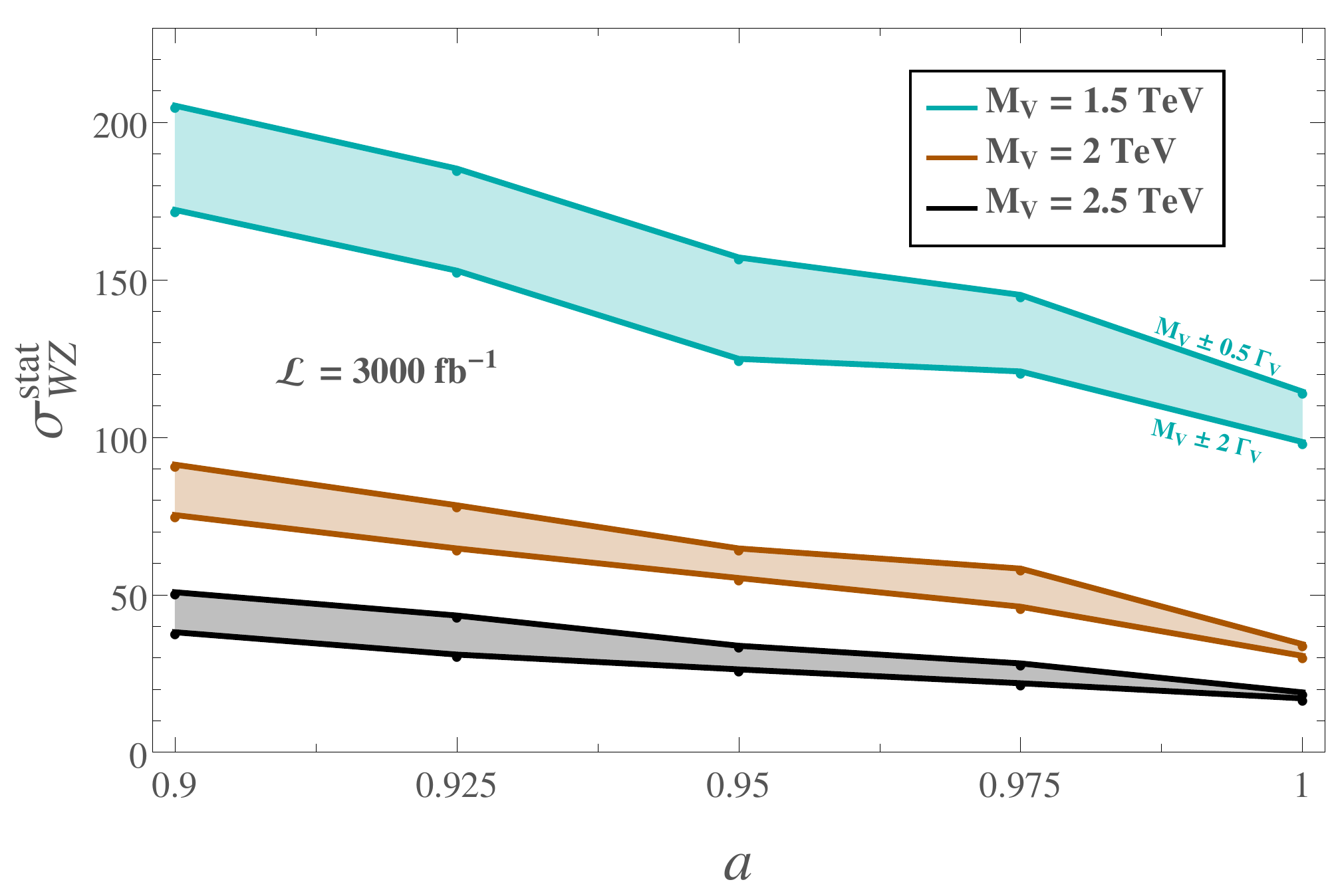}
\caption{Predictions for the number of events, ${\rm N}^{\rm IAM-MC}_{WZ}$ (left panel), and the statistical significance, $\significance_{WZ}$ (right panel), as a function of the parameter $a$ for $\mL=3000~{\rm fb}^{-1}$. The marked points correspond to our  selected benchmark points in \figref{fig:contourMW}. The two lines for each mass are computed by summing events within $\pm 0.5\,\Gamma_V $ and $\pm 2\,\Gamma_V$, respectively.}
\label{fig:asensitivityWZ}
\end{center}
\end{figure}

The above predictions in \tabref{TablasigmasWZ-paper} are for the selected reference scenarios with the values of the $a$ parameter fixed to the borders of the considered interval $(0.9, 1)$. In order to study further the sensitivity at the LHC to different values of the $a$ parameter within this interval, we have also performed the computation of predicted $W^+Zjj$ events, for the additional benchmark points specified in \figref{fig:contourMW}. The results for these new BP's are collected in \figref{fig:asensitivityWZ}. It shows both the predicted  event rates, ${\rm N}^{\rm IAM-MC}_{WZ}$, and statistical significances, $\significance_{WZ}$, as a function of the $a$ parameter, taken  within the interval $(0.9,1)$, for an integrated luminosity of $\mL=3000~{\rm fb}^{-1}$. The corresponding rates and significances for the other two luminosities considered here can be easily scaled from these results of $\mL=3000~{\rm fb}^{-1}$. The marked points correspond to our selected BP's of \figref{fig:contourMW}. As in \tabref{TablasigmasWZ-paper}, the two lines displayed for each $M_V$ value correspond, respectively,  to summing events in the bins contained in the interval of $\pm 0.5\,\Gamma_V$ and  $\pm 2\,\Gamma_V$ around each resonance mass.
From this \figref{fig:asensitivityWZ} it is clear that the high luminosity LHC with  $\mL=3000~{\rm fb}^{-1}$ would be sensitive to all values of $a$ in $(0.9,1)$ through the study of vector resonances with masses of $1.5$, $2$ and $2.5$~TeV. Actually, for this $WZ$ final state, these same conclusions apply to the other two luminosities considered, $\mL=1000~{\rm fb}^{-1}$  and $\mL=300~{\rm fb}^{-1}$.

\begin{figure}[t!]
\begin{center}
\includegraphics[width=.49\textwidth]{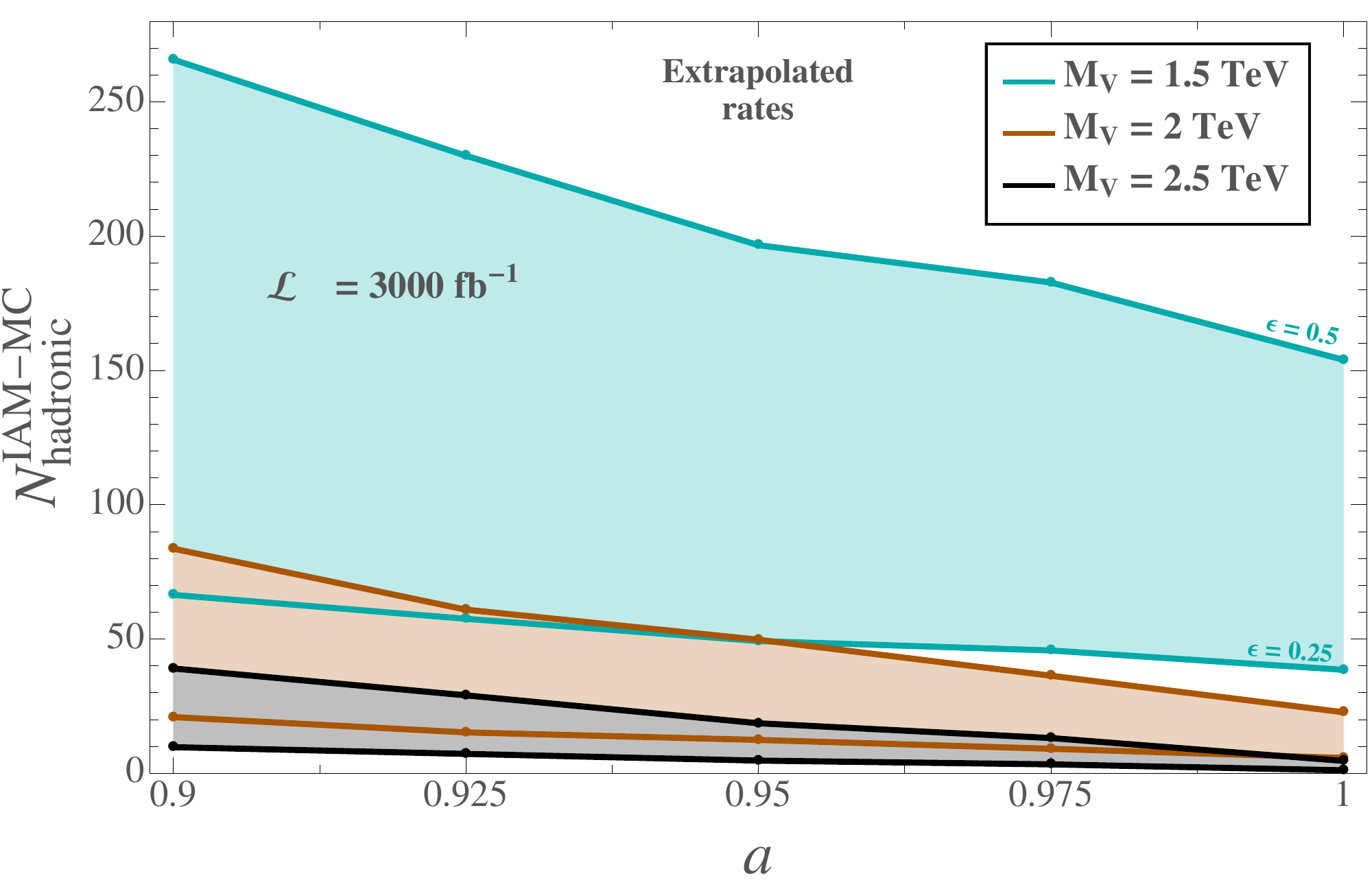}
\includegraphics[width=.48\textwidth]{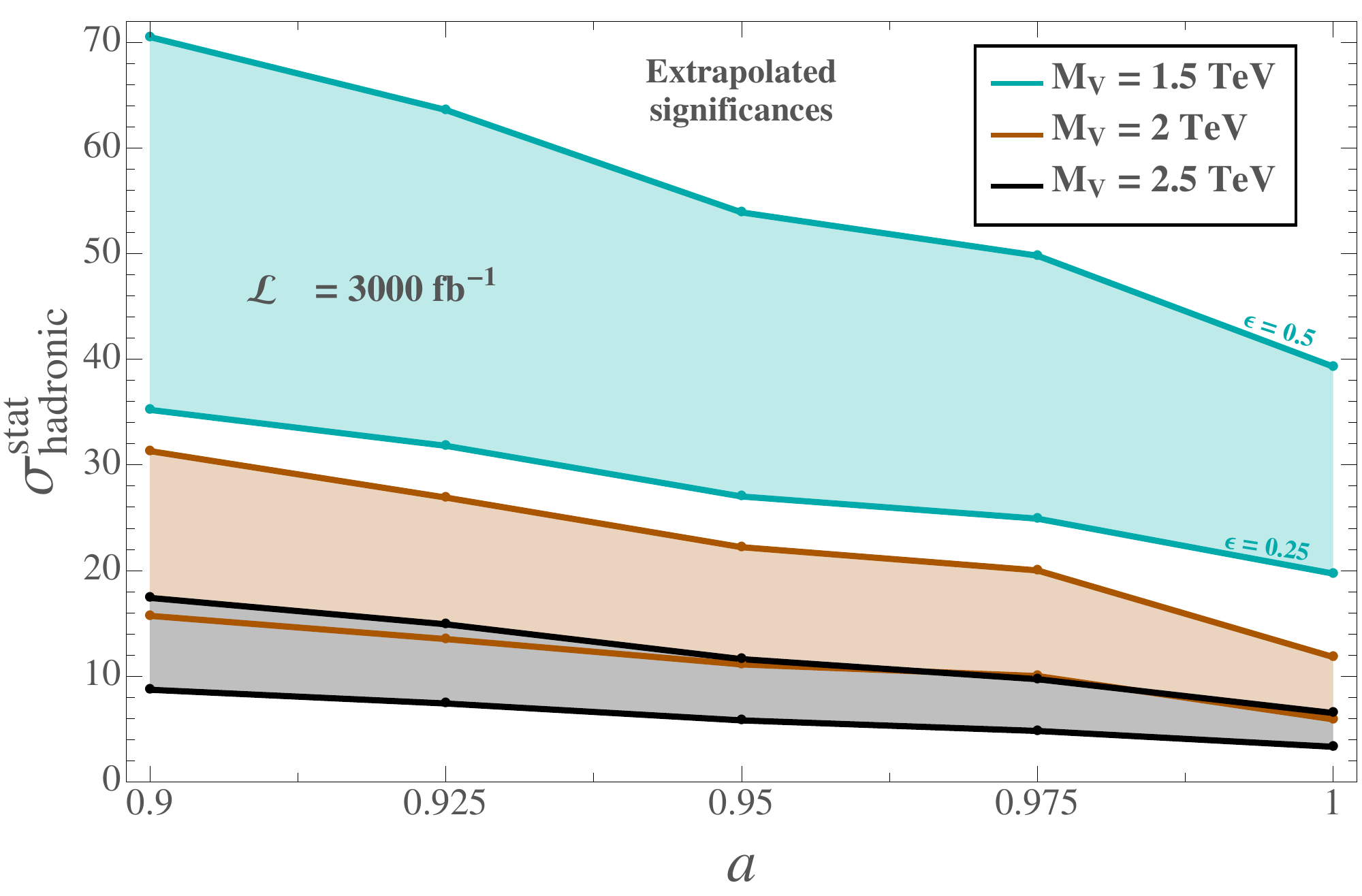}
\caption{Extrapolated $JJjj$ signal event rates from 
\figref{fig:asensitivityWZ} (for $\pm 0.5\,\Gamma_V $), ${\rm N}^{\rm IAM-MC}_{\rm hadronic}$ (left panel), and their corresponding  extrapolated statistical significances (right panel), $\significance_{\rm hadronic}$. The two lines shown for each resonance mass correspond, respectively,  assuming an efficiency in the reconstruction of $W$'s and $Z$'s from the `fat jets' of $\epsilon=0.5$ (upper line) and $\epsilon=0.25$ (lower line).} 
\label{fig:asensitivityhad}
\end{center}
\end{figure}

The previous results for the statistical significances of $W^+Zjj$ events are really encouraging. The high statistical significances found show that the resonances would be visible if the $W^+$ and $Z$ gauge bosons could be detected as final state particles. However, this is not the real case at colliders, and  one has to reconstruct $W$'s and $Z$'s from their decay products. In particular, the study of the so called `fat jets' in the final state, coming from the hadronic decays of boosted gauge bosons, could lead to a reasonably good reconstruction of the $W^+$ and the $Z$. The typical signatures of these hadronic events would then consist of four hadronic jets, two thin ones $jj$ triggering the VBS, and two fat ones $JJ$ triggering the final $WZ$.  If these type of signal events were able to be extracted from the QCD backgrounds, the predicted resonances that we show in \figref{fig:WZdistributions} could be very easily discovered. For a fast estimation of the number of signal events and significances that will be obtained by analyzing these kind of hadronic channels with `fat jets' we have performed a naive extrapolation from our results for $WZjj$ events by assuming two hypothetical efficiencies $\epsilon$  for the $W/Z$ reconstruction from `fat jets', which we take from the literature~\cite{Khachatryan:2014hpa, atlas2015identification,Aad:2015owa,heinrich2014reconstruction}, and are usually referred to as `medium' with $\epsilon=0.5$, and `tight' with  $\epsilon=0.25$. The corresponding $JJjj$ signal event rates can be extracted simply by \cite{heinrich2014reconstruction}:
\begin{equation} 
{\rm N}^{\rm IAM-MC}_{\rm hadronic}= {\rm N}^{\rm IAM-MC}_{WZ} \times
{\rm BR}(W \to {\rm hadrons}) \times {\rm BR}(Z \to {\rm hadrons})\times 
\epsilon_W \times \epsilon_Z .
\end{equation}
We show in \figref{fig:asensitivityhad} our predictions for these naively extrapolated number of events and statistical significances. These results are very encouraging and clearly indicate that with a more devoted study of the $W$ and $Z$ hadronic decays leading to `fat jets' the vector resonances of our selected scenarios would all be visible at the high luminosity option of the LHC with $\mL=3000~{\rm fb}^{-1}$. Looking at the scaled results for other luminosities, one can see that some of the resonances could be seen already for $\mL=300~{\rm fb}^{-1}$. Concretely, we find that resonances of $M_V\sim 1.5$ TeV could be observed at the LHC with this luminosity with statistical significances larger than 11 (6) for all values of the $a$ parameter if a medium (tight) reconstruction efficiency is assumed. A medium reconstruction efficiency would also allow to find heavier resonances of $M_V\sim$2 (2.5) TeV for values of $a<$0.975 (0.925). The case of $\mL=1000~{\rm fb}^{-1}$, is also very interesting. For this luminosity, the resonances with $M_V$=1.5 TeV and $M_V$=2 TeV could all be seen for any value of the $a$ parameter between 0.9 and 1 and for the two efficiencies considered. The heaviest ones, with masses of $\sim$2.5 TeV, would have significances larger than 3, and therefore could be used to probe values of $a$ in the whole interval studied in this work, if a medium efficiency is assumed. For a tight efficiency, one could still be sensitive to values of the $a$ parameter between 0.9 and 0.95. 

On the other hand, the alternative semileptonic channels  where one final EW gauge boson goes to leptons and the other one to hadrons observed as one fat jet, will also lead to interesting signatures like
$\ell \nu J jj$ and $\ell \ell Jjj$ and are also very promising, with comparable statistics to the previous hadronic channels, as our corresponding naively extrapolated rates (not shown) indicate.  The potential of these semileptonic channels can also be inferred from the studies in \cite{Aaboud:2016uuk}, where they have been used to notably improve the experimental constraints on $a_4$ and $a_5$ by roughly one order of magnitude, with respect to their previous constraints based on the pure leptonic decays \cite{Aad:2014zda}.  Nevertheless, our previous estimates of event rates involving `fat jets' although really encouraging are yet too naive and deserve further studies for a more precise conclusion.  A more realistic and precise computation is needed, but it would require a fully simulated MC analysis of the events with `fat jets' and a good control of the QCD backgrounds and other reducible backgrounds, which is far beyond the scope of this work.

Therefore, from now on,  we will focus on the cleanest decays of the $W^+$ and $Z$, which are the pure leptonic ones, leading to a final state from the $WZ$ pair with three leptons and one neutrino. Concretely, to unsure a good efficiency in the detection of the final particles we consider just the two first leptonic generations. Therefore, all together, we propose to explore at the LHC events of the type $(\ell_1^+\ell_1^-\ell_2^+\slashed{p}_T j_1j_2)$, with ${\ell_{1,2}}$ being either a muon or an electron,  $\slashed{p}_T$ the missing transverse momentum coming from the neutrino, and $j_{1,2}$ the two emergent jets from the final quarks that are key to tag the VBS configuration. The event rates in these leptonic channels suffer from a suppression factor of ${\rm BR}(WZ \to \ell\ell\ell\nu) \simeq 0.014$, but have the advantage of allowing us to reconstruct the invariant mass of the $WZ$ pair in the transverse plane, and also to provide a good reconstruction of the $Z$.

For the present study of the leptonic channels we apply the set of cuts that are partially extracted from Ref.~\cite{Szleper:2014xxa} and optimized as described in the previous background subsection, to make the selection of VBS processes more efficient when having leptons in the final state. These contain all the previous VBS cuts and others, and are summarized by:
\begin{eqnarray}
&2<|\eta_{j_{1,2}}|<5\,, \label{rapjlep} \nn \\
&\eta_{j_1} \cdot \eta_{j_2} < 0\,,\label{difrap} \nn \\
&p_T^{j_1,j_2}>20 ~{\rm GeV} \,, \nn \\
&M_{jj}>500 ~{\rm GeV}\,, \nn \\
&M_Z-10~{\rm GeV} < M_{\ell^+_Z \ell^-_Z} < M_Z+10~{\rm GeV}\,,
\nn \\
&M^T_{WZ}\equiv M^T_{\ell\ell\ell\nu}>500 ~{\rm GeV}\,, \nn \\
&\slashed{p}_T>75 ~{\rm GeV}\,, \nn \\
& p_T^\ell>100 ~{\rm GeV}\,,
\label{lepfin}
\end{eqnarray}
where $ \eta_{j_{1,2}}$ are the pseudorapidities of the jets,
$M_{jj}$ is the invariant mass of the jet pair,
$M_{\ell^+_Z \ell^-_Z}$ the invariant mass of the lepton pair coming from the Z decay (this means at least one of the two $\ell^+\ell^-$ combinations in the case of $\ell^+\ell^-\ell^+\nu$ with the same lepton flavor),
$\slashed{p}_T$ the transverse missing momentum, $p_T^\ell$ the transverse momentum of the final leptons, and $M^T_{WZ}$ the transverse invariant mass of the $WZ$ pair defined as follows in terms of the final lepton variables:
\begin{equation}
M^T_{WZ}\equiv M^T_{\ell\ell\ell\nu}=\sqrt{\Big(\sqrt{M^2(\ell\ell\ell)+p_T^2(\ell\ell\ell)} +|\slashed{p}_T|  \Big)^2-\big(\vec{p_T}(\ell\ell\ell)+\vec{\slashed{p}_T}\big)^2}\,,
\label{Mtrans}
\end{equation}
with $M(\ell\ell\ell)$ and $\vec{p_T}(\ell\ell\ell)$ being the invariant mass and the transverse momentum of the three final leptons respectively, and $\vec{\slashed{p}_T}$ the transverse momentum of the neutrino.

\begin{figure}[t!]
\begin{center}
\includegraphics[width=.49\textwidth]{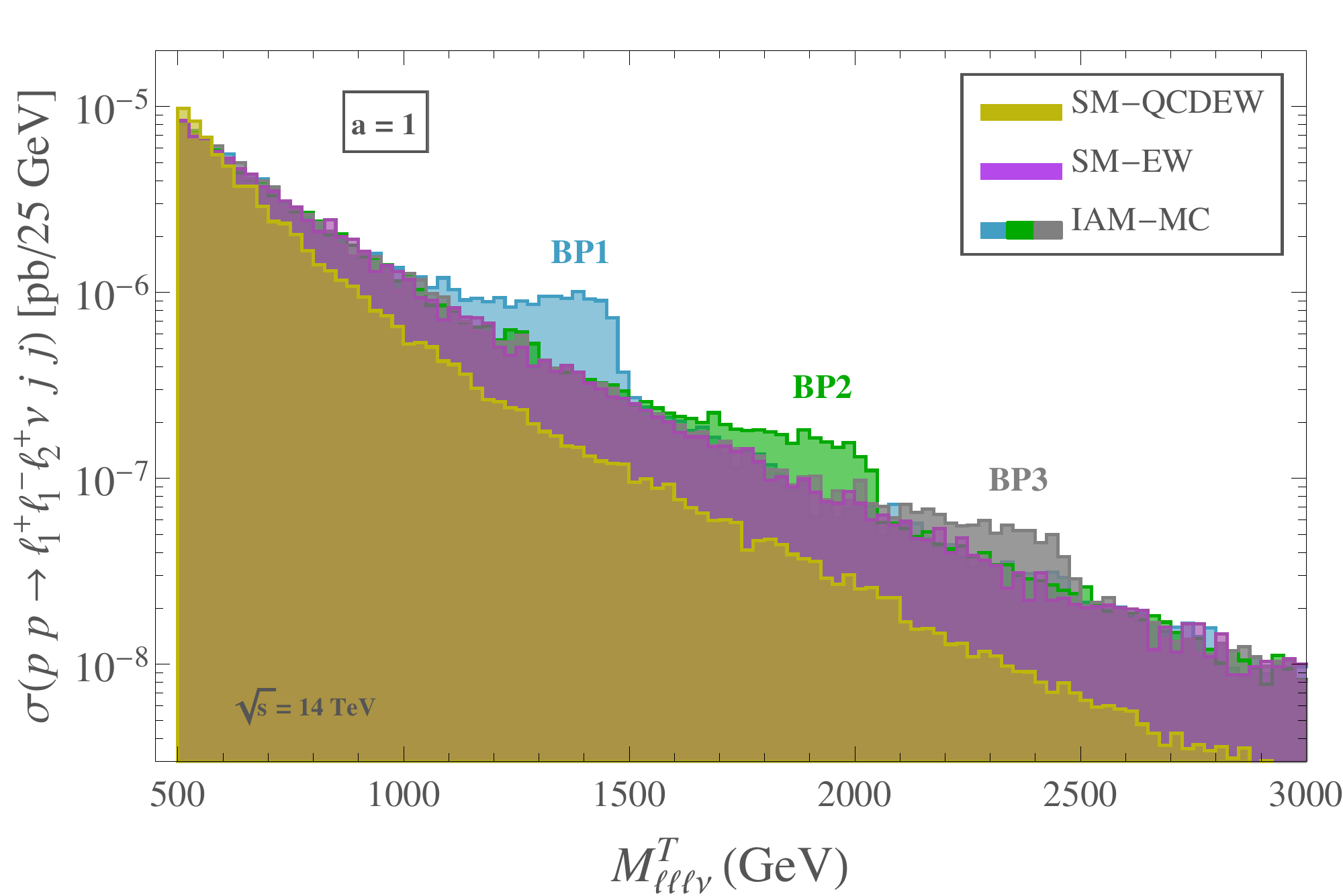}
\includegraphics[width=.49\textwidth]{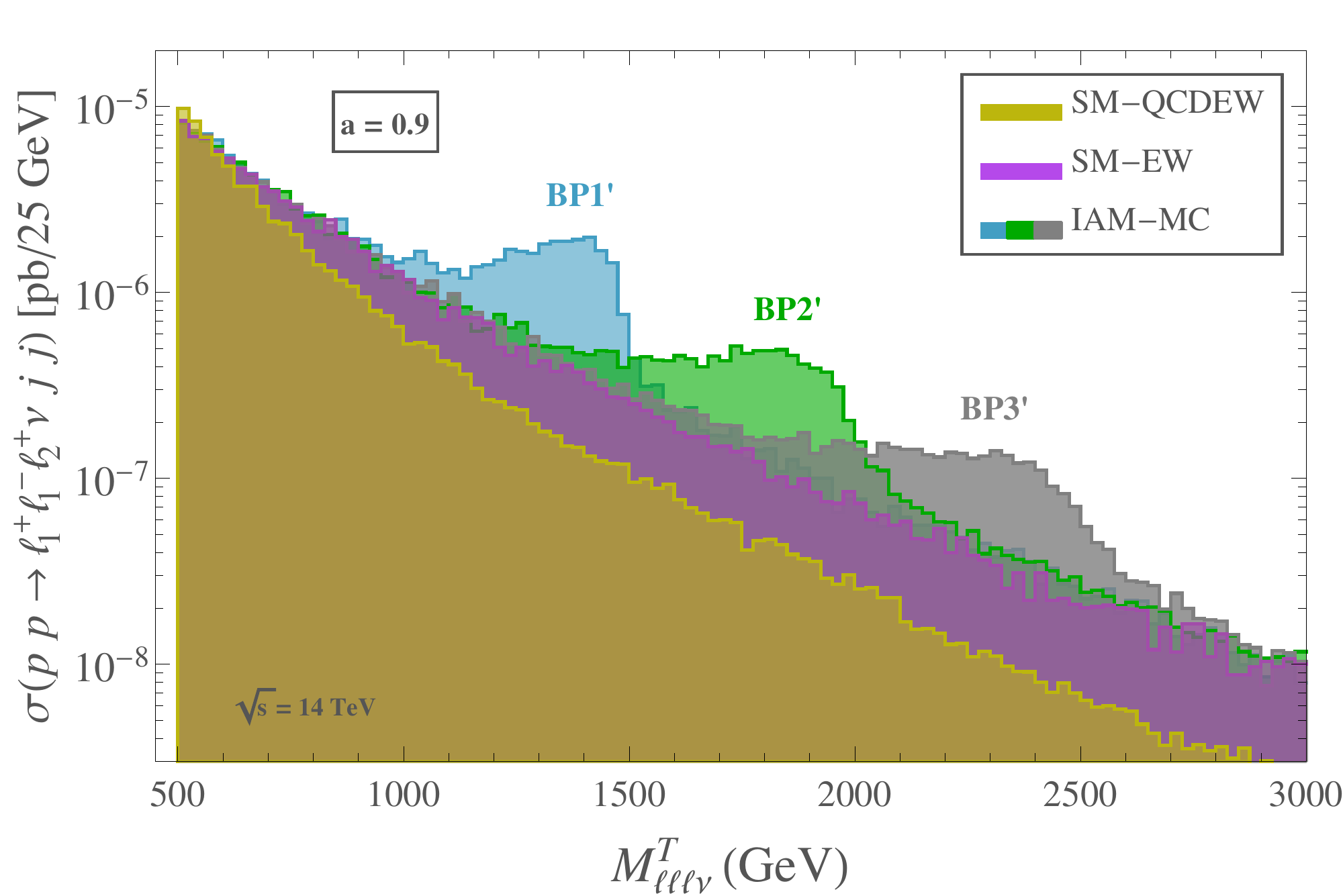}\\
\caption{Predictions of the $\sigma(pp \to \ell_1^+\ell_1^-\ell_2^+ \nu jj)$ distributions with the transverse invariant mass, $M^T_{\ell\ell\ell \nu}$,  for the selected benchmark points of the IAM-MC model BP1 (blue), BP2 (green), BP3 (gray) in the left panel and BP1' (blue), BP2' (green), BP3' (gray) in the right panel, and for the two main SM backgrounds, SM-QCDEW (yellow) and SM-EW (purple). The cuts in \eqref{lepfin} have been applied.}
\label{fig:leptondistributions}
\end{center}
\end{figure}

As before, we generate all the signal, IAM-MC, and background, SM-QCDEW and SM-EW,  events with MadGraph5. The results obtained, after applying the previous cuts in \eqref{lepfin}, are displayed in \figref{fig:leptondistributions}, where the total cross section per bin has been plotted as a function of the transverse invariant mass of the $WZ$ pair as defined in \eqref{Mtrans} .
From this figure we can conclude that the peaks, although smoother, are again clearly seen over the SM backgrounds, specially for the lighter resonances. The shape of the emergent peaks is different than in \figref{fig:WZdistributions}, typically smaller and broader, as corresponding to distributions with the transverse invariant mass, having the maximum at bit lower values, and getting spread in a wider invariant mass range.

Finally, in order to quantify the statistical significance of these emergent peaks, we have computed the quantity $\significance_\ell$, defined in terms of the predicted number of events from the IAM-MC,
${\rm N}(pp\to \ell_1^+\ell_1^-\ell_2^+ \slashed{p}_T jj)^{\footnotesize \rm IAM-MC}$, and the background events,
${\rm N}(pp\to \ell_1^+\ell_1^-\ell_2^+ \slashed{p}_T jj)^{\footnotesize \rm SM}$,  as follows:

\begin{align}
\significance_{\ell} &=\frac{S_{\ell}}{\sqrt{B_{\ell}}}\,,
\label{SS-l}
\end{align}
with,
\begin{align}
S_{\ell}&={\rm N}(pp\to \ell_1^+\ell_1^-\ell_2^+ \slashed{p}_T jj)^{\footnotesize \rm IAM-MC}- {\rm N}(pp\to \ell_1^+\ell_1^-\ell_2^+ \slashed{p} jj)^{\footnotesize \rm SM}\,, \nn \\
B_{\ell}&={\rm N}(pp\to \ell_1^+\ell_1^-\ell_2^+ \slashed{p}_T jj)^{\footnotesize \rm SM}\,.
\end{align}

The final numerical results for $\significance_\ell$ are collected in \tabref{Tablasigmaslep-paper}. Again, we have considered three
 different LHC luminosities: $\mL=300 ~{\rm fb}^{-1}$,
 $\mL=1000 ~{\rm fb}^{-1}$ and $\mL=3000 ~{\rm fb}^{-1}$.
 The numbers of events presented are the results after summing over the intervals in which we have found the largest  statistical significance with at least one IAM-MC event for $\mL=3000 ~{\rm fb}^{-1}$.  
 In particular we consider the following ranges of $M^T_{\ell\ell\ell\nu}$:
\begin{align}
&{\rm BP1:}~1325-1450~{\rm GeV}\,,
&& {\rm BP2:}~1875-2025~{\rm GeV}\,,
&& {\rm BP3:}~2300-2425~{\rm GeV}\,,\nn \\
&{\rm BP1':}~1250-1475~{\rm GeV}\,,
&& {\rm BP2':}~1675-2000~{\rm GeV}\,,
&&{\rm BP3':}~2050-2475~{\rm GeV}\,.
\label{MTintervals}
\end{align} 

\begin{table}[t!]
\begin{center}
\begin{tabular}{cc|c|c|c|c|c|c|}
\hhline{~~|-|-|-|-|-|-|}
& & \cellcolor{gray! 50}BP1 & \cellcolor{gray! 50}BP2 &\cellcolor{gray! 50} BP3 & \cellcolor{gray! 50}BP1' & \cellcolor{gray! 50}BP2' & \cellcolor{gray! 50}BP3' \\
\hhline{~|-|-|-|-|-|-|-|}
\multicolumn{1}{c|}{\multirow{ 3}{*}{ \begin{sideways}$\mL=300\,{\rm fb}^{-1}$\end{sideways} }}&\cellcolor{gray! 15} ${\rm N}^{\rm IAM-MC}_{\ell}$ & 2 & 0.5 & 0.1 & 5 & 2 & 0.7 \\[0.5ex]
\hhline{~|-|-|-|-|-|-|-|}
\multicolumn{1}{c|}{} & \cellcolor{gray! 15}${\rm N}^{\rm SM}_{\ell}$ & 1 & 0.4 & 0.1 & 2 &0.6 &0.3 \\[0.5ex]
\hhline{~|-|-|-|-|-|-|-|}
\multicolumn{1}{c|}{} &\cellcolor{gray! 15} $\significance_{\ell}$ & 0.9 & - &- &2.8 &1.4 &- \\[0.5ex]
\hhline{~|-|-|-|-|-|-|-|}\\[-4ex] \hhline{~|-|-|-|-|-|-|-|}
\multicolumn{1}{c|}{\multirow{ 3}{*}{  \begin{sideways} $\mL=1000\,{\rm fb}^{-1}$\end{sideways}}} &\cellcolor{gray! 15}${\rm N}^{\rm IAM-MC}_{\ell}$ &7 & 2 & 0.4 &18 &5 &2 \\[0.5ex]
\hhline{~|-|-|-|-|-|-|-|}
\multicolumn{1}{c|}{}&\cellcolor{gray! 15} ${\rm N}^{\rm SM}_{\ell}$ &4 & 1 &0.3 &6 &2 &1 \\[0.5ex]
\hhline{~|-|-|-|-|-|-|-|}
\multicolumn{1}{c|}{}& \cellcolor{gray! 15}$\significance_{\ell}$&1.6 & 0.3 & -& 5.1&2.5 &1.4 \\[0.5ex]
\hhline{~|-|-|-|-|-|-|-|}\\[-4ex] \hhline{~|-|-|-|-|-|-|-|}
\multicolumn{1}{c|}{\multirow{ 3}{*}{   \begin{sideways}$\mL=3000\,{\rm fb}^{-1}$\end{sideways}}} &\cellcolor{gray! 15} ${\rm N}^{\rm IAM-MC}_{\ell}$ & 22& 5 &1 &53 &16 &7 \\[0.5ex]
\hhline{~|-|-|-|-|-|-|-|}
\multicolumn{1}{c|}{}& \cellcolor{gray! 15}${\rm N}^{\rm SM}_{\ell}$ &12 &4  &1 &17 &6 &3 \\[0.5ex]
\hhline{~|-|-|-|-|-|-|-|}
\multicolumn{1}{c|}{}& \cellcolor{gray! 15}$\significance_{\ell}$ & 2.7& 0.6 &0.3 &8.9 &4.4 &2.4 \\[0.5ex]
\hhline{~|-|-|-|-|-|-|-|}\end{tabular}
\caption{ Predicted number of $pp\to \ell_1^+\ell_1^-\ell_2^+\nu j j$ events of the IAM-MC,  ${\rm N}^{\rm IAM-MC}_{\ell}$, and of the SM background (EW+QCDEW), ${\rm N}^{\rm SM}_{\ell}$,  at 14 TeV,  for different LHC luminosities: $\mL=300 ~{\rm fb}^{-1}$, $\mL=1000 ~{\rm fb}^{-1}$ and $\mL=3000 ~{\rm fb}^{-1}$.  We also present the corresponding statistical significances, $\significance_{\ell}$, calculated according to \eqref{SS-l} after summing events in the intervals collected in \eqref{MTintervals}.
 We only display the value of $\significance_{\ell}$ for the cases in which there is at least one IAM-MC event. The cuts in \eqref{lepfin} have been applied.}
\label{Tablasigmaslep-paper}
\end{center}
\end{table}

 

 As we can see in this \tabref{Tablasigmaslep-paper}, these more realistic statistical significances
 for the leptonic channels, $\significance_\ell$ are considerably smaller than
 the previous $\significance_{WZ}$.  However, we still get scenarios
 with sizable $\significance_\ell$ larger than 3. 
Concretely, the scenarios with $a=0.9$ leading to vector resonance masses at and below 2 TeV, could be seen in these leptonic channels at the LHC in its forthcoming high luminosity stages. Particularly, for BP1' with $M_V=1.5$ TeV we get sizeable significances around 3, 5, and 9 for luminosities of 300, 1000 and 3000 ${\rm fb}^{-1}$ respectively, whereas for BP2' with $M_V=2$ TeV the significances are lower, close to 3 for 1000 ${\rm fb}^{-1}$ and slightly above 4 for 3000 ${\rm fb}^{-1}$. The scenarios with $a=1$ have comparatively smaller significances, and only the lightest resonances with $M_V=1.5$ TeV , like BP1,   lead to a significance of around 3 for the highest studied luminosity of 3000 ${\rm fb}^{-1}$.  
Notice that there are some cases that we do not consider in our discussion because of the lack of statistics.
The scenarios with heavier resonance masses, at and above 2.5 TeV seem to be very difficult to observe, due to the poor statistics for these masses in the leptonic channels. Only our benchmark point BP3' gets a significance larger that 2 for  3000 ${\rm fb}^{-1}$. Therefore, in order to get more sizable significances in those cases  one would have to perform a more devoted study in other channels like the semileptonic and hadronic ones of the final $WZ$ pair, as we have already commented above.  

\begin{figure}[t!]
\begin{center}
\includegraphics[width=.49\textwidth]{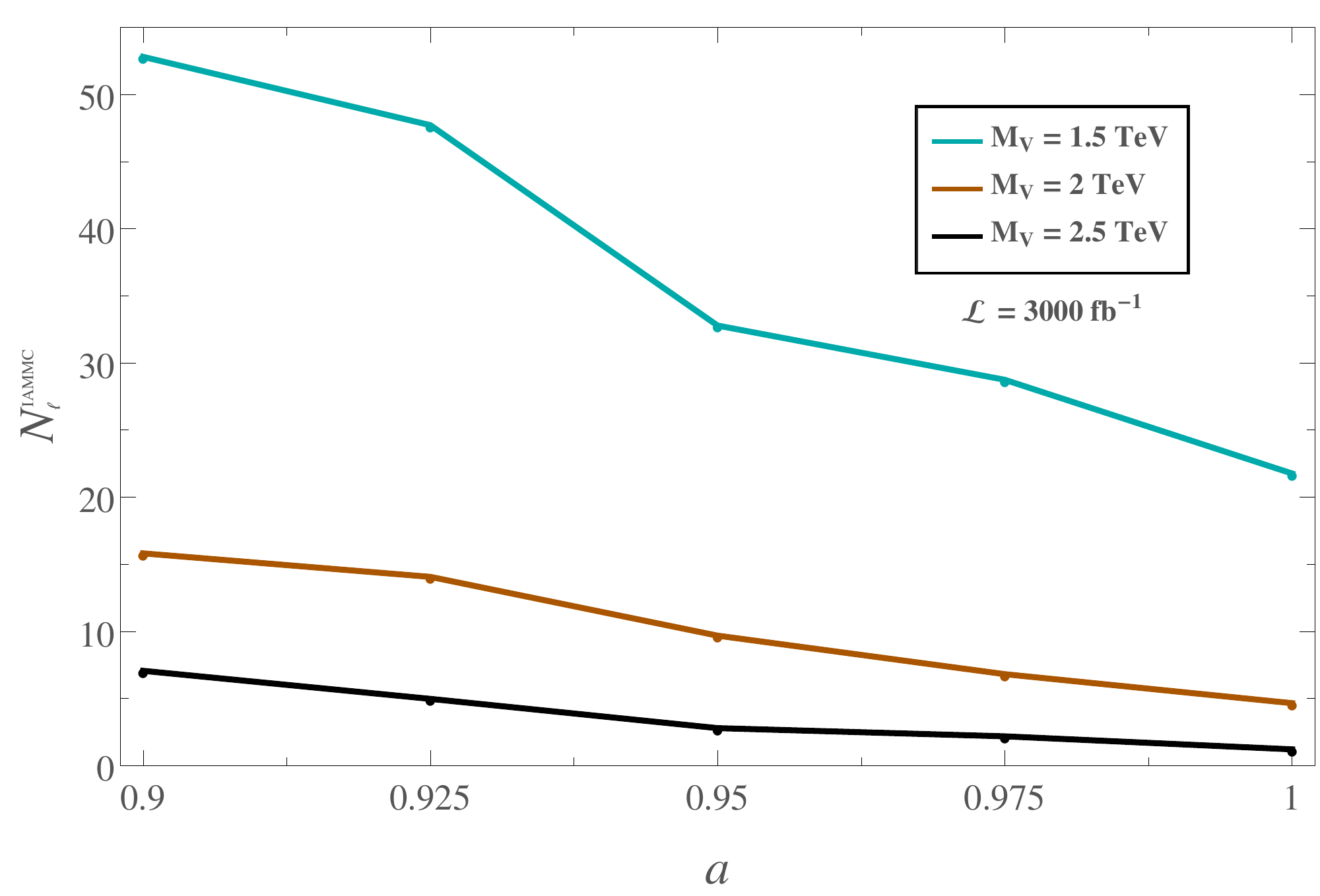}
\includegraphics[width=.48\textwidth]{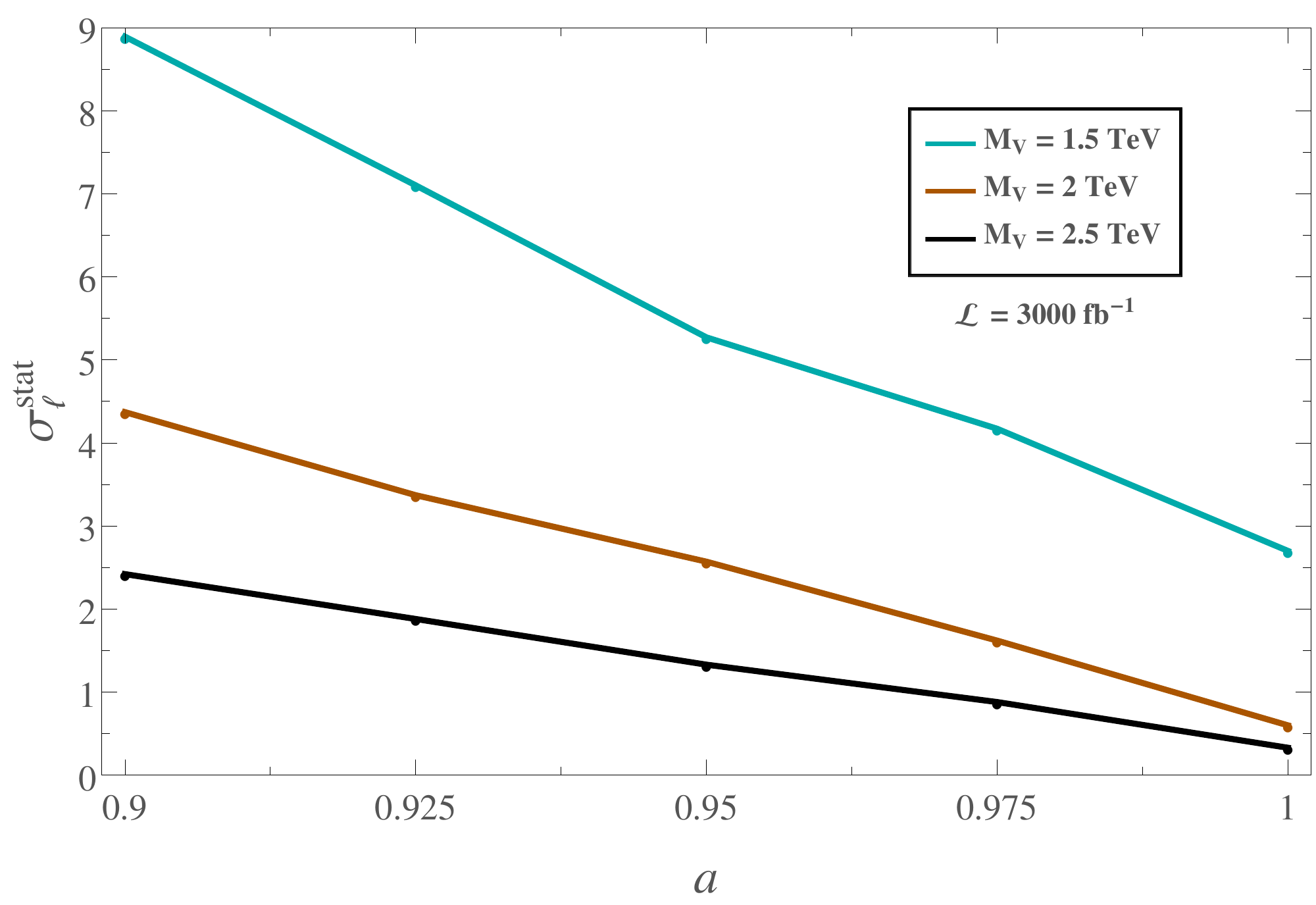}\\
\caption{Predictions for the number of $pp\to  \ell_1^+\ell_1^-\ell_2^+\nu j j$ events, ${\rm N}_{\ell}^{\footnotesize \rm IAM-MC}$, (left panel) and the statistical significance, $\significance_{\ell}$, (right panel) as a function of the parameter $a$ for $\mL=3000~{\rm fb}^{-1}$. Marked points correspond to our selected benchmark points in \figref{fig:contourMW}. The cuts in \eqref{lepfin} have been applied.}
\label{fig:asensitivitylep}
\end{center}
\end{figure}

Finally, we have also explored the additional BP points with different values of the $a$ parameter and studied the sensitivities to this parameter in the leptonic channels. The results of the predicted $pp\to  \ell_1^+\ell_1^-\ell_2^+\nu j j$ event rates, ${\rm N}^{\rm IAM-MC}_{\ell}$, and statistical significances, $\significance_\ell$, in terms  of the parameter  $a$, within the interval $(0.9,1)$ are displayed in \figref{fig:asensitivitylep}. From this figure we can clearly conclude that, for the highest luminosity  $\mL=3000~{\rm fb}^{-1}$, and for $M_V=1.5$ TeV, there will be good sensitivity to the $a$ parameter, 
with $\significance_\ell$ larger than 3, in the full interval $(0.9,1)$, except for the limiting value of $a=1$ where $\significance_\ell$ is slightly below 3.
For the heavier resonances, we find lower sensitivities, with  $\significance_\ell$ larger than 3 only for $M_V=2$ TeV and $a$ below around 0.94. The case $M_V=2.5$ TeV is not very promising to learn about the parameter $a$ in the fully leptonic channel except, perhaps, for the scenario with the lowest considered value of $a=0.9$ where, as said above for BP3', $\significance_\ell$ gets larger than 2. Nevertheless, this would be strongly improved by exploring other decay channels, as we mentioned before.
 
\section{Conclusions}
In this work we have explored
the production and sensitivity to vector resonances at the LHC. We have worked under the framework of the EChL supplemented by another  effective chiral Lagrangian to describe the vector resonances that have the same properties as the dynamically generated resonances found by the IAM. 
This approach provides unitary amplitudes and effectively takes into account the re-summation of the infinite re-scattering bubbles of the longitudinal gauge bosons which are the dominant ones in the case of a strongly EWSB scenario. 
We have then built our IAM-MC model that uses this Lagrangian framework and  mimics the resonant behavior of the IAM amplitudes. We believe that this IAM-MC framework, where the VBS amplitudes are built from Feynman rules, is the proper one for a MonteCarlo analysis  like the one we have done in the present work with MadGraph5. For that purpose we have built the needed UFO file with our IAM-MC model which is ready for other users, upon request. Our IAM-MC model for the vector resonance production at LHC provides unitary VBS amplitudes (we have checked indeed, that the LHC cross sections respect the Froissart bound given by \eqref{froisbound}), and therefore does not require unphysical {\it ad hoc} cuts to respect unitarity in the study of the signal versus background events.
We also wish to emphasize that our predictions presented here for both the amplitudes and the cross sections are for massive $W$ and $Z$ gauge bosons and are complete in the sense that they are not obtained from the lowest partial waves but from a complete tree level diagrammatic computation. 

Concretely, we have focused on the $pp\to W^+Zjj$ channel which is the most relevant one if one is interested in the study of charged vector resonances from a strongly interacting EWSB. This particular channel is also appealing because it suffers from less sever backgrounds than other channels with two EW vector bosons and two jets in the final state like, for instance, $pp\to W^+W^-jj$ and $pp\to ZZjj$. With the selection of the proper optimal VBS cuts, the process, $pp\to W^+Zjj$, proceeds mainly via the scattering subprocess $W^+Z \to W^+ Z$ and it is in this VBS where the resonances of our interest manifest.

We have selected specific benchmark points in the IAM-MC model parameter space which have vector resonances emerging at mass and width values that are of phenomenological interest for the searches at the LHC. Concretely, the fifteen scenarios that we have chosen, summarised in \figref{fig:contourMW}, have their respective resonance masses placed at $M_V=$ 1.5 , 2 and 2.5 TeV, and they correspond in our approach to specific values of the relevant EChL parameters, $a$, $a_4$ and $a_5$ in the experimentally allowed region. Specifically, we have considered the intervals $a \in (0.9,1)$ and $a_4$, $a_5$ $\in {\cal O}(10^{-4},10^{-3})$ and set our first six reference scenarios in the borders of the $a$ interval: BP1, BP2, and BP3 with $a=1$ and BP1', BP2', BP3' with $a=0.9$. These scenarios are used to perform the full study of the MC generated events. The remaining nine scenarios have been used to further explore the sensitivity to the $a$ parameter by trying other values in the allowed $(0.9,1)$ interval.     
  
We have fully analyzed the $W^+Zjj$ event distributions of both the signal and main SM background events with respect to the $M_{WZ}$ invariant mass by using MadGraph5, and we have seen clearly the emergence of the vector resonances in all these distributions on top of the SM backgrounds with extremely high statistical significances. Our numerical results are summarised  in \figref{fig:WZdistributions}, \tabref{TablasigmasWZ-paper} and in \figref{fig:asensitivityWZ}.  We have found, indeed,  great sensitivity in all the studied scenarios, with masses at  $M_V=$ 1.5, 2 an 2.5 TeV, and with values of the $a$ parameter in the allowed interval $(0.9,1)$.  The largest significances are obtained for the lightest resonances with   
 $M_V=$ 1.5 TeV and the lowest studied values of $a=0.9$, corresponding to our BP1' scenario,  which lead  to  $\significance_{WZ}$ as large as 65, 118 and 205 for respective luminosities of $\mL=300 ~{\rm fb}^{-1}$, $1000 ~{\rm fb}^{-1}$ and $3000 ~{\rm fb}^{-1}$.  The lowest significances are obtained for the heaviest resonances with  $M_V=$ 2.5 TeV and the highest studied value of $a=1$,  corresponding to our BP3 scenario, but they are yet quite sizable, 6, 11 and 19, again for $\mL=300 ~{\rm fb}^{-1}$, $1000 ~{\rm fb}^{-1}$ and $3000 ~{\rm fb}^{-1}$, respectively.
 
These encouraging results for $W^+Zjj$ events are assuming that the $W$ and the $Z$ can be fully detected. However, this is not the real case  at colliders and one has to rely instead on the partial reconstruction of the final $W$ and $Z$ from their decay products. Thus, in order to profit from the largest rates, we have first discussed the case of the hadronic channels where each EW gauge boson decays into hadrons measured as `fat jets', leading to total signatures of type $JJjj$ with four jets,  two thin ones $jj$ triggering the VBS, and two fat ones $JJ$ triggering the final $WZ$. We have performed a fast estimate of the event rates and significances of these hadronic channels by a naive extrapolation from our results of $WZjj$ events. This is done by using the corresponding decay ratios to hadrons and by assuming two hypothetical efficiencies 
$\epsilon$  for the $W/Z$ reconstruction from `fat jets',   'Medium' with $\epsilon=0.5$, and 'Tight' with  $\epsilon=0.25$ following \cite{Khachatryan:2014hpa, atlas2015identification,Aad:2015owa,heinrich2014reconstruction}.  Our results in \figref{fig:asensitivityhad}  show the big potential of these hadronic channels in the future discovery of these vector resonances, leading to extrapolated significances larger than 3 for all the studied scenarios with masses $M_V=$ 1.5, 2 an 2.5 TeV, and values of the $a$ parameter in the allowed interval $(0.9,1)$, if  the highest luminosity option for the LHC with $\mL=3000~{\rm fb}^{-1}$ is assumed.  Looking into other luminosities, one can see that some of the resonances could be seen already for $\mL=300~{\rm fb}^{-1}$. Concretely, we find that resonances of $M_V\sim 1.5$ TeV could be observed at the LHC with this later luminosity with statistical significances larger than 11 (6) for all values of the $a$ parameter if a medium (tight) reconstruction efficiency is assumed. At this luminosity, a medium reconstruction efficiency would also allow to find heavier resonances of $M_V\sim$2 (2.5) TeV for values of $a<$0.975 (0.925). 
For $\mL=1000~{\rm fb}^{-1}$, the resonances with $M_V$=1.5 TeV and $M_V$=2 TeV could all be seen for any value of the $a$ parameter between 0.9 and 1 and for the two efficiencies considered. The heaviest ones, with masses of $\sim$2.5 TeV, would have significances larger than 3, and therefore could be used to probe values of $a$ in the whole interval considered, if a medium efficiency is assumed. For a tight efficiency, one could still be sensitive to values of the $a$ parameter between 0.9 and 0.95. 
We have also commented on the comparable statistics that we get for the extrapolated rates in the case of semileptonic channels of the final $WZ$ leading to signatures like $\ell \nu J jj$ and $\ell \ell Jjj$, showing also the big potential of these channels.

Nevertheless, our previous estimates of event rates involving `fat jets' although really encouraging are not sufficiently precise and we  have emphasized that a more realistic and precise computation is needed.  This would require a fully simulated MC analysis of the events with `fat jets' and a good control of the QCD backgrounds and other reducible backgrounds, which is far beyond the scope of this work. Instead, we have preferred to study here in full detail the cleanest channels where the final $W$ and $Z$ decay into leptons and to provide our most realistic predictions in those leptonic channels, with lowest rates but with cleanest signatures.

We have then fully studied the golden leptonic $W$ and $Z$ decay channels, i.e., the channels leading to a final state with $\ell_1^+\ell_1^-\ell_2^+\nu jj$, $\ell=e,\mu$, and we have presented the results of the appearing resonances in terms of an experimentally measurable variable, the transverse invariant mass of the $\ell_1^+\ell_1^-\ell_2^+\nu$ final leptons. As it is clearly illustrated in \figref{fig:leptondistributions}, the shape of the peaks is softened as expected with respect to the final $W$ and $Z$ case, but they are still visible. Our numerical evaluation of the future event rates and sensitivities are summarized in  \tabref{Tablasigmaslep-paper} and in \figref{fig:asensitivitylep}. 

The results in \tabref{Tablasigmaslep-paper} demonstrate that with a luminosity of $300~ {\rm fb}^{-1}$ a first hint (with $\significance_\ell$ around 3) of resonances with mass around 1.5 TeV for the case $a=0.9$ could be seen in the leptonic channels. For the first stage of the high luminosity LHC, with $1000~ {\rm fb}^{-1}$, we estimate that these scenarios could  be tested with a high statistical significance larger than 5 and a discovery of these resonances with masses close to 1.5 TeV, like in BP1',  could be done. Interestingly, for the last luminosity considered, $3000~ {\rm fb}^{-1}$,  all the studied scenarios with  resonance masses at and below 2 TeV and with $a=0.9$ could be seen. Concretely, for BP1' and BP2'  we get  $\significance_\ell$ close to 9 and 4 respectively.   For the heaviest studied resonances, with masses around 2.5 TeV, small hints with $\significance_\ell$  slightly larger than 2  might as well show up in the highest luminosity stage. The sensitivities to other values of $a$ in the interval $(0.9,1)$  have also been explored.  Our numerical results in \figref{fig:asensitivitylep} show that for the highest luminosity  $\mL=3000~{\rm fb}^{-1}$, and for $M_V=1.5$ TeV, there will be good sensitivity to the $a$ parameter in the leptonic channels, 
with $\significance_\ell$ larger than 3, in the full interval $(0.9,1)$ except for the limiting value of $a=1$ where $\significance_\ell$ is slightly below 3. For the heavier resonances, we find lower sensitivities, with  $\significance_\ell$ larger than 3 only for $M_V=2$ TeV and $a$ below around 0.94. The case $M_V=2.5$ TeV does not show appreciable sensitivity to $a$, except for the lowest considered value of $a=0.9$ where,   $\significance_\ell$ gets larger than 2.  Therefore, a
fully efficient study of charged vector resonances with masses at (and heavier than)  2.5 TeV would imply to analyze the hadronic and semileptonic  channels of the $WZ$ final gauge bosons, as we have already indicated  above. 
\label{conclusions}

\section*{Acknowledgments}
We thank P. Arnan for providing us with the FORTRAN code to localize the IAM resonances and for his help at the early stages of this work. 
A.D. thanks F.J. Llanes-Estrada for previous collaboration. 
This work is supported by the European Union through 
the ITN ELUSIVES H2020-MSCA-ITN-2015//674896 
and the RISE INVISIBLESPLUS H2020-MSCA-RISE-2015//690575, 
by the Spanish MINECO through the projects FPA2013-46570-C2-1-P, FPA2014-53375-C2-1-P, FPA2016-75654-C2-1-P, FPA2016-76005-C2-1-P,  FPA2016-78645-P
(MINECO/ FEDER, EU),   
by the Spanish Consolider-Ingenio 2010 Programme CPAN (CSD2007-00042) 
and by the Spanish MINECO's ``Centro de Excelencia Severo Ochoa''  
Programme under grants SEV-2012-0249 and SEV-2016-0597 
and the ``Mar\'ia de Maeztu'' Programme under grant MDM-2014-0369.  
X.M. is supported through the Spanish MINECO ``Ram\'on y Cajal'' Programme (RYC-2015-17173). 
R.L.D is supported by the Spanish MINECO grant MINECO:BES-2012-056054, the MINECO project FIS2013-41716-P and the ``Ram\'on Areces'' Foundation. We also acknowledge 8000 hours of computer time granted at a small departamental cluster at the UCM.


%
%
%
%
%
%
%
%

\newpage
\begin{appendix}
\section*{Appendices}
\section{Relevant Feynman rules for $\boldsymbol{A(WZ\to WZ)^{\rm SM}}$} \label{FR-SM}
\begin{figure}[H]
\begin{center}
\includegraphics[scale=0.94]{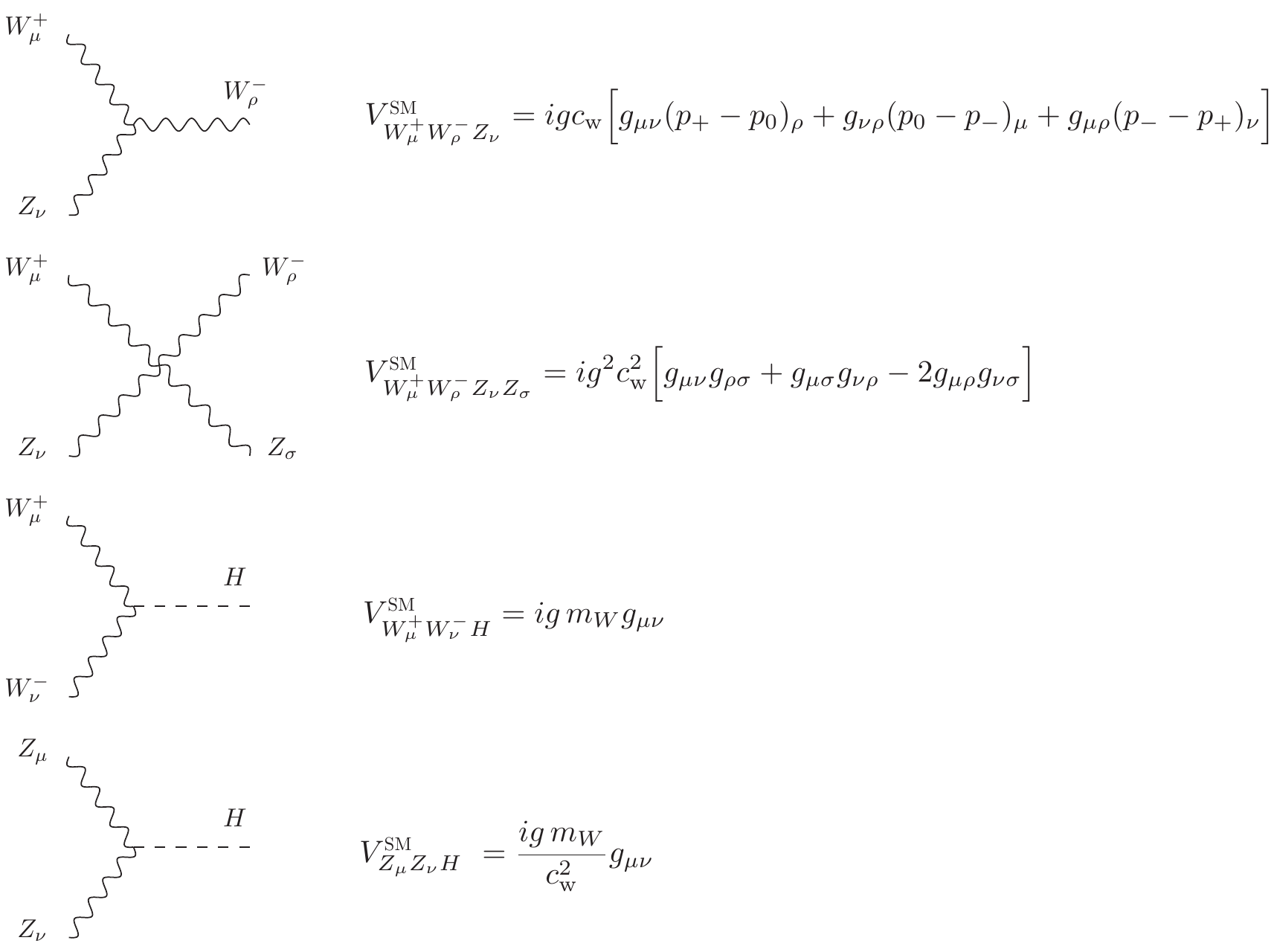}\\
\caption{Relevant Feynman rules for the $WZ\to WZ$ process in the SM. We take all momenta as incoming.}\label{fig:FR}
\end{center}
\end{figure}

In this appendix we collect the relevant Feynman rules, \figref{fig:FR}, for the computation of the $A(WZ\to WZ)$ scattering amplitude in the SM at the tree level. Notice that our conventions here for the SM Feynman rules are the same as in FeynRules~\cite{Alloul:2013bka}, except for the sign in the vertex $V^{\rm SM}_{W^+W^-Z}$ that is opposite. However, this will not give any difference in the predicted amplitudes nor in the predicted events with MadGraph5 (which uses the FeynRules conventions), since this particular vertex always appears squared in all quantities predicted in the present work.
We use here and in the following the short notation $\cosw=\cos\theta_W$. We also label the momenta according to the charge of the associated particle. This way, $p_{\pm,0}$ refers to an incoming $W^\pm$ or a $Z$ respectively. 
\vspace{0.2cm}
\newpage
\section{Relevant Feynman rules for $\boldsymbol{A(WZ\to WZ)^{\rm EChL}}$} \label{FR-EChL}
\begin{figure}[H]
\begin{center}
\includegraphics[scale=0.94]{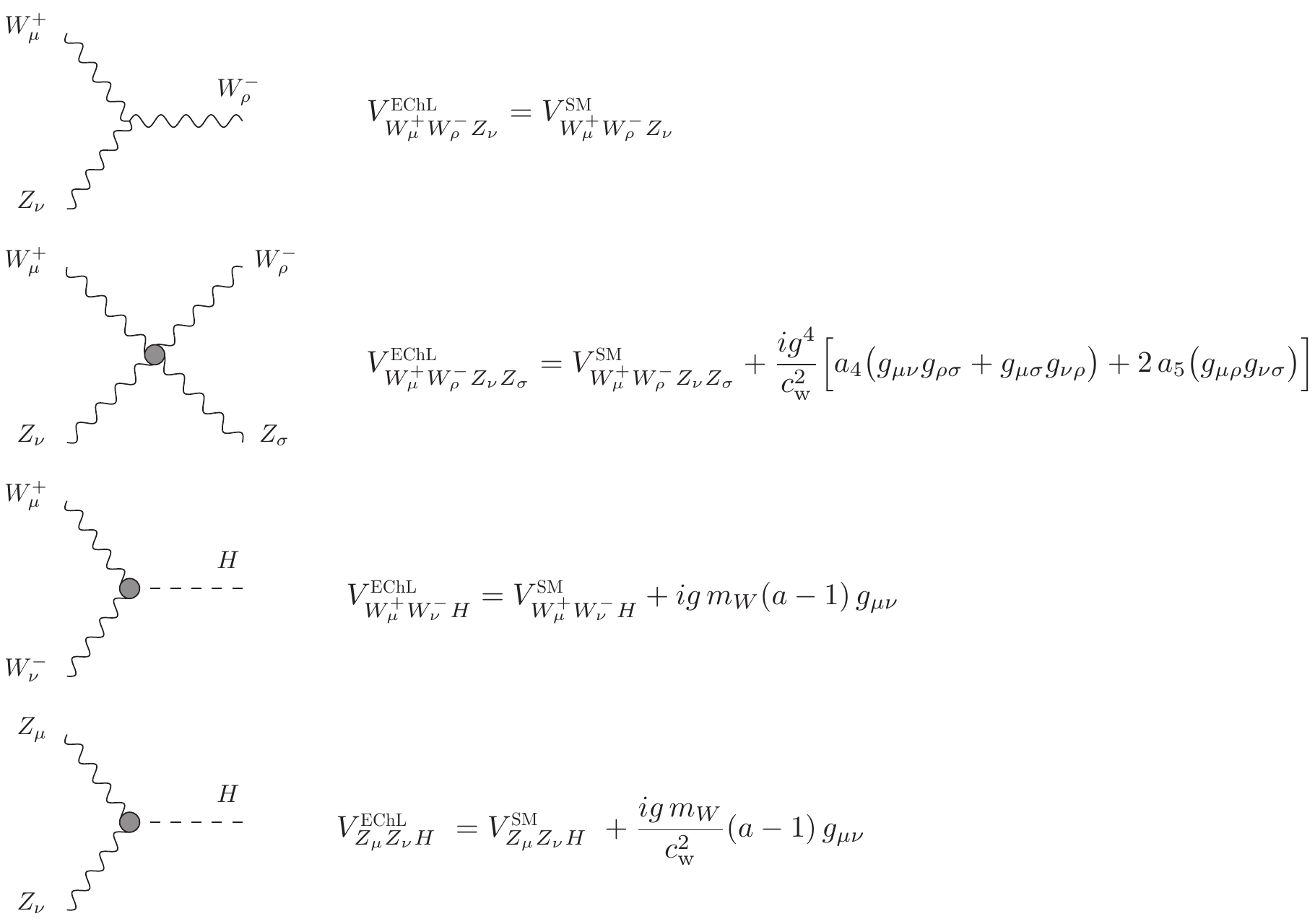}\\
\vspace{1cm}
\caption{Relevant Feynman rules for the $WZ\to WZ$ process in the EChL. Gray circles represent vertices that are sensitive to the chiral parameters $a, a_4$ and $a_5$ of our simplified scenario.
We take all momenta as incoming.}
\label{fig:FREChL}
\end{center}
\end{figure}

\vspace{0.2cm}
In this appendix we summarize the relevant EChL Feynman rules, \figref{fig:FREChL}, for the computation of the $A(WZ\to WZ)$ scattering amplitude at the tree level. These rules come from $\mL_2$, defined in \eqref{eq.L2}, and $\mL_4$, defined in \eqref{eq.L4}, as we are computing up to order $\mO(p^4)$. We signal with a gray circle the vertices that receive contributions from the chiral parameters that we consider in this work, $a, a_4$ and $a_5$. We also present these Feynman rules with the SM common part singled out  for an easier comparison.
\vspace{1cm}

\newpage
\section{Relevant Feynman rules for $\boldsymbol{A(WZ\to WZ)^{\rm IAM-MC}}$} \label{FR-IAM-MC}
\begin{figure}[H]
\begin{center}
\includegraphics[scale=0.94]{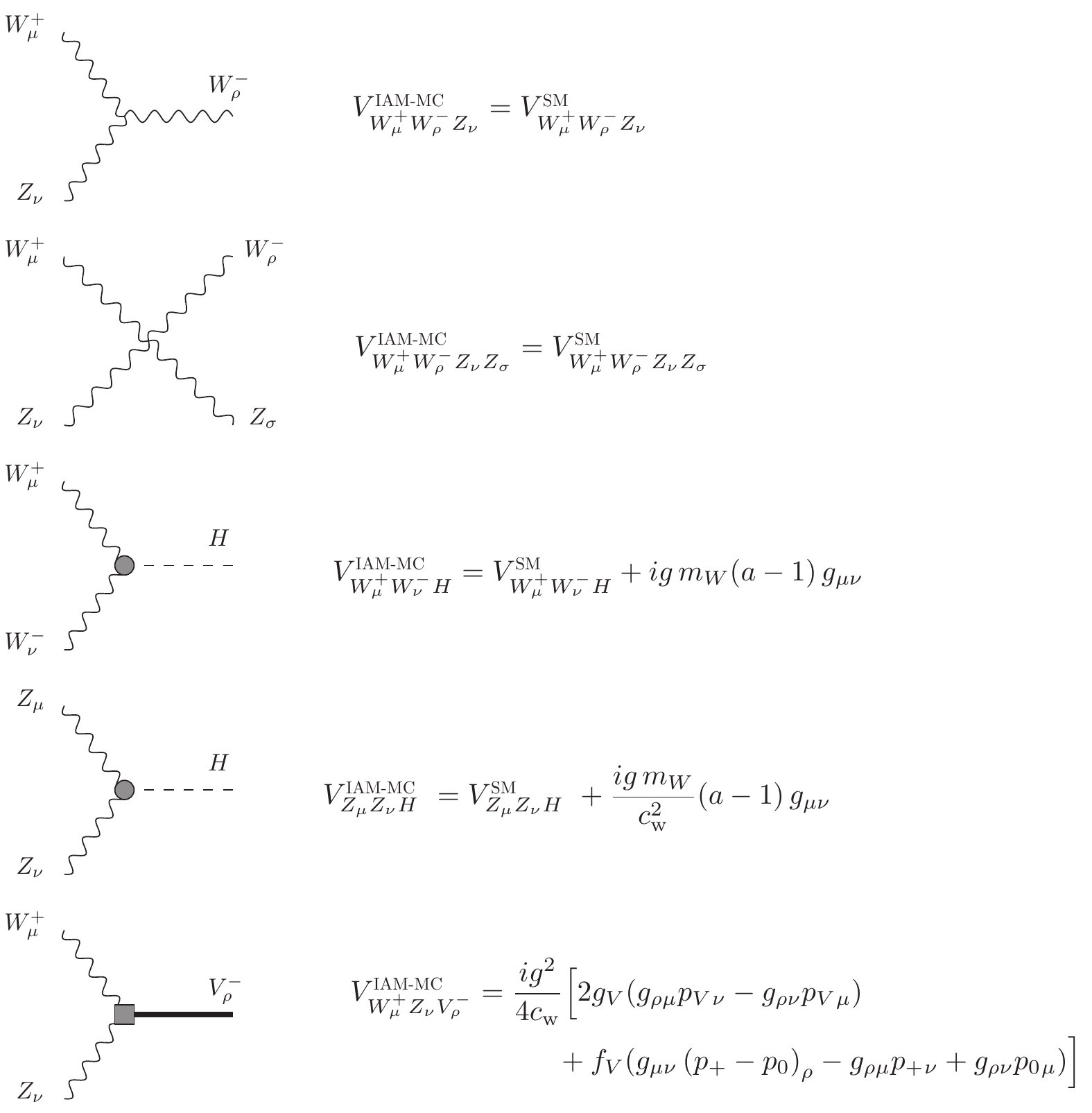}\\
\vspace{0.5cm}
\caption{Relevant Feynman rules for the $WZ\to WZ$ process in the IAM-MC. Gray circles represent vertices that are sensitive to the chiral parameter $a$. The gray square shows the vertex with contributions from $\mL_V$.
We take all momenta as incoming.}
\label{fig:FRIAMMC}
\end{center}
\end{figure}

\vspace{0.2cm}
In this appendix we summarize the relevant Feynman rules, \figref{fig:FRIAMMC}, for the computation of the $A(WZ\to WZ)$ scattering amplitude in our IAM-MC at the tree level. These rules come from $\mL_2$, defined in \eqref{eq.L2}, and from $\mL_V$ in \eqref{LVugauge}. We signal with a gray circle the vertices that receive contributions from the chiral parameter $a$, and with a gray square the one that involves the charged resonance, $V^\pm$, and therefore $g_V$. We also show, for completeness, the terms involving $f_V$ from $\mL_V$, although in all the numerical estimates in this work we set it to 0. 
\vspace{0.5cm}

\section{Analytical expressions for $\boldsymbol{A(WZ \to WZ)^{\rm SM}_{\rm tree}}$}
\begin{figure}[b!]
\begin{center}
\includegraphics[width=.6\textwidth]{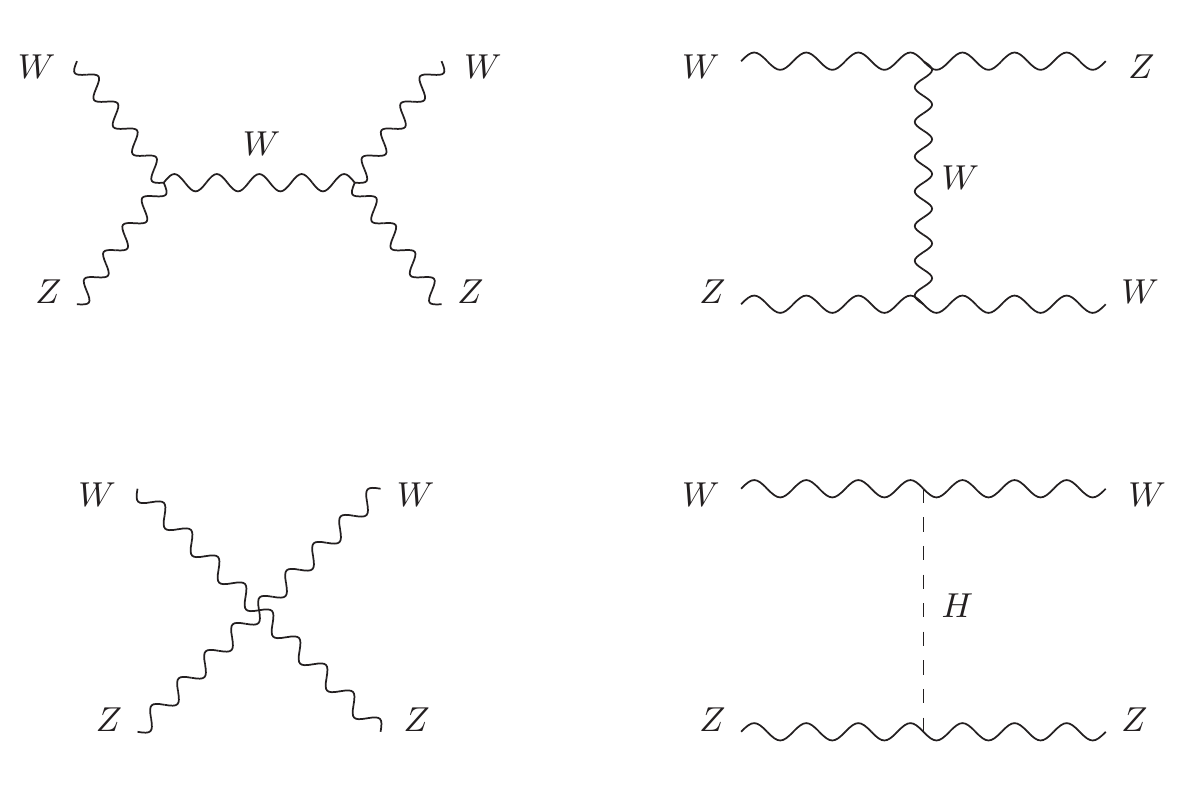}\\
\caption{Feynman diagrams that contribute to the $A(WZ \to WZ)^{\rm SM}_{\rm tree}$ amplitude in the SM at the tree level and in the unitary gauge. }\label{fig:DiagramsSM}
\end{center}
\end{figure}

The total amplitude $A(W^+(k_1, \euno)Z(k_2, \edos)\to W^+(k_3, \varepsilon_3)Z(k_4, \varepsilon_4))^{\rm SM}_{\rm tree}$ reads:
\begin{equation}
A(WZ \to WZ)^{\rm SM}_{\rm tree}=A_c^{\rm SM}+A_{sW}^{\rm SM}+A_{tH}^{\rm SM}+A_{uW}^{\rm SM}\,,
\end{equation}
where we have used a shorthand notation to name the amplitude of each of the diagrams that contribute to the process, depicted in \figref{fig:DiagramsSM}: contact, $A_c^{\rm SM}$, $s$-channel with a propagating $W$, $A_{sW}^{\rm SM}$, $t$-channel with a propagating Higgs, $A_{tH}^{\rm SM}$, and $u$-channel with a propagating $W$, $A_{uW}^{\rm SM}$.
We find the following analytical results for the varios contributions to the amplitude:
\begin{align}
A_c^{\rm SM}=&\,g^2 \cosw^2 \, \Big[(\varepsilon_1\cdot \varepsilon^*_4) (\varepsilon_2\cdot \varepsilon^*_3) + (\varepsilon_1\cdot \varepsilon_2) (\varepsilon^*_3\cdot \varepsilon^*_4)- 2 (\varepsilon_1\cdot \varepsilon^*_3)(\varepsilon_2\cdot \varepsilon^*_4) \Big]\,,\label{AcSM}\\
A_{sW}
^{\rm SM}=&-\dfrac{g^2}{\cosw^2}\dfrac{1}{s-m_W^2} \bigg[
(\varepsilon_1\cdot \varepsilon_2)(\varepsilon^*_3\cdot \varepsilon^*_4)\Big(\sinw^4 \, m_W^2 + \cosw^4(t - u)\Big)\nn \\
&\hspace{2.5cm}+4\cosw^4%
  (\varepsilon_2\cdot k_1)\Big[(\varepsilon_1\cdot k_3)(\varepsilon^*_3\cdot \varepsilon^*_4) %
+(\varepsilon_1\cdot \varepsilon^*_4)(\varepsilon^*_3\cdot k_4) %
- (\varepsilon_1\cdot \varepsilon^*_3)(\varepsilon^*_4\cdot k_3)\Big] \nn \\
&\hspace{2.5cm} 
- 4\cosw^4(\varepsilon_1\cdot k_2)\Big[(\varepsilon_2\cdot k_3)(\varepsilon^*_3\cdot \varepsilon^*_4) + (\varepsilon_2\cdot \varepsilon^*_4)(\varepsilon^*_3\cdot k_4) - (\varepsilon_2\cdot \varepsilon^*_3)(\varepsilon^*_4\cdot k_3)\Big]\nn \\
&\hspace{2.5cm} 
- 4\cosw^4(\varepsilon_1\cdot \varepsilon_2)\Big[(\varepsilon^*_3\cdot k_4)(\varepsilon^*_4\cdot k_1))-(\varepsilon^*_3\cdot k_1)(\varepsilon^*_4\cdot k_3)\Big]
\bigg]\,,
\\
A_{tH}^{\rm SM}=&-\dfrac{g^2}{\cosw^2}  \dfrac{m_W^2}{t-m_H^2}\, (\varepsilon_1\cdot \varepsilon^*_3) (\varepsilon_2\cdot \varepsilon^*_4)\,,
\\
A_{uW}^{\rm SM}=&-\frac{g^2}{\cosw^2}\frac{1}{u-m_W^2}\bigg[%
(\varepsilon_1\cdot \varepsilon^*_4)(\varepsilon_2\cdot \varepsilon^*_3)\Big(\sinw^4\, m_W^2 + \cosw^4(t-s)\Big) \nn\\
&\hspace{2.5cm}- 4\cosw^4(\varepsilon^*_4\cdot k_1)\Big[(\varepsilon_1\cdot \varepsilon^*_3)(\varepsilon_2\cdot k_3) + (\varepsilon_1\cdot \varepsilon_2)(\varepsilon^*_3\cdot k_2)-(\varepsilon_1\cdot k_2)(\varepsilon_2\cdot \varepsilon^*_3)\Big]\nn \\
&\hspace{2.5cm}+4\cosw^4(\varepsilon_1\cdot k_4)\Big[(\varepsilon_2\cdot k_3)(\varepsilon^*_3\cdot \varepsilon^*_4) + (\varepsilon_2\cdot \varepsilon^*_4)(\varepsilon^*_3\cdot k_2) - (\varepsilon_2\cdot \varepsilon^*_3)(\varepsilon^*_4\cdot k_2)\Big]\nn \\
& \hspace{2.5cm}
-4\cosw^4(\varepsilon_1\cdot \varepsilon^*_4)\Big[(\varepsilon_2\cdot k_3)(\varepsilon^*_3\cdot k_1) + (\varepsilon_2\cdot k_1)(\varepsilon^*_3\cdot k_2)\Big]%
\bigg]\,.\label{AuSM}
\end{align}

\section{Analytical expressions for  $\boldsymbol{A(WZ \to WZ)^{\rm EChL}_{\rm tree}}$}
\begin{figure}[t!]
\begin{center}
\includegraphics[width=.6\textwidth]{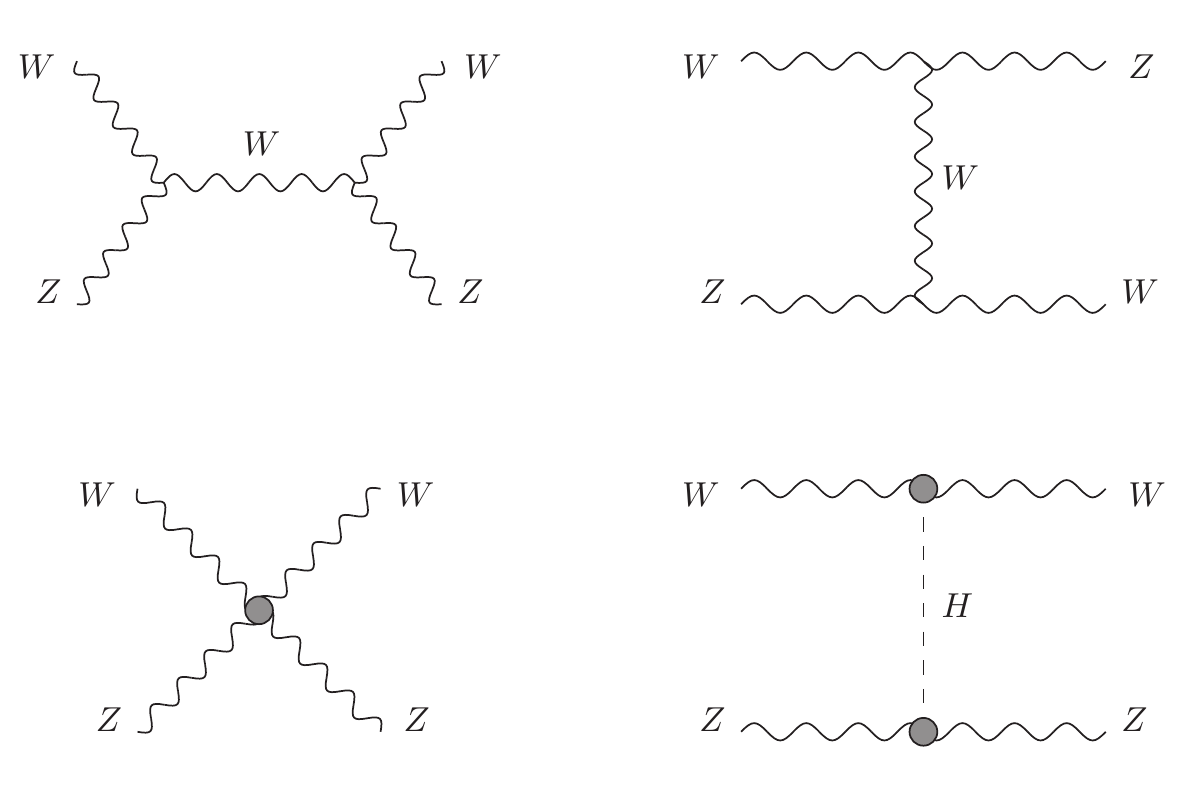}\\
\caption{Feynman diagrams that contribute to the $A(WZ \to WZ)^{\rm EChL}_{\rm tree}$ amplitude in the EChL and in the unitary gauge. Gray circles represent vertices that are sensitive to the chiral parameters $a, a_4$ and $a_5$ of our simplified scenario. }\label{fig:DiagramsEChL}
\end{center}
\end{figure}

The total amplitude,  $A(W^+(k_1, \euno)Z(k_2, \edos)\to W^+(k_3, \varepsilon_3)Z(k_4, \varepsilon_4))^{\rm EChL}_{\rm tree}$, computed with the EChL at the tree level is:
\begin{equation}
A(WZ \to WZ)^{\rm EChL}_{\rm tree}=A_c^{\rm EChL}+A_{sW}^{\rm EChL}+A_{tH}^{\rm EChL}+A_{uW}^{\rm EChL}\,,
\end{equation}
 quantified in the gray dots of the diagrams in \figref{fig:DiagramsEChL}, which are again the ones that contribute to the process of interest.
We find the following results for the various contributions to the amplitude:
\begin{align}
A_c^{\rm EChL}&=A_c^{\rm SM}+\dfrac{g^4}{\cosw^2}\bigg[a_4~\Big((\varepsilon_1\cdot \varepsilon_4^*)(\varepsilon_2\cdot \varepsilon_3^*)+(\varepsilon_1\cdot \varepsilon_2)(\varepsilon_3^*\cdot \varepsilon_4^*)\Big)
+2a_5(\varepsilon_1\cdot \varepsilon_3^*)(\varepsilon_2\cdot \varepsilon_4^*)\bigg]\,,\\
A_{sW}^{\rm EChL}&=A_{sW}^{\rm SM}\,,\\
A_t^{\rm EChL}&= A_{tH}^{\rm SM}\ a^2\,,\\
A_{uW}^{\rm EChL}&=A_{uW}^{\rm SM}\,.
\end{align}
\section{Analytical expressions for $\boldsymbol{A(WZ \to WZ)^{\rm IAM-MC}}$}

\begin{figure}[t!]
\begin{center}
\includegraphics[width=.6\textwidth]{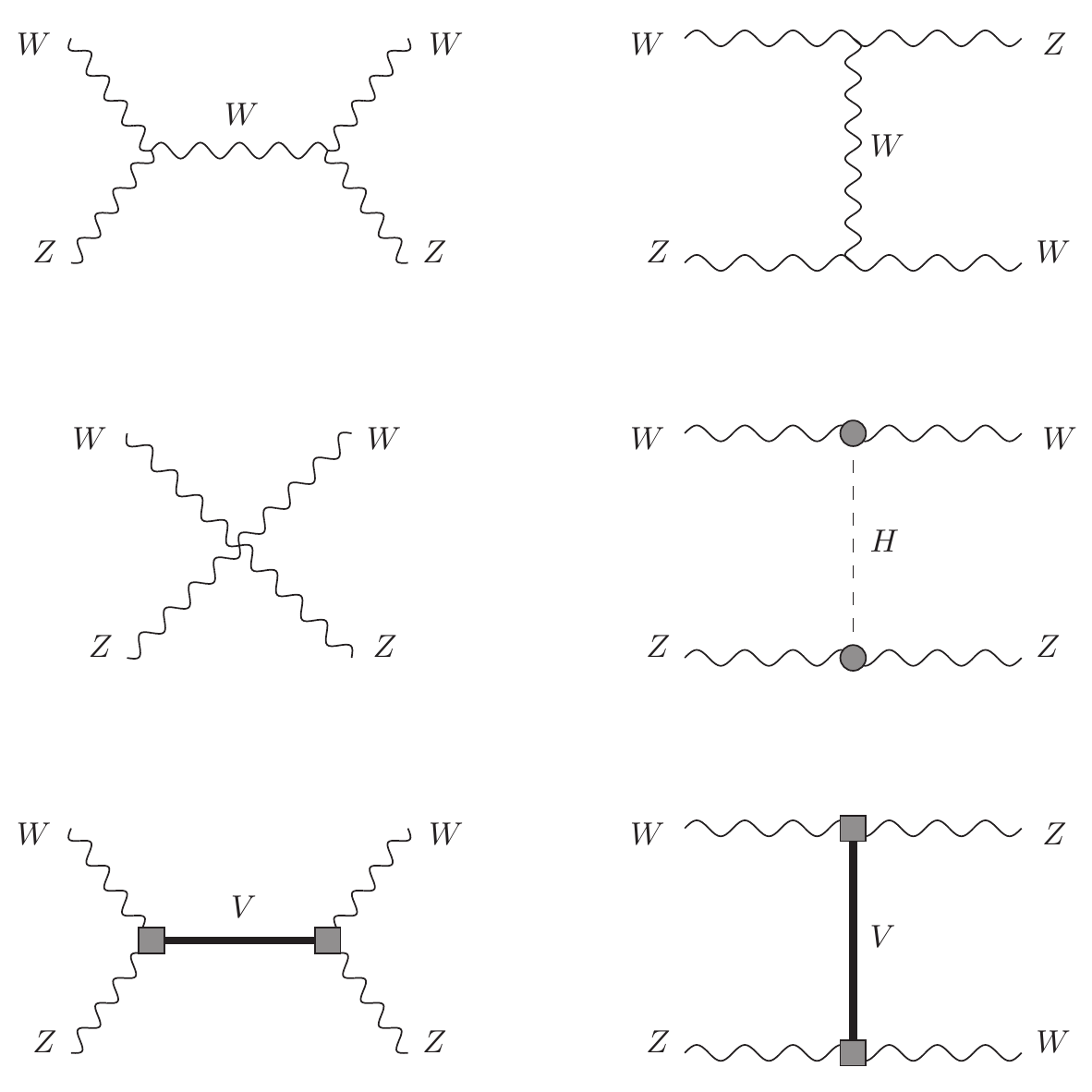}\\
\caption{Feynman diagrams that contribute to the $A(WZ \to WZ)^{\rm IAM-MC}_{\rm tree}$ amplitude in the IAM-MC and in the unitary gauge. Gray circles represent vertices that are sensitive to the chiral parameter $a$.
Gray squares show vertices with contributions from $\mL_V$. }
\label{fig:DiagramsIAMMC}
\end{center}
\end{figure}

Finally, we present the amplitudes that allow to compute the total prediction,  $A(W^+(k_1, \euno)Z(k_2, \edos)\to W^+(k_3, \varepsilon_3)Z(k_4, \varepsilon_4))^{\rm IAM-MC}_{\rm tree}$, of our model, the IAM-MC. In this case we have: \begin{equation}
A(WZ \to WZ)^{\rm IAM-MC}_{\rm tree}=A_c^{\rm IAM-MC}+A_{sW}^{\rm IAM-MC}+A_{tH}^{\rm IAM-MC}+A_{uW}^{\rm IAM-MC}+A_{sV}^{\rm IAM-MC}+A_{uV}^{\rm IAM-MC}\,,\label{IAMMCamp}
\end{equation}
because of the two extra diagrams involving the resonance, as shown in \figref{fig:DiagramsIAMMC}. Here the deviations from the SM are encoded in the gray dots (contributions from $a\neq 1$ in $\mL_2$) and in the gray squares (resonance couplings) of the above diagrams.
We find the following analytical results for the various contributions to the amplitude:

\begin{align}
A_c^{\rm IAM-MC}&=A_c^{\rm SM}\,,\\
%
A_{sW}^{\rm IAM-MC}&=A_{sW}^{\rm SM}\,,\\
A_{tH}^{\rm IAM-MC}&=A_{tH}^{\rm EChL}\,,\\
A_{uW}^{\rm IAM-MC}&=A_{uW}^{\rm SM}\,,\\
A_{sV}^{\rm IAM-MC}&=~\dfrac{g^4}{4\cosw^2}\dfrac{g_V^2}{s-M_V^2+iM_V\Gamma_V}\bigg[(\edos\cdot k_1)\Big[(\euno\cdot\etres)(\ecuatro\cdot k_3)-(\euno\cdot\ecuatro)(\etres\cdot k_4)\Big]\nn\\
&\hspace{4.2cm}+(\euno\cdot k_2)\Big[(\edos\cdot\ecuatro)(\etres\cdot k_4)-(\edos\cdot\etres)(\ecuatro\cdot k_3)\Big]\bigg]\nn\\
&+\dfrac{g^4}{16\cosw^2}\dfrac{f_V^2}{s-M_V^2+iM_V\Gamma_V}\bigg[\Big(u-t-\dfrac{\sinw^4 }{\cosw^4 }\dfrac{m_W^4}{M_V^2}\Big)\,(\euno\cdot \edos)(\etres\cdot \ecuatro)\nn\\
&\hspace{4cm}+2\,(\euno\cdot \edos)\Big[(\etres\cdot k_4)(\ecuatro\cdot k_1)-(\etres\cdot k_1)(\ecuatro\cdot k_3)\Big]\nn\\
&\hspace{4cm}+2\,(\etres\cdot \ecuatro)\Big[(\euno\cdot k_2)(\edos\cdot k_3)-(\euno\cdot k_3)(\edos\cdot k_1)\Big]\nn\\
&\hspace{4cm}+~~(\edos\cdot k_1)\Big[(\euno\cdot \etres)(\ecuatro\cdot k_3)-(\euno\cdot \ecuatro)(\etres\cdot k_4)\Big]\nn\\
&\hspace{4cm}+~~(\euno\cdot k_2)\Big[(\edos\cdot\ecuatro)(\etres\cdot k_4)-(\edos\cdot\etres)(\ecuatro\cdot k_3)\Big]\bigg]\nn\\
&+\dfrac{g^4}{4\cosw^2}\dfrac{g_V f_V}{s-M_V^2+iM_V\Gamma_V}\bigg[
(\edos\cdot k_1)\Big[(\euno\cdot\etres)(\ecuatro\cdot k_3)-(\euno\cdot\ecuatro)(\etres\cdot k_4)\Big]\nn\\
&\hspace{4cm}+(\euno\cdot k_2)\Big[(\edos\cdot\ecuatro)(\etres\cdot k_4)-(\edos\cdot\etres)(\ecuatro\cdot k_3)\Big]\nn\\
&\hspace{4cm}+(\euno\cdot\edos)\Big[(\etres\cdot k_4)(\ecuatro\cdot k_1)-(\etres\cdot k_1)(\ecuatro\cdot k_3)\Big]\nn\\
&\hspace{4cm}+(\etres\cdot\ecuatro)\Big[(\euno\cdot k_2)(\edos\cdot k_3)-(\euno\cdot k_3)(\edos\cdot k_1)\Big]\bigg]\,,\\
A_{uV}^{\rm IAM-MC}&=A_{sV}^{\rm IAM-MC} \big(k_2 \leftrightarrow -k_4,~ \edos\leftrightarrow\ecuatro\big)\,.
\end{align}

It must be noticed that when computing $A_{uV}^{\rm IAM-MC}$ the width is  not appearing in the propagator.

Finally, we present the four point vertex, $\Gamma_{W^+_\mu Z_\nu W^+_\sigma Z_\lambda}^{\rm IAM-MC}$ shown schematically in \figref{fig:diagppWZjj}. It corresponds to the total IAM-MC amplitude  coming from the computation of the diagrams displayed in \figref{fig:DiagramsIAMMC}, i.e., the formula presented in \eqref{IAMMCamp}, with the polarization vectors factored out. It receives contributions from $\mL_2$ and $\mL_V$, as defined in the section \ref{sec-model},
\begin{equation}
-i\,\Gamma_{W^+_\mu Z_\nu W^+_\sigma Z_\lambda}^{\rm IAM-MC} =
-i\,\Gamma_{W^+_\mu Z_\nu W^+_\sigma Z_\lambda}^{\mL_2} 
-i\,\Gamma_{W^+_\mu Z_\nu W^+_\sigma Z_\lambda}^{\mL_V}\,,
\label{fpfUFOMJ}
\end{equation}
or, equivalently, extracting the SM amplitude out,
\begin{equation}
-i\,\Gamma_{W^+_\mu Z_\nu W^+_\sigma Z_\lambda}^{\rm IAM-MC} =
-i\,\Gamma_{W^+_\mu Z_\nu W^+_\sigma Z_\lambda}^{\rm SM} 
-i\,\Gamma_{W^+_\mu Z_\nu W^+_\sigma Z_\lambda}^{ (a-1)}
-i\,\Gamma_{W^+_\mu Z_\nu W^+_\sigma Z_\lambda}^{\mL_V}\,.
\label{fpfUFO}
\end{equation}
Here $\Gamma^{\rm SM}$ comes from the diagrams in \figref{fig:DiagramsSM}, $\Gamma^{(a-1)}$ denotes the new effects introduced by  $\mL_2$ with $a\neq1$ with respect to the SM and $\Gamma^{\mL_V}$ accounts for the new contributions from the dynamically generated resonance. The decomposition defined in \eqref{fpfUFO} turns out to be very convenient to introduce our model in MadGraph, as one can use the SM default model as the basic tool to build the UFO. In this way, we just add up to the SM model files the $\Gamma^{(a-1)}$ and $\Gamma^{\mL_V}$ as four point effective vertices given by:
\begin{align}
-i\,\Gamma_{W^+_\mu Z_\nu W^+_\sigma Z_\lambda}^{(a-1)} &= -\dfrac{g^2}{\cosw^2}\,\dfrac{m_W^2}{t-m_H^2}\,
\big(a^2-1\big)\, g_{\mu\sigma} g_{\nu\lambda}\,,\\
-i\,\Gamma_{W^+_\mu Z_\nu W^+_\sigma Z_\lambda}^{\mL_V} &=\, \dfrac{g^4}{4\, {\rm c}^2_{\rm w}} \Bigg[\dfrac{g_V^2(s)}{s-M_V^2+iM_V\Gamma_V}\Big[h_\nu h_\lambda g_{\mu\sigma}-h_\nu h_\sigma g_{\mu\lambda}-h_\mu h_\lambda g_{\nu\sigma}+h_\mu h_\sigma g_{\nu\lambda}\Big] \nn \\
&\hspace{1.3cm} +\dfrac{g_V^2(u)}{u-M_V^2}\Big[l_\nu l_\lambda g_{\mu\sigma}-l_\lambda h_\sigma g_{\mu\nu} -l_\mu l_\nu g_{\lambda\sigma} + l_\mu l_\sigma g_{\nu\lambda} \Big]\Bigg]\,,\label{fpfUFOparts}
\end{align}
where $h=k_1+k_2$ and $l=k_1-k_4$. The energy dependent couplings $g_V(s)$ and $g_V(u)$ are the ones defined in \eqrefs{gvenergy}-(\ref{gvenergytu}).

\end{appendix}

\bibliography{bibliography}

\begin{thebibliography}{83}%
\makeatletter
\providecommand \@ifxundefined [1]{%
 \@ifx{#1\undefined}
}%
\providecommand \@ifnum [1]{%
 \ifnum #1\expandafter \@firstoftwo
 \else \expandafter \@secondoftwo
 \fi
}%
\providecommand \@ifx [1]{%
 \ifx #1\expandafter \@firstoftwo
 \else \expandafter \@secondoftwo
 \fi
}%
\providecommand \natexlab [1]{#1}%
\providecommand \enquote  [1]{``#1''}%
\providecommand \bibnamefont  [1]{#1}%
\providecommand \bibfnamefont [1]{#1}%
\providecommand \citenamefont [1]{#1}%
\providecommand \href@noop [0]{\@secondoftwo}%
\providecommand \href [0]{\begingroup \@sanitize@url \@href}%
\providecommand \@href[1]{\@@startlink{#1}\@@href}%
\providecommand \@@href[1]{\endgroup#1\@@endlink}%
\providecommand \@sanitize@url [0]{\catcode `\\12\catcode `\$12\catcode
  `\&12\catcode `\#12\catcode `\^12\catcode `\_12\catcode `\%12\relax}%
\providecommand \@@startlink[1]{}%
\providecommand \@@endlink[0]{}%
\providecommand \url  [0]{\begingroup\@sanitize@url \@url }%
\providecommand \@url [1]{\endgroup\@href {#1}{\urlprefix }}%
\providecommand \urlprefix  [0]{URL }%
\providecommand \Eprint [0]{\href }%
\providecommand \doibase [0]{http://dx.doi.org/}%
\providecommand \selectlanguage [0]{\@gobble}%
\providecommand \bibinfo  [0]{\@secondoftwo}%
\providecommand \bibfield  [0]{\@secondoftwo}%
\providecommand \translation [1]{[#1]}%
\providecommand \BibitemOpen [0]{}%
\providecommand \bibitemStop [0]{}%
\providecommand \bibitemNoStop [0]{.\EOS\space}%
\providecommand \EOS [0]{\spacefactor3000\relax}%
\providecommand \BibitemShut  [1]{\csname bibitem#1\endcsname}%
\let\auto@bib@innerbib\@empty
\bibitem [{\citenamefont {Agashe}\ \emph {et~al.}(2005)\citenamefont {Agashe},
  \citenamefont {Contino},\ and\ \citenamefont {Pomarol}}]{Agashe:2004rs}%
  \BibitemOpen
  \bibfield  {author} {\bibinfo {author} {\bibfnamefont {K.}~\bibnamefont
  {Agashe}}, \bibinfo {author} {\bibfnamefont {R.}~\bibnamefont {Contino}}, \
  and\ \bibinfo {author} {\bibfnamefont {A.}~\bibnamefont {Pomarol}},\ }\href
  {\doibase 10.1016/j.nuclphysb.2005.04.035} {\bibfield  {journal} {\bibinfo
  {journal} {Nucl. Phys.}\ }\textbf {\bibinfo {volume} {B719}},\ \bibinfo
  {pages} {165} (\bibinfo {year} {2005})},\ \Eprint
  {http://arxiv.org/abs/hep-ph/0412089} {arXiv:hep-ph/0412089 [hep-ph]}
  \BibitemShut {NoStop}%
\bibitem [{\citenamefont {Appelquist}\ and\ \citenamefont
  {Bernard}(1980)}]{Appelquist:1980vg}%
  \BibitemOpen
  \bibfield  {author} {\bibinfo {author} {\bibfnamefont {T.}~\bibnamefont
  {Appelquist}}\ and\ \bibinfo {author} {\bibfnamefont {C.~W.}\ \bibnamefont
  {Bernard}},\ }\href {\doibase 10.1103/PhysRevD.22.200} {\bibfield  {journal}
  {\bibinfo  {journal} {Phys. Rev.}\ }\textbf {\bibinfo {volume} {D22}},\
  \bibinfo {pages} {200} (\bibinfo {year} {1980})}\BibitemShut {NoStop}%
\bibitem [{\citenamefont {Longhitano}(1980)}]{Longhitano:1980iz}%
  \BibitemOpen
  \bibfield  {author} {\bibinfo {author} {\bibfnamefont {A.~C.}\ \bibnamefont
  {Longhitano}},\ }\href {\doibase 10.1103/PhysRevD.22.1166} {\bibfield
  {journal} {\bibinfo  {journal} {Phys. Rev.}\ }\textbf {\bibinfo {volume}
  {D22}},\ \bibinfo {pages} {1166} (\bibinfo {year} {1980})}\BibitemShut
  {NoStop}%
\bibitem [{\citenamefont {Longhitano}(1981)}]{Longhitano:1980tm}%
  \BibitemOpen
  \bibfield  {author} {\bibinfo {author} {\bibfnamefont {A.~C.}\ \bibnamefont
  {Longhitano}},\ }\href {\doibase 10.1016/0550-3213(81)90109-7} {\bibfield
  {journal} {\bibinfo  {journal} {Nucl. Phys.}\ }\textbf {\bibinfo {volume}
  {B188}},\ \bibinfo {pages} {118} (\bibinfo {year} {1981})}\BibitemShut
  {NoStop}%
\bibitem [{\citenamefont {Chanowitz}\ and\ \citenamefont
  {Gaillard}(1985)}]{Chanowitz:1985hj}%
  \BibitemOpen
  \bibfield  {author} {\bibinfo {author} {\bibfnamefont {M.~S.}\ \bibnamefont
  {Chanowitz}}\ and\ \bibinfo {author} {\bibfnamefont {M.~K.}\ \bibnamefont
  {Gaillard}},\ }\href {\doibase 10.1016/0550-3213(85)90580-2} {\bibfield
  {journal} {\bibinfo  {journal} {Nucl. Phys.}\ }\textbf {\bibinfo {volume}
  {B261}},\ \bibinfo {pages} {379} (\bibinfo {year} {1985})}\BibitemShut
  {NoStop}%
\bibitem [{\citenamefont {Cheyette}\ and\ \citenamefont
  {Gaillard}(1987)}]{Cheyette:1987jf}%
  \BibitemOpen
  \bibfield  {author} {\bibinfo {author} {\bibfnamefont {O.}~\bibnamefont
  {Cheyette}}\ and\ \bibinfo {author} {\bibfnamefont {M.~K.}\ \bibnamefont
  {Gaillard}},\ }\href {\doibase 10.1016/0370-2693(87)90369-8} {\bibfield
  {journal} {\bibinfo  {journal} {Phys. Lett.}\ }\textbf {\bibinfo {volume}
  {B197}},\ \bibinfo {pages} {205} (\bibinfo {year} {1987})}\BibitemShut
  {NoStop}%
\bibitem [{\citenamefont {Dobado}\ and\ \citenamefont
  {Herrero}(1989{\natexlab{a}})}]{Dobado:1989ax}%
  \BibitemOpen
  \bibfield  {author} {\bibinfo {author} {\bibfnamefont {A.}~\bibnamefont
  {Dobado}}\ and\ \bibinfo {author} {\bibfnamefont {M.~J.}\ \bibnamefont
  {Herrero}},\ }\href {\doibase 10.1016/0370-2693(89)90981-7} {\bibfield
  {journal} {\bibinfo  {journal} {Phys. Lett.}\ }\textbf {\bibinfo {volume}
  {B228}},\ \bibinfo {pages} {495} (\bibinfo {year}
  {1989}{\natexlab{a}})}\BibitemShut {NoStop}%
\bibitem [{\citenamefont {Dobado}\ and\ \citenamefont
  {Herrero}(1989{\natexlab{b}})}]{Dobado:1989ue}%
  \BibitemOpen
  \bibfield  {author} {\bibinfo {author} {\bibfnamefont {A.}~\bibnamefont
  {Dobado}}\ and\ \bibinfo {author} {\bibfnamefont {M.~J.}\ \bibnamefont
  {Herrero}},\ }\href {\doibase 10.1016/0370-2693(89)91349-X} {\bibfield
  {journal} {\bibinfo  {journal} {Phys. Lett.}\ }\textbf {\bibinfo {volume}
  {B233}},\ \bibinfo {pages} {505} (\bibinfo {year}
  {1989}{\natexlab{b}})}\BibitemShut {NoStop}%
\bibitem [{\citenamefont {Weinberg}(1979)}]{Weinberg:1978kz}%
  \BibitemOpen
  \bibfield  {author} {\bibinfo {author} {\bibfnamefont {S.}~\bibnamefont
  {Weinberg}},\ }\href {\doibase 10.1016/0378-4371(79)90223-1} {\bibfield
  {journal} {\bibinfo  {journal} {Physica}\ }\textbf {\bibinfo {volume}
  {A96}},\ \bibinfo {pages} {327} (\bibinfo {year} {1979})}\BibitemShut
  {NoStop}%
\bibitem [{\citenamefont {Gasser}\ and\ \citenamefont
  {Leutwyler}(1985)}]{Gasser:1984gg}%
  \BibitemOpen
  \bibfield  {author} {\bibinfo {author} {\bibfnamefont {J.}~\bibnamefont
  {Gasser}}\ and\ \bibinfo {author} {\bibfnamefont {H.}~\bibnamefont
  {Leutwyler}},\ }\href {\doibase 10.1016/0550-3213(85)90492-4} {\bibfield
  {journal} {\bibinfo  {journal} {Nucl. Phys.}\ }\textbf {\bibinfo {volume}
  {B250}},\ \bibinfo {pages} {465} (\bibinfo {year} {1985})}\BibitemShut
  {NoStop}%
\bibitem [{\citenamefont {Gasser}\ and\ \citenamefont
  {Leutwyler}(1984)}]{Gasser:1983yg}%
  \BibitemOpen
  \bibfield  {author} {\bibinfo {author} {\bibfnamefont {J.}~\bibnamefont
  {Gasser}}\ and\ \bibinfo {author} {\bibfnamefont {H.}~\bibnamefont
  {Leutwyler}},\ }\href {\doibase 10.1016/0003-4916(84)90242-2} {\bibfield
  {journal} {\bibinfo  {journal} {Annals Phys.}\ }\textbf {\bibinfo {volume}
  {158}},\ \bibinfo {pages} {142} (\bibinfo {year} {1984})}\BibitemShut
  {NoStop}%
\bibitem [{\citenamefont {Dobado}\ \emph
  {et~al.}(1991{\natexlab{a}})\citenamefont {Dobado}, \citenamefont {Espriu},\
  and\ \citenamefont {Herrero}}]{Dobado:1990zh}%
  \BibitemOpen
  \bibfield  {author} {\bibinfo {author} {\bibfnamefont {A.}~\bibnamefont
  {Dobado}}, \bibinfo {author} {\bibfnamefont {D.}~\bibnamefont {Espriu}}, \
  and\ \bibinfo {author} {\bibfnamefont {M.~J.}\ \bibnamefont {Herrero}},\
  }\href {\doibase 10.1016/0370-2693(91)90786-P} {\bibfield  {journal}
  {\bibinfo  {journal} {Phys. Lett.}\ }\textbf {\bibinfo {volume} {B255}},\
  \bibinfo {pages} {405} (\bibinfo {year} {1991}{\natexlab{a}})}\BibitemShut
  {NoStop}%
\bibitem [{\citenamefont {Espriu}\ and\ \citenamefont
  {Herrero}(1992)}]{Espriu:1991vm}%
  \BibitemOpen
  \bibfield  {author} {\bibinfo {author} {\bibfnamefont {D.}~\bibnamefont
  {Espriu}}\ and\ \bibinfo {author} {\bibfnamefont {M.~J.}\ \bibnamefont
  {Herrero}},\ }\href {\doibase 10.1016/0550-3213(92)90452-H} {\bibfield
  {journal} {\bibinfo  {journal} {Nucl. Phys.}\ }\textbf {\bibinfo {volume}
  {B373}},\ \bibinfo {pages} {117} (\bibinfo {year} {1992})}\BibitemShut
  {NoStop}%
\bibitem [{\citenamefont {Dobado}\ \emph
  {et~al.}(1991{\natexlab{b}})\citenamefont {Dobado}, \citenamefont {Herrero},\
  and\ \citenamefont {Terron}}]{Dobado:1990jy}%
  \BibitemOpen
  \bibfield  {author} {\bibinfo {author} {\bibfnamefont {A.}~\bibnamefont
  {Dobado}}, \bibinfo {author} {\bibfnamefont {M.~J.}\ \bibnamefont {Herrero}},
  \ and\ \bibinfo {author} {\bibfnamefont {J.}~\bibnamefont {Terron}},\ }\href
  {\doibase 10.1007/BF01474075} {\bibfield  {journal} {\bibinfo  {journal} {Z.
  Phys.}\ }\textbf {\bibinfo {volume} {C50}},\ \bibinfo {pages} {205} (\bibinfo
  {year} {1991}{\natexlab{b}})}\BibitemShut {NoStop}%
\bibitem [{\citenamefont {Dobado}\ \emph
  {et~al.}(1991{\natexlab{c}})\citenamefont {Dobado}, \citenamefont {Herrero},\
  and\ \citenamefont {Terron}}]{Dobado:1990am}%
  \BibitemOpen
  \bibfield  {author} {\bibinfo {author} {\bibfnamefont {A.}~\bibnamefont
  {Dobado}}, \bibinfo {author} {\bibfnamefont {M.~J.}\ \bibnamefont {Herrero}},
  \ and\ \bibinfo {author} {\bibfnamefont {J.}~\bibnamefont {Terron}},\ }\href
  {\doibase 10.1007/BF01551458} {\bibfield  {journal} {\bibinfo  {journal} {Z.
  Phys.}\ }\textbf {\bibinfo {volume} {C50}},\ \bibinfo {pages} {465} (\bibinfo
  {year} {1991}{\natexlab{c}})}\BibitemShut {NoStop}%
\bibitem [{\citenamefont {Dobado}\ \emph {et~al.}(1995)\citenamefont {Dobado},
  \citenamefont {Herrero}, \citenamefont {Pelaez}, \citenamefont
  {Ruiz~Morales},\ and\ \citenamefont {Urdiales}}]{Dobado:1995qy}%
  \BibitemOpen
  \bibfield  {author} {\bibinfo {author} {\bibfnamefont {A.}~\bibnamefont
  {Dobado}}, \bibinfo {author} {\bibfnamefont {M.~J.}\ \bibnamefont {Herrero}},
  \bibinfo {author} {\bibfnamefont {J.~R.}\ \bibnamefont {Pelaez}}, \bibinfo
  {author} {\bibfnamefont {E.}~\bibnamefont {Ruiz~Morales}}, \ and\ \bibinfo
  {author} {\bibfnamefont {M.~T.}\ \bibnamefont {Urdiales}},\ }\href {\doibase
  10.1016/0370-2693(95)00431-J} {\bibfield  {journal} {\bibinfo  {journal}
  {Phys. Lett.}\ }\textbf {\bibinfo {volume} {B352}},\ \bibinfo {pages} {400}
  (\bibinfo {year} {1995})},\ \Eprint {http://arxiv.org/abs/hep-ph/9502309}
  {arXiv:hep-ph/9502309 [hep-ph]} \BibitemShut {NoStop}%
\bibitem [{\citenamefont {Dobado}\ \emph {et~al.}(2000)\citenamefont {Dobado},
  \citenamefont {Herrero}, \citenamefont {Pelaez},\ and\ \citenamefont
  {Ruiz~Morales}}]{Dobado:1999xb}%
  \BibitemOpen
  \bibfield  {author} {\bibinfo {author} {\bibfnamefont {A.}~\bibnamefont
  {Dobado}}, \bibinfo {author} {\bibfnamefont {M.~J.}\ \bibnamefont {Herrero}},
  \bibinfo {author} {\bibfnamefont {J.~R.}\ \bibnamefont {Pelaez}}, \ and\
  \bibinfo {author} {\bibfnamefont {E.}~\bibnamefont {Ruiz~Morales}},\ }\href
  {\doibase 10.1103/PhysRevD.62.055011} {\bibfield  {journal} {\bibinfo
  {journal} {Phys. Rev.}\ }\textbf {\bibinfo {volume} {D62}},\ \bibinfo {pages}
  {055011} (\bibinfo {year} {2000})},\ \Eprint
  {http://arxiv.org/abs/hep-ph/9912224} {arXiv:hep-ph/9912224 [hep-ph]}
  \BibitemShut {NoStop}%
\bibitem [{\citenamefont {Alonso}\ \emph {et~al.}(2013)\citenamefont {Alonso},
  \citenamefont {Gavela}, \citenamefont {Merlo}, \citenamefont {Rigolin},\ and\
  \citenamefont {Yepes}}]{Alonso:2012px}%
  \BibitemOpen
  \bibfield  {author} {\bibinfo {author} {\bibfnamefont {R.}~\bibnamefont
  {Alonso}}, \bibinfo {author} {\bibfnamefont {M.~B.}\ \bibnamefont {Gavela}},
  \bibinfo {author} {\bibfnamefont {L.}~\bibnamefont {Merlo}}, \bibinfo
  {author} {\bibfnamefont {S.}~\bibnamefont {Rigolin}}, \ and\ \bibinfo
  {author} {\bibfnamefont {J.}~\bibnamefont {Yepes}},\ }\href {\doibase
  10.1016/j.physletb.2013.04.037, 10.1016/j.physletb.2013.09.028} {\bibfield
  {journal} {\bibinfo  {journal} {Phys. Lett.}\ }\textbf {\bibinfo {volume}
  {B722}},\ \bibinfo {pages} {330} (\bibinfo {year} {2013})},\ \bibinfo {note}
  {[Erratum: Phys. Lett.B726,926(2013)]},\ \Eprint
  {http://arxiv.org/abs/1212.3305} {arXiv:1212.3305 [hep-ph]} \BibitemShut
  {NoStop}%
\bibitem [{\citenamefont {Buchalla}\ \emph {et~al.}(2014)\citenamefont
  {Buchalla}, \citenamefont {Cat{\`a}},\ and\ \citenamefont
  {Krause}}]{Buchalla:2013rka}%
  \BibitemOpen
  \bibfield  {author} {\bibinfo {author} {\bibfnamefont {G.}~\bibnamefont
  {Buchalla}}, \bibinfo {author} {\bibfnamefont {O.}~\bibnamefont {Cat{\`a}}},
  \ and\ \bibinfo {author} {\bibfnamefont {C.}~\bibnamefont {Krause}},\ }\href
  {\doibase 10.1016/j.nuclphysb.2016.09.010, 10.1016/j.nuclphysb.2014.01.018}
  {\bibfield  {journal} {\bibinfo  {journal} {Nucl. Phys.}\ }\textbf {\bibinfo
  {volume} {B880}},\ \bibinfo {pages} {552} (\bibinfo {year} {2014})},\
  \bibinfo {note} {[Erratum: Nucl. Phys.B913,475(2016)]},\ \Eprint
  {http://arxiv.org/abs/1307.5017} {arXiv:1307.5017 [hep-ph]} \BibitemShut
  {NoStop}%
\bibitem [{\citenamefont {Espriu}\ and\ \citenamefont
  {Yencho}(2013)}]{Espriu:2012ih}%
  \BibitemOpen
  \bibfield  {author} {\bibinfo {author} {\bibfnamefont {D.}~\bibnamefont
  {Espriu}}\ and\ \bibinfo {author} {\bibfnamefont {B.}~\bibnamefont
  {Yencho}},\ }\href {\doibase 10.1103/PhysRevD.87.055017} {\bibfield
  {journal} {\bibinfo  {journal} {Phys. Rev.}\ }\textbf {\bibinfo {volume}
  {D87}},\ \bibinfo {pages} {055017} (\bibinfo {year} {2013})},\ \Eprint
  {http://arxiv.org/abs/1212.4158} {arXiv:1212.4158 [hep-ph]} \BibitemShut
  {NoStop}%
\bibitem [{\citenamefont {Delgado}\ \emph
  {et~al.}(2014{\natexlab{a}})\citenamefont {Delgado}, \citenamefont {Dobado},\
  and\ \citenamefont {Llanes-Estrada}}]{Delgado:2013loa}%
  \BibitemOpen
  \bibfield  {author} {\bibinfo {author} {\bibfnamefont {R.~L.}\ \bibnamefont
  {Delgado}}, \bibinfo {author} {\bibfnamefont {A.}~\bibnamefont {Dobado}}, \
  and\ \bibinfo {author} {\bibfnamefont {F.~J.}\ \bibnamefont
  {Llanes-Estrada}},\ }\href {\doibase 10.1088/0954-3899/41/2/025002}
  {\bibfield  {journal} {\bibinfo  {journal} {J. Phys.}\ }\textbf {\bibinfo
  {volume} {G41}},\ \bibinfo {pages} {025002} (\bibinfo {year}
  {2014}{\natexlab{a}})},\ \Eprint {http://arxiv.org/abs/1308.1629}
  {arXiv:1308.1629 [hep-ph]} \BibitemShut {NoStop}%
\bibitem [{\citenamefont {Delgado}\ \emph
  {et~al.}(2014{\natexlab{b}})\citenamefont {Delgado}, \citenamefont {Dobado},\
  and\ \citenamefont {Llanes-Estrada}}]{Delgado:2013hxa}%
  \BibitemOpen
  \bibfield  {author} {\bibinfo {author} {\bibfnamefont {R.~L.}\ \bibnamefont
  {Delgado}}, \bibinfo {author} {\bibfnamefont {A.}~\bibnamefont {Dobado}}, \
  and\ \bibinfo {author} {\bibfnamefont {F.~J.}\ \bibnamefont
  {Llanes-Estrada}},\ }\href {\doibase 10.1007/JHEP02(2014)121} {\bibfield
  {journal} {\bibinfo  {journal} {JHEP}\ }\textbf {\bibinfo {volume} {02}},\
  \bibinfo {pages} {121} (\bibinfo {year} {2014}{\natexlab{b}})},\ \Eprint
  {http://arxiv.org/abs/1311.5993} {arXiv:1311.5993 [hep-ph]} \BibitemShut
  {NoStop}%
\bibitem [{\citenamefont {Brivio}\ \emph {et~al.}(2014)\citenamefont {Brivio},
  \citenamefont {Corbett}, \citenamefont {{\'E}boli}, \citenamefont {Gavela},
  \citenamefont {Gonzalez-Fraile}, \citenamefont {Gonzalez-Garcia},
  \citenamefont {Merlo},\ and\ \citenamefont {Rigolin}}]{Brivio:2013pma}%
  \BibitemOpen
  \bibfield  {author} {\bibinfo {author} {\bibfnamefont {I.}~\bibnamefont
  {Brivio}}, \bibinfo {author} {\bibfnamefont {T.}~\bibnamefont {Corbett}},
  \bibinfo {author} {\bibfnamefont {O.~J.~P.}\ \bibnamefont {{\'E}boli}},
  \bibinfo {author} {\bibfnamefont {M.~B.}\ \bibnamefont {Gavela}}, \bibinfo
  {author} {\bibfnamefont {J.}~\bibnamefont {Gonzalez-Fraile}}, \bibinfo
  {author} {\bibfnamefont {M.~C.}\ \bibnamefont {Gonzalez-Garcia}}, \bibinfo
  {author} {\bibfnamefont {L.}~\bibnamefont {Merlo}}, \ and\ \bibinfo {author}
  {\bibfnamefont {S.}~\bibnamefont {Rigolin}},\ }\href {\doibase
  10.1007/JHEP03(2014)024} {\bibfield  {journal} {\bibinfo  {journal} {JHEP}\
  }\textbf {\bibinfo {volume} {03}},\ \bibinfo {pages} {024} (\bibinfo {year}
  {2014})},\ \Eprint {http://arxiv.org/abs/1311.1823} {arXiv:1311.1823
  [hep-ph]} \BibitemShut {NoStop}%
\bibitem [{\citenamefont {Espriu}\ \emph {et~al.}(2013)\citenamefont {Espriu},
  \citenamefont {Mescia},\ and\ \citenamefont {Yencho}}]{Espriu:2013fia}%
  \BibitemOpen
  \bibfield  {author} {\bibinfo {author} {\bibfnamefont {D.}~\bibnamefont
  {Espriu}}, \bibinfo {author} {\bibfnamefont {F.}~\bibnamefont {Mescia}}, \
  and\ \bibinfo {author} {\bibfnamefont {B.}~\bibnamefont {Yencho}},\ }\href
  {\doibase 10.1103/PhysRevD.88.055002} {\bibfield  {journal} {\bibinfo
  {journal} {Phys. Rev.}\ }\textbf {\bibinfo {volume} {D88}},\ \bibinfo {pages}
  {055002} (\bibinfo {year} {2013})},\ \Eprint {http://arxiv.org/abs/1307.2400}
  {arXiv:1307.2400 [hep-ph]} \BibitemShut {NoStop}%
\bibitem [{\citenamefont {Espriu}\ and\ \citenamefont
  {Mescia}(2014)}]{Espriu:2014jya}%
  \BibitemOpen
  \bibfield  {author} {\bibinfo {author} {\bibfnamefont {D.}~\bibnamefont
  {Espriu}}\ and\ \bibinfo {author} {\bibfnamefont {F.}~\bibnamefont
  {Mescia}},\ }\href {\doibase 10.1103/PhysRevD.90.015035} {\bibfield
  {journal} {\bibinfo  {journal} {Phys. Rev.}\ }\textbf {\bibinfo {volume}
  {D90}},\ \bibinfo {pages} {015035} (\bibinfo {year} {2014})},\ \Eprint
  {http://arxiv.org/abs/1403.7386} {arXiv:1403.7386 [hep-ph]} \BibitemShut
  {NoStop}%
\bibitem [{\citenamefont {Delgado}\ \emph
  {et~al.}(2014{\natexlab{c}})\citenamefont {Delgado}, \citenamefont {Dobado},
  \citenamefont {Herrero},\ and\ \citenamefont
  {Sanz-Cillero}}]{Delgado:2014jda}%
  \BibitemOpen
  \bibfield  {author} {\bibinfo {author} {\bibfnamefont {R.~L.}\ \bibnamefont
  {Delgado}}, \bibinfo {author} {\bibfnamefont {A.}~\bibnamefont {Dobado}},
  \bibinfo {author} {\bibfnamefont {M.~J.}\ \bibnamefont {Herrero}}, \ and\
  \bibinfo {author} {\bibfnamefont {J.~J.}\ \bibnamefont {Sanz-Cillero}},\
  }\href {\doibase 10.1007/JHEP07(2014)149} {\bibfield  {journal} {\bibinfo
  {journal} {JHEP}\ }\textbf {\bibinfo {volume} {07}},\ \bibinfo {pages} {149}
  (\bibinfo {year} {2014}{\natexlab{c}})},\ \Eprint
  {http://arxiv.org/abs/1404.2866} {arXiv:1404.2866 [hep-ph]} \BibitemShut
  {NoStop}%
\bibitem [{\citenamefont {Buchalla}\ \emph {et~al.}(2016)\citenamefont
  {Buchalla}, \citenamefont {Cata}, \citenamefont {Celis},\ and\ \citenamefont
  {Krause}}]{Buchalla:2015qju}%
  \BibitemOpen
  \bibfield  {author} {\bibinfo {author} {\bibfnamefont {G.}~\bibnamefont
  {Buchalla}}, \bibinfo {author} {\bibfnamefont {O.}~\bibnamefont {Cata}},
  \bibinfo {author} {\bibfnamefont {A.}~\bibnamefont {Celis}}, \ and\ \bibinfo
  {author} {\bibfnamefont {C.}~\bibnamefont {Krause}},\ }\href {\doibase
  10.1140/epjc/s10052-016-4086-9} {\bibfield  {journal} {\bibinfo  {journal}
  {Eur. Phys. J.}\ }\textbf {\bibinfo {volume} {C76}},\ \bibinfo {pages} {233}
  (\bibinfo {year} {2016})},\ \Eprint {http://arxiv.org/abs/1511.00988}
  {arXiv:1511.00988 [hep-ph]} \BibitemShut {NoStop}%
\bibitem [{\citenamefont {Arnan}\ \emph {et~al.}(2016)\citenamefont {Arnan},
  \citenamefont {Espriu},\ and\ \citenamefont {Mescia}}]{Arnan:2015csa}%
  \BibitemOpen
  \bibfield  {author} {\bibinfo {author} {\bibfnamefont {P.}~\bibnamefont
  {Arnan}}, \bibinfo {author} {\bibfnamefont {D.}~\bibnamefont {Espriu}}, \
  and\ \bibinfo {author} {\bibfnamefont {F.}~\bibnamefont {Mescia}},\ }\href
  {\doibase 10.1103/PhysRevD.93.015020} {\bibfield  {journal} {\bibinfo
  {journal} {Phys. Rev.}\ }\textbf {\bibinfo {volume} {D93}},\ \bibinfo {pages}
  {015020} (\bibinfo {year} {2016})},\ \Eprint
  {http://arxiv.org/abs/1508.00174} {arXiv:1508.00174 [hep-ph]} \BibitemShut
  {NoStop}%
\bibitem [{\citenamefont {de~Florian}\ \emph {et~al.}(2016)\citenamefont
  {de~Florian} \emph {et~al.}}]{deFlorian:2016spz}%
  \BibitemOpen
  \bibfield  {author} {\bibinfo {author} {\bibfnamefont {D.}~\bibnamefont
  {de~Florian}} \emph {et~al.} (\bibinfo {collaboration} {LHC Higgs Cross
  Section Working Group}),\ }\href@noop {} {\  (\bibinfo {year} {2016})},\
  \Eprint {http://arxiv.org/abs/1610.07922} {arXiv:1610.07922 [hep-ph]}
  \BibitemShut {NoStop}%
\bibitem [{\citenamefont {Falkowski}\ \emph {et~al.}(2013)\citenamefont
  {Falkowski}, \citenamefont {Riva},\ and\ \citenamefont
  {Urbano}}]{Falkowski:2013dza}%
  \BibitemOpen
  \bibfield  {author} {\bibinfo {author} {\bibfnamefont {A.}~\bibnamefont
  {Falkowski}}, \bibinfo {author} {\bibfnamefont {F.}~\bibnamefont {Riva}}, \
  and\ \bibinfo {author} {\bibfnamefont {A.}~\bibnamefont {Urbano}},\ }\href
  {\doibase 10.1007/JHEP11(2013)111} {\bibfield  {journal} {\bibinfo  {journal}
  {JHEP}\ }\textbf {\bibinfo {volume} {11}},\ \bibinfo {pages} {111} (\bibinfo
  {year} {2013})},\ \Eprint {http://arxiv.org/abs/1303.1812} {arXiv:1303.1812
  [hep-ph]} \BibitemShut {NoStop}%
\bibitem [{\citenamefont {Khachatryan}\ \emph {et~al.}(2015)\citenamefont
  {Khachatryan} \emph {et~al.}}]{Khachatryan:2014jba}%
  \BibitemOpen
  \bibfield  {author} {\bibinfo {author} {\bibfnamefont {V.}~\bibnamefont
  {Khachatryan}} \emph {et~al.} (\bibinfo {collaboration} {CMS}),\ }\href
  {\doibase 10.1140/epjc/s10052-015-3351-7} {\bibfield  {journal} {\bibinfo
  {journal} {Eur. Phys. J.}\ }\textbf {\bibinfo {volume} {C75}},\ \bibinfo
  {pages} {212} (\bibinfo {year} {2015})},\ \Eprint
  {http://arxiv.org/abs/1412.8662} {arXiv:1412.8662 [hep-ex]} \BibitemShut
  {NoStop}%
\bibitem [{\citenamefont {Aad}\ \emph {et~al.}(2014)\citenamefont {Aad} \emph
  {et~al.}}]{Aad:2014zda}%
  \BibitemOpen
  \bibfield  {author} {\bibinfo {author} {\bibfnamefont {G.}~\bibnamefont
  {Aad}} \emph {et~al.} (\bibinfo {collaboration} {ATLAS}),\ }\href {\doibase
  10.1103/PhysRevLett.113.141803} {\bibfield  {journal} {\bibinfo  {journal}
  {Phys. Rev. Lett.}\ }\textbf {\bibinfo {volume} {113}},\ \bibinfo {pages}
  {141803} (\bibinfo {year} {2014})},\ \Eprint {http://arxiv.org/abs/1405.6241}
  {arXiv:1405.6241 [hep-ex]} \BibitemShut {NoStop}%
\bibitem [{\citenamefont {{The ATLAS collaboration}}(2014)}]{ATLAS:2014yka}%
  \BibitemOpen
  \bibfield  {author} {\bibinfo {author} {\bibnamefont {{The ATLAS
  collaboration}}},\ }\href@noop {} {\bibfield  {journal} {\bibinfo  {journal}
  {ATLAS-CONF-2014-009}\ } (\bibinfo {year} {2014})}\BibitemShut {NoStop}%
\bibitem [{\citenamefont {Fabbrichesi}\ \emph {et~al.}(2016)\citenamefont
  {Fabbrichesi}, \citenamefont {Pinamonti}, \citenamefont {Tonero},\ and\
  \citenamefont {Urbano}}]{Fabbrichesi:2015hsa}%
  \BibitemOpen
  \bibfield  {author} {\bibinfo {author} {\bibfnamefont {M.}~\bibnamefont
  {Fabbrichesi}}, \bibinfo {author} {\bibfnamefont {M.}~\bibnamefont
  {Pinamonti}}, \bibinfo {author} {\bibfnamefont {A.}~\bibnamefont {Tonero}}, \
  and\ \bibinfo {author} {\bibfnamefont {A.}~\bibnamefont {Urbano}},\ }\href
  {\doibase 10.1103/PhysRevD.93.015004} {\bibfield  {journal} {\bibinfo
  {journal} {Phys. Rev.}\ }\textbf {\bibinfo {volume} {D93}},\ \bibinfo {pages}
  {015004} (\bibinfo {year} {2016})},\ \Eprint
  {http://arxiv.org/abs/1509.06378} {arXiv:1509.06378 [hep-ph]} \BibitemShut
  {NoStop}%
\bibitem [{\citenamefont {Aaboud}\ \emph {et~al.}(2017)\citenamefont {Aaboud}
  \emph {et~al.}}]{Aaboud:2016uuk}%
  \BibitemOpen
  \bibfield  {author} {\bibinfo {author} {\bibfnamefont {M.}~\bibnamefont
  {Aaboud}} \emph {et~al.} (\bibinfo {collaboration} {ATLAS}),\ }\href
  {\doibase 10.1103/PhysRevD.95.032001} {\bibfield  {journal} {\bibinfo
  {journal} {Phys. Rev.}\ }\textbf {\bibinfo {volume} {D95}},\ \bibinfo {pages}
  {032001} (\bibinfo {year} {2017})},\ \Eprint
  {http://arxiv.org/abs/1609.05122} {arXiv:1609.05122 [hep-ex]} \BibitemShut
  {NoStop}%
\bibitem [{\citenamefont {Pich}\ \emph {et~al.}(2012)\citenamefont {Pich},
  \citenamefont {Rosell},\ and\ \citenamefont {Sanz-Cillero}}]{Pich:2012jv}%
  \BibitemOpen
  \bibfield  {author} {\bibinfo {author} {\bibfnamefont {A.}~\bibnamefont
  {Pich}}, \bibinfo {author} {\bibfnamefont {I.}~\bibnamefont {Rosell}}, \ and\
  \bibinfo {author} {\bibfnamefont {J.~J.}\ \bibnamefont {Sanz-Cillero}},\
  }\href {\doibase 10.1007/JHEP08(2012)106} {\bibfield  {journal} {\bibinfo
  {journal} {JHEP}\ }\textbf {\bibinfo {volume} {08}},\ \bibinfo {pages} {106}
  (\bibinfo {year} {2012})},\ \Eprint {http://arxiv.org/abs/1206.3454}
  {arXiv:1206.3454 [hep-ph]} \BibitemShut {NoStop}%
\bibitem [{\citenamefont {Pich}\ \emph {et~al.}(2013)\citenamefont {Pich},
  \citenamefont {Rosell},\ and\ \citenamefont {Sanz-Cillero}}]{Pich:2012dv}%
  \BibitemOpen
  \bibfield  {author} {\bibinfo {author} {\bibfnamefont {A.}~\bibnamefont
  {Pich}}, \bibinfo {author} {\bibfnamefont {I.}~\bibnamefont {Rosell}}, \ and\
  \bibinfo {author} {\bibfnamefont {J.~J.}\ \bibnamefont {Sanz-Cillero}},\
  }\href {\doibase 10.1103/PhysRevLett.110.181801} {\bibfield  {journal}
  {\bibinfo  {journal} {Phys. Rev. Lett.}\ }\textbf {\bibinfo {volume} {110}},\
  \bibinfo {pages} {181801} (\bibinfo {year} {2013})},\ \Eprint
  {http://arxiv.org/abs/1212.6769} {arXiv:1212.6769 [hep-ph]} \BibitemShut
  {NoStop}%
\bibitem [{\citenamefont {Pich}\ \emph {et~al.}(2014)\citenamefont {Pich},
  \citenamefont {Rosell},\ and\ \citenamefont {Sanz-Cillero}}]{Pich:2013fea}%
  \BibitemOpen
  \bibfield  {author} {\bibinfo {author} {\bibfnamefont {A.}~\bibnamefont
  {Pich}}, \bibinfo {author} {\bibfnamefont {I.}~\bibnamefont {Rosell}}, \ and\
  \bibinfo {author} {\bibfnamefont {J.~J.}\ \bibnamefont {Sanz-Cillero}},\
  }\href {\doibase 10.1007/JHEP01(2014)157} {\bibfield  {journal} {\bibinfo
  {journal} {JHEP}\ }\textbf {\bibinfo {volume} {01}},\ \bibinfo {pages} {157}
  (\bibinfo {year} {2014})},\ \Eprint {http://arxiv.org/abs/1310.3121}
  {arXiv:1310.3121 [hep-ph]} \BibitemShut {NoStop}%
\bibitem [{\citenamefont {Pich}\ \emph {et~al.}(2016)\citenamefont {Pich},
  \citenamefont {Rosell}, \citenamefont {Santos},\ and\ \citenamefont
  {Sanz-Cillero}}]{Pich:2015kwa}%
  \BibitemOpen
  \bibfield  {author} {\bibinfo {author} {\bibfnamefont {A.}~\bibnamefont
  {Pich}}, \bibinfo {author} {\bibfnamefont {I.}~\bibnamefont {Rosell}},
  \bibinfo {author} {\bibfnamefont {J.}~\bibnamefont {Santos}}, \ and\ \bibinfo
  {author} {\bibfnamefont {J.~J.}\ \bibnamefont {Sanz-Cillero}},\ }\href
  {\doibase 10.1103/PhysRevD.93.055041} {\bibfield  {journal} {\bibinfo
  {journal} {Phys. Rev.}\ }\textbf {\bibinfo {volume} {D93}},\ \bibinfo {pages}
  {055041} (\bibinfo {year} {2016})},\ \Eprint
  {http://arxiv.org/abs/1510.03114} {arXiv:1510.03114 [hep-ph]} \BibitemShut
  {NoStop}%
\bibitem [{\citenamefont {Pich}\ \emph {et~al.}(2017)\citenamefont {Pich},
  \citenamefont {Rosell}, \citenamefont {Santos},\ and\ \citenamefont
  {Sanz-Cillero}}]{Pich:2016lew}%
  \BibitemOpen
  \bibfield  {author} {\bibinfo {author} {\bibfnamefont {A.}~\bibnamefont
  {Pich}}, \bibinfo {author} {\bibfnamefont {I.}~\bibnamefont {Rosell}},
  \bibinfo {author} {\bibfnamefont {J.}~\bibnamefont {Santos}}, \ and\ \bibinfo
  {author} {\bibfnamefont {J.~J.}\ \bibnamefont {Sanz-Cillero}},\ }\href
  {\doibase 10.1007/JHEP04(2017)012} {\bibfield  {journal} {\bibinfo  {journal}
  {JHEP}\ }\textbf {\bibinfo {volume} {04}},\ \bibinfo {pages} {012} (\bibinfo
  {year} {2017})},\ \Eprint {http://arxiv.org/abs/1609.06659} {arXiv:1609.06659
  [hep-ph]} \BibitemShut {NoStop}%
\bibitem [{\citenamefont {Ecker}\ \emph {et~al.}(1989)\citenamefont {Ecker},
  \citenamefont {Gasser}, \citenamefont {Leutwyler}, \citenamefont {Pich},\
  and\ \citenamefont {de~Rafael}}]{Ecker:1989yg}%
  \BibitemOpen
  \bibfield  {author} {\bibinfo {author} {\bibfnamefont {G.}~\bibnamefont
  {Ecker}}, \bibinfo {author} {\bibfnamefont {J.}~\bibnamefont {Gasser}},
  \bibinfo {author} {\bibfnamefont {H.}~\bibnamefont {Leutwyler}}, \bibinfo
  {author} {\bibfnamefont {A.}~\bibnamefont {Pich}}, \ and\ \bibinfo {author}
  {\bibfnamefont {E.}~\bibnamefont {de~Rafael}},\ }\href {\doibase
  10.1016/0370-2693(89)91627-4} {\bibfield  {journal} {\bibinfo  {journal}
  {Phys. Lett.}\ }\textbf {\bibinfo {volume} {B223}},\ \bibinfo {pages} {425}
  (\bibinfo {year} {1989})}\BibitemShut {NoStop}%
\bibitem [{\citenamefont {Alboteanu}\ \emph {et~al.}(2008)\citenamefont
  {Alboteanu}, \citenamefont {Kilian},\ and\ \citenamefont
  {Reuter}}]{Alboteanu:2008my}%
  \BibitemOpen
  \bibfield  {author} {\bibinfo {author} {\bibfnamefont {A.}~\bibnamefont
  {Alboteanu}}, \bibinfo {author} {\bibfnamefont {W.}~\bibnamefont {Kilian}}, \
  and\ \bibinfo {author} {\bibfnamefont {J.}~\bibnamefont {Reuter}},\ }\href
  {\doibase 10.1088/1126-6708/2008/11/010} {\bibfield  {journal} {\bibinfo
  {journal} {JHEP}\ }\textbf {\bibinfo {volume} {11}},\ \bibinfo {pages} {010}
  (\bibinfo {year} {2008})},\ \Eprint {http://arxiv.org/abs/0806.4145}
  {arXiv:0806.4145 [hep-ph]} \BibitemShut {NoStop}%
\bibitem [{\citenamefont {Delgado}\ \emph
  {et~al.}(2015{\natexlab{a}})\citenamefont {Delgado}, \citenamefont {Dobado},\
  and\ \citenamefont {Llanes-Estrada}}]{Delgado:2014dxa}%
  \BibitemOpen
  \bibfield  {author} {\bibinfo {author} {\bibfnamefont {R.~L.}\ \bibnamefont
  {Delgado}}, \bibinfo {author} {\bibfnamefont {A.}~\bibnamefont {Dobado}}, \
  and\ \bibinfo {author} {\bibfnamefont {F.~J.}\ \bibnamefont
  {Llanes-Estrada}},\ }\href {\doibase 10.1103/PhysRevLett.114.221803}
  {\bibfield  {journal} {\bibinfo  {journal} {Phys. Rev. Lett.}\ }\textbf
  {\bibinfo {volume} {114}},\ \bibinfo {pages} {221803} (\bibinfo {year}
  {2015}{\natexlab{a}})},\ \Eprint {http://arxiv.org/abs/1408.1193}
  {arXiv:1408.1193 [hep-ph]} \BibitemShut {NoStop}%
\bibitem [{\citenamefont {Dobado}\ \emph {et~al.}(2015)\citenamefont {Dobado},
  \citenamefont {Guo},\ and\ \citenamefont {Llanes-Estrada}}]{Dobado:2015hha}%
  \BibitemOpen
  \bibfield  {author} {\bibinfo {author} {\bibfnamefont {A.}~\bibnamefont
  {Dobado}}, \bibinfo {author} {\bibfnamefont {F.-K.}\ \bibnamefont {Guo}}, \
  and\ \bibinfo {author} {\bibfnamefont {F.~J.}\ \bibnamefont
  {Llanes-Estrada}},\ }\href@noop {} {\bibfield  {journal} {\bibinfo  {journal}
  {Commun. Theor. Phys.}\ }\textbf {\bibinfo {volume} {64}},\ \bibinfo {pages}
  {701} (\bibinfo {year} {2015})},\ \Eprint {http://arxiv.org/abs/1508.03544}
  {arXiv:1508.03544 [hep-ph]} \BibitemShut {NoStop}%
\bibitem [{\citenamefont {Corbett}\ \emph {et~al.}(2016)\citenamefont
  {Corbett}, \citenamefont {{\'E}boli},\ and\ \citenamefont
  {Gonzalez-Garcia}}]{Corbett:2015lfa}%
  \BibitemOpen
  \bibfield  {author} {\bibinfo {author} {\bibfnamefont {T.}~\bibnamefont
  {Corbett}}, \bibinfo {author} {\bibfnamefont {O.~J.~P.}\ \bibnamefont
  {{\'E}boli}}, \ and\ \bibinfo {author} {\bibfnamefont {M.~C.}\ \bibnamefont
  {Gonzalez-Garcia}},\ }\href {\doibase 10.1103/PhysRevD.93.015005} {\bibfield
  {journal} {\bibinfo  {journal} {Phys. Rev.}\ }\textbf {\bibinfo {volume}
  {D93}},\ \bibinfo {pages} {015005} (\bibinfo {year} {2016})},\ \Eprint
  {http://arxiv.org/abs/1509.01585} {arXiv:1509.01585 [hep-ph]} \BibitemShut
  {NoStop}%
\bibitem [{\citenamefont {Buarque~Franzosi}\ and\ \citenamefont
  {Ferrarese}(2017)}]{BuarqueFranzosi:2017prc}%
  \BibitemOpen
  \bibfield  {author} {\bibinfo {author} {\bibfnamefont {D.}~\bibnamefont
  {Buarque~Franzosi}}\ and\ \bibinfo {author} {\bibfnamefont {P.}~\bibnamefont
  {Ferrarese}},\ }\href@noop {} {\  (\bibinfo {year} {2017})},\ \Eprint
  {http://arxiv.org/abs/1705.02787} {arXiv:1705.02787 [hep-ph]} \BibitemShut
  {NoStop}%
\bibitem [{\citenamefont {Truong}(1988)}]{Truong:1988zp}%
  \BibitemOpen
  \bibfield  {author} {\bibinfo {author} {\bibfnamefont {T.~N.}\ \bibnamefont
  {Truong}},\ }\href {\doibase 10.1103/PhysRevLett.61.2526} {\bibfield
  {journal} {\bibinfo  {journal} {Phys. Rev. Lett.}\ }\textbf {\bibinfo
  {volume} {61}},\ \bibinfo {pages} {2526} (\bibinfo {year}
  {1988})}\BibitemShut {NoStop}%
\bibitem [{\citenamefont {Dobado}\ \emph
  {et~al.}(1990{\natexlab{a}})\citenamefont {Dobado}, \citenamefont {Herrero},\
  and\ \citenamefont {Truong}}]{Dobado:1989qm}%
  \BibitemOpen
  \bibfield  {author} {\bibinfo {author} {\bibfnamefont {A.}~\bibnamefont
  {Dobado}}, \bibinfo {author} {\bibfnamefont {M.~J.}\ \bibnamefont {Herrero}},
  \ and\ \bibinfo {author} {\bibfnamefont {T.~N.}\ \bibnamefont {Truong}},\
  }\href {\doibase 10.1016/0370-2693(90)90109-J} {\bibfield  {journal}
  {\bibinfo  {journal} {Phys. Lett.}\ }\textbf {\bibinfo {volume} {B235}},\
  \bibinfo {pages} {134} (\bibinfo {year} {1990}{\natexlab{a}})}\BibitemShut
  {NoStop}%
\bibitem [{\citenamefont {Dobado}\ and\ \citenamefont
  {Pelaez}(1993)}]{Dobado:1992ha}%
  \BibitemOpen
  \bibfield  {author} {\bibinfo {author} {\bibfnamefont {A.}~\bibnamefont
  {Dobado}}\ and\ \bibinfo {author} {\bibfnamefont {J.~R.}\ \bibnamefont
  {Pelaez}},\ }\href {\doibase 10.1103/PhysRevD.47.4883} {\bibfield  {journal}
  {\bibinfo  {journal} {Phys. Rev.}\ }\textbf {\bibinfo {volume} {D47}},\
  \bibinfo {pages} {4883} (\bibinfo {year} {1993})},\ \Eprint
  {http://arxiv.org/abs/hep-ph/9301276} {arXiv:hep-ph/9301276 [hep-ph]}
  \BibitemShut {NoStop}%
\bibitem [{\citenamefont {Hannah}(1995)}]{Hannah:1995si}%
  \BibitemOpen
  \bibfield  {author} {\bibinfo {author} {\bibfnamefont {T.}~\bibnamefont
  {Hannah}},\ }\href {\doibase 10.1103/PhysRevD.51.103} {\bibfield  {journal}
  {\bibinfo  {journal} {Phys. Rev.}\ }\textbf {\bibinfo {volume} {D51}},\
  \bibinfo {pages} {103} (\bibinfo {year} {1995})}\BibitemShut {NoStop}%
\bibitem [{\citenamefont {Dobado}\ \emph
  {et~al.}(1990{\natexlab{b}})\citenamefont {Dobado}, \citenamefont {Herrero},\
  and\ \citenamefont {Truong}}]{Dobado:1989gr}%
  \BibitemOpen
  \bibfield  {author} {\bibinfo {author} {\bibfnamefont {A.}~\bibnamefont
  {Dobado}}, \bibinfo {author} {\bibfnamefont {M.~J.}\ \bibnamefont {Herrero}},
  \ and\ \bibinfo {author} {\bibfnamefont {T.~N.}\ \bibnamefont {Truong}},\
  }\href {\doibase 10.1016/0370-2693(90)90108-I} {\bibfield  {journal}
  {\bibinfo  {journal} {Phys. Lett.}\ }\textbf {\bibinfo {volume} {B235}},\
  \bibinfo {pages} {129} (\bibinfo {year} {1990}{\natexlab{b}})}\BibitemShut
  {NoStop}%
\bibitem [{\citenamefont {Cornwall}\ \emph {et~al.}(1974)\citenamefont
  {Cornwall}, \citenamefont {Levin},\ and\ \citenamefont
  {Tiktopoulos}}]{Cornwall:1974km}%
  \BibitemOpen
  \bibfield  {author} {\bibinfo {author} {\bibfnamefont {J.~M.}\ \bibnamefont
  {Cornwall}}, \bibinfo {author} {\bibfnamefont {D.~N.}\ \bibnamefont {Levin}},
  \ and\ \bibinfo {author} {\bibfnamefont {G.}~\bibnamefont {Tiktopoulos}},\
  }\href {\doibase 10.1103/PhysRevD.10.1145, 10.1103/PhysRevD.11.972}
  {\bibfield  {journal} {\bibinfo  {journal} {Phys. Rev.}\ }\textbf {\bibinfo
  {volume} {D10}},\ \bibinfo {pages} {1145} (\bibinfo {year} {1974})},\
  \bibinfo {note} {[Erratum: Phys. Rev.D11,972(1975)]}\BibitemShut {NoStop}%
\bibitem [{\citenamefont {Vayonakis}(1976)}]{Vayonakis:1976vz}%
  \BibitemOpen
  \bibfield  {author} {\bibinfo {author} {\bibfnamefont {C.~E.}\ \bibnamefont
  {Vayonakis}},\ }\href {\doibase 10.1007/BF02746538} {\bibfield  {journal}
  {\bibinfo  {journal} {Lett. Nuovo Cim.}\ }\textbf {\bibinfo {volume} {17}},\
  \bibinfo {pages} {383} (\bibinfo {year} {1976})}\BibitemShut {NoStop}%
\bibitem [{\citenamefont {Lee}\ \emph {et~al.}(1977)\citenamefont {Lee},
  \citenamefont {Quigg},\ and\ \citenamefont {Thacker}}]{Lee:1977eg}%
  \BibitemOpen
  \bibfield  {author} {\bibinfo {author} {\bibfnamefont {B.~W.}\ \bibnamefont
  {Lee}}, \bibinfo {author} {\bibfnamefont {C.}~\bibnamefont {Quigg}}, \ and\
  \bibinfo {author} {\bibfnamefont {H.~B.}\ \bibnamefont {Thacker}},\ }\href
  {\doibase 10.1103/PhysRevD.16.1519} {\bibfield  {journal} {\bibinfo
  {journal} {Phys. Rev.}\ }\textbf {\bibinfo {volume} {D16}},\ \bibinfo {pages}
  {1519} (\bibinfo {year} {1977})}\BibitemShut {NoStop}%
\bibitem [{\citenamefont {Gounaris}\ \emph {et~al.}(1986)\citenamefont
  {Gounaris}, \citenamefont {Kogerler},\ and\ \citenamefont
  {Neufeld}}]{Gounaris:1986cr}%
  \BibitemOpen
  \bibfield  {author} {\bibinfo {author} {\bibfnamefont {G.~J.}\ \bibnamefont
  {Gounaris}}, \bibinfo {author} {\bibfnamefont {R.}~\bibnamefont {Kogerler}},
  \ and\ \bibinfo {author} {\bibfnamefont {H.}~\bibnamefont {Neufeld}},\ }\href
  {\doibase 10.1103/PhysRevD.34.3257} {\bibfield  {journal} {\bibinfo
  {journal} {Phys. Rev.}\ }\textbf {\bibinfo {volume} {D34}},\ \bibinfo {pages}
  {3257} (\bibinfo {year} {1986})}\BibitemShut {NoStop}%
\bibitem [{\citenamefont {Alwall}\ \emph {et~al.}(2014)\citenamefont {Alwall},
  \citenamefont {Frederix}, \citenamefont {Frixione}, \citenamefont {Hirschi},
  \citenamefont {Maltoni}, \citenamefont {Mattelaer}, \citenamefont {Shao},
  \citenamefont {Stelzer}, \citenamefont {Torrielli},\ and\ \citenamefont
  {Zaro}}]{Alwall:2014hca}%
  \BibitemOpen
  \bibfield  {author} {\bibinfo {author} {\bibfnamefont {J.}~\bibnamefont
  {Alwall}}, \bibinfo {author} {\bibfnamefont {R.}~\bibnamefont {Frederix}},
  \bibinfo {author} {\bibfnamefont {S.}~\bibnamefont {Frixione}}, \bibinfo
  {author} {\bibfnamefont {V.}~\bibnamefont {Hirschi}}, \bibinfo {author}
  {\bibfnamefont {F.}~\bibnamefont {Maltoni}}, \bibinfo {author} {\bibfnamefont
  {O.}~\bibnamefont {Mattelaer}}, \bibinfo {author} {\bibfnamefont {H.~S.}\
  \bibnamefont {Shao}}, \bibinfo {author} {\bibfnamefont {T.}~\bibnamefont
  {Stelzer}}, \bibinfo {author} {\bibfnamefont {P.}~\bibnamefont {Torrielli}},
  \ and\ \bibinfo {author} {\bibfnamefont {M.}~\bibnamefont {Zaro}},\ }\href
  {\doibase 10.1007/JHEP07(2014)079} {\bibfield  {journal} {\bibinfo  {journal}
  {JHEP}\ }\textbf {\bibinfo {volume} {07}},\ \bibinfo {pages} {079} (\bibinfo
  {year} {2014})},\ \Eprint {http://arxiv.org/abs/1405.0301} {arXiv:1405.0301
  [hep-ph]} \BibitemShut {NoStop}%
\bibitem [{\citenamefont {Haywood}\ \emph {et~al.}(1999)\citenamefont {Haywood}
  \emph {et~al.}}]{Haywood:1999qg}%
  \BibitemOpen
  \bibfield  {author} {\bibinfo {author} {\bibfnamefont {S.}~\bibnamefont
  {Haywood}} \emph {et~al.},\ }\href@noop {} {\  (\bibinfo {year} {1999})},\
  \Eprint {http://arxiv.org/abs/hep-ph/0003275} {arXiv:hep-ph/0003275 [hep-ph]}
  \BibitemShut {NoStop}%
\bibitem [{\citenamefont {Doroba}\ \emph {et~al.}(2012)\citenamefont {Doroba},
  \citenamefont {Kalinowski}, \citenamefont {Kuczmarski}, \citenamefont
  {Pokorski}, \citenamefont {Rosiek}, \citenamefont {Szleper},\ and\
  \citenamefont {Tkaczyk}}]{Doroba:2012pd}%
  \BibitemOpen
  \bibfield  {author} {\bibinfo {author} {\bibfnamefont {K.}~\bibnamefont
  {Doroba}}, \bibinfo {author} {\bibfnamefont {J.}~\bibnamefont {Kalinowski}},
  \bibinfo {author} {\bibfnamefont {J.}~\bibnamefont {Kuczmarski}}, \bibinfo
  {author} {\bibfnamefont {S.}~\bibnamefont {Pokorski}}, \bibinfo {author}
  {\bibfnamefont {J.}~\bibnamefont {Rosiek}}, \bibinfo {author} {\bibfnamefont
  {M.}~\bibnamefont {Szleper}}, \ and\ \bibinfo {author} {\bibfnamefont
  {S.}~\bibnamefont {Tkaczyk}},\ }\href {\doibase 10.1103/PhysRevD.86.036011}
  {\bibfield  {journal} {\bibinfo  {journal} {Phys. Rev.}\ }\textbf {\bibinfo
  {volume} {D86}},\ \bibinfo {pages} {036011} (\bibinfo {year} {2012})},\
  \Eprint {http://arxiv.org/abs/1201.2768} {arXiv:1201.2768 [hep-ph]}
  \BibitemShut {NoStop}%
\bibitem [{\citenamefont {Szleper}(2014)}]{Szleper:2014xxa}%
  \BibitemOpen
  \bibfield  {author} {\bibinfo {author} {\bibfnamefont {M.}~\bibnamefont
  {Szleper}},\ }\href@noop {} {\  (\bibinfo {year} {2014})},\ \Eprint
  {http://arxiv.org/abs/1412.8367} {arXiv:1412.8367 [hep-ph]} \BibitemShut
  {NoStop}%
\bibitem [{\citenamefont {Aad}\ \emph {et~al.}(2016)\citenamefont {Aad} \emph
  {et~al.}}]{Aad:2016ett}%
  \BibitemOpen
  \bibfield  {author} {\bibinfo {author} {\bibfnamefont {G.}~\bibnamefont
  {Aad}} \emph {et~al.} (\bibinfo {collaboration} {ATLAS}),\ }\href {\doibase
  10.1103/PhysRevD.93.092004} {\bibfield  {journal} {\bibinfo  {journal} {Phys.
  Rev.}\ }\textbf {\bibinfo {volume} {D93}},\ \bibinfo {pages} {092004}
  (\bibinfo {year} {2016})},\ \Eprint {http://arxiv.org/abs/1603.02151}
  {arXiv:1603.02151 [hep-ex]} \BibitemShut {NoStop}%
\bibitem [{\citenamefont {Herrero}\ and\ \citenamefont
  {Ruiz~Morales}(1994)}]{Herrero:1993nc}%
  \BibitemOpen
  \bibfield  {author} {\bibinfo {author} {\bibfnamefont {M.~J.}\ \bibnamefont
  {Herrero}}\ and\ \bibinfo {author} {\bibfnamefont {E.}~\bibnamefont
  {Ruiz~Morales}},\ }\href {\doibase 10.1016/0550-3213(94)90525-8} {\bibfield
  {journal} {\bibinfo  {journal} {Nucl. Phys.}\ }\textbf {\bibinfo {volume}
  {B418}},\ \bibinfo {pages} {431} (\bibinfo {year} {1994})},\ \Eprint
  {http://arxiv.org/abs/hep-ph/9308276} {arXiv:hep-ph/9308276 [hep-ph]}
  \BibitemShut {NoStop}%
\bibitem [{\citenamefont {Herrero}\ and\ \citenamefont
  {Ruiz~Morales}(1995)}]{Herrero:1994iu}%
  \BibitemOpen
  \bibfield  {author} {\bibinfo {author} {\bibfnamefont {M.~J.}\ \bibnamefont
  {Herrero}}\ and\ \bibinfo {author} {\bibfnamefont {E.}~\bibnamefont
  {Ruiz~Morales}},\ }\href {\doibase 10.1016/0550-3213(94)00589-7} {\bibfield
  {journal} {\bibinfo  {journal} {Nucl. Phys.}\ }\textbf {\bibinfo {volume}
  {B437}},\ \bibinfo {pages} {319} (\bibinfo {year} {1995})},\ \Eprint
  {http://arxiv.org/abs/hep-ph/9411207} {arXiv:hep-ph/9411207 [hep-ph]}
  \BibitemShut {NoStop}%
\bibitem [{\citenamefont {Dobado}\ and\ \citenamefont
  {Pel{\'a}ez}(1994)}]{Dobado:1993dg}%
  \BibitemOpen
  \bibfield  {author} {\bibinfo {author} {\bibfnamefont {A.}~\bibnamefont
  {Dobado}}\ and\ \bibinfo {author} {\bibfnamefont {J.~R.}\ \bibnamefont
  {Pel{\'a}ez}},\ }\href {\doibase 10.1016/0550-3213(94)90174-0,
  10.1016/0550-3213(94)00533-K} {\bibfield  {journal} {\bibinfo  {journal}
  {Nucl. Phys.}\ }\textbf {\bibinfo {volume} {B425}},\ \bibinfo {pages} {110}
  (\bibinfo {year} {1994})},\ \bibinfo {note} {[Erratum: Nucl.
  Phys.B434,475(1995)]},\ \Eprint {http://arxiv.org/abs/hep-ph/9401202}
  {arXiv:hep-ph/9401202 [hep-ph]} \BibitemShut {NoStop}%
\bibitem [{\citenamefont {Dobado}\ and\ \citenamefont
  {Pelaez}(1994)}]{Dobado:1994vr}%
  \BibitemOpen
  \bibfield  {author} {\bibinfo {author} {\bibfnamefont {A.}~\bibnamefont
  {Dobado}}\ and\ \bibinfo {author} {\bibfnamefont {J.~R.}\ \bibnamefont
  {Pelaez}},\ }\href {\doibase 10.1016/0370-2693(94)90392-1,
  10.1016/0370-2693(94)91092-8} {\bibfield  {journal} {\bibinfo  {journal}
  {Phys. Lett.}\ }\textbf {\bibinfo {volume} {B329}},\ \bibinfo {pages} {469}
  (\bibinfo {year} {1994})},\ \bibinfo {note} {[Addendum: Phys.
  Lett.B335,554(1994)]},\ \Eprint {http://arxiv.org/abs/hep-ph/9404239}
  {arXiv:hep-ph/9404239 [hep-ph]} \BibitemShut {NoStop}%
\bibitem [{\citenamefont {Dobado}\ \emph {et~al.}(1997)\citenamefont {Dobado},
  \citenamefont {Pelaez},\ and\ \citenamefont {Urdiales}}]{Dobado:1997fv}%
  \BibitemOpen
  \bibfield  {author} {\bibinfo {author} {\bibfnamefont {A.}~\bibnamefont
  {Dobado}}, \bibinfo {author} {\bibfnamefont {J.~R.}\ \bibnamefont {Pelaez}},
  \ and\ \bibinfo {author} {\bibfnamefont {M.~T.}\ \bibnamefont {Urdiales}},\
  }\href {\doibase 10.1103/PhysRevD.56.7133} {\bibfield  {journal} {\bibinfo
  {journal} {Phys. Rev.}\ }\textbf {\bibinfo {volume} {D56}},\ \bibinfo {pages}
  {7133} (\bibinfo {year} {1997})},\ \Eprint
  {http://arxiv.org/abs/hep-ph/9702206} {arXiv:hep-ph/9702206 [hep-ph]}
  \BibitemShut {NoStop}%
\bibitem [{\citenamefont {He}\ \emph {et~al.}(1994)\citenamefont {He},
  \citenamefont {Kuang},\ and\ \citenamefont {Li}}]{He:1993qa}%
  \BibitemOpen
  \bibfield  {author} {\bibinfo {author} {\bibfnamefont {H.-J.}\ \bibnamefont
  {He}}, \bibinfo {author} {\bibfnamefont {Y.-P.}\ \bibnamefont {Kuang}}, \
  and\ \bibinfo {author} {\bibfnamefont {X.-y.}\ \bibnamefont {Li}},\ }\href
  {\doibase 10.1016/0370-2693(94)90772-2} {\bibfield  {journal} {\bibinfo
  {journal} {Phys. Lett.}\ }\textbf {\bibinfo {volume} {B329}},\ \bibinfo
  {pages} {278} (\bibinfo {year} {1994})},\ \Eprint
  {http://arxiv.org/abs/hep-ph/9403283} {arXiv:hep-ph/9403283 [hep-ph]}
  \BibitemShut {NoStop}%
\bibitem [{\citenamefont {Hahn}(2001)}]{Hahn:2000kx}%
  \BibitemOpen
  \bibfield  {author} {\bibinfo {author} {\bibfnamefont {T.}~\bibnamefont
  {Hahn}},\ }\href {\doibase 10.1016/S0010-4655(01)00290-9} {\bibfield
  {journal} {\bibinfo  {journal} {Comput. Phys. Commun.}\ }\textbf {\bibinfo
  {volume} {140}},\ \bibinfo {pages} {418} (\bibinfo {year} {2001})},\ \Eprint
  {http://arxiv.org/abs/hep-ph/0012260} {arXiv:hep-ph/0012260 [hep-ph]}
  \BibitemShut {NoStop}%
\bibitem [{\citenamefont {Hahn}\ and\ \citenamefont
  {Perez-Victoria}(1999)}]{Hahn:1998yk}%
  \BibitemOpen
  \bibfield  {author} {\bibinfo {author} {\bibfnamefont {T.}~\bibnamefont
  {Hahn}}\ and\ \bibinfo {author} {\bibfnamefont {M.}~\bibnamefont
  {Perez-Victoria}},\ }\href {\doibase 10.1016/S0010-4655(98)00173-8}
  {\bibfield  {journal} {\bibinfo  {journal} {Comput. Phys. Commun.}\ }\textbf
  {\bibinfo {volume} {118}},\ \bibinfo {pages} {153} (\bibinfo {year}
  {1999})},\ \Eprint {http://arxiv.org/abs/hep-ph/9807565}
  {arXiv:hep-ph/9807565 [hep-ph]} \BibitemShut {NoStop}%
\bibitem [{\citenamefont {Dawson}\ and\ \citenamefont
  {Valencia}(1990)}]{Dawson:1990cp}%
  \BibitemOpen
  \bibfield  {author} {\bibinfo {author} {\bibfnamefont {S.}~\bibnamefont
  {Dawson}}\ and\ \bibinfo {author} {\bibfnamefont {G.}~\bibnamefont
  {Valencia}},\ }\href {\doibase 10.1016/0370-2693(90)91324-5} {\bibfield
  {journal} {\bibinfo  {journal} {Phys. Lett.}\ }\textbf {\bibinfo {volume}
  {B246}},\ \bibinfo {pages} {156} (\bibinfo {year} {1990})}\BibitemShut
  {NoStop}%
\bibitem [{\citenamefont {Pelaez}(2015)}]{Pelaez:2015qba}%
  \BibitemOpen
  \bibfield  {author} {\bibinfo {author} {\bibfnamefont {J.~R.}\ \bibnamefont
  {Pelaez}},\ }\href@noop {} {\  (\bibinfo {year} {2015})},\ \Eprint
  {http://arxiv.org/abs/1510.00653} {arXiv:1510.00653 [hep-ph]} \BibitemShut
  {NoStop}%
\bibitem [{\citenamefont {Delgado}\ \emph
  {et~al.}(2015{\natexlab{b}})\citenamefont {Delgado}, \citenamefont {Dobado},\
  and\ \citenamefont {Llanes-Estrada}}]{Delgado:2015kxa}%
  \BibitemOpen
  \bibfield  {author} {\bibinfo {author} {\bibfnamefont {R.~L.}\ \bibnamefont
  {Delgado}}, \bibinfo {author} {\bibfnamefont {A.}~\bibnamefont {Dobado}}, \
  and\ \bibinfo {author} {\bibfnamefont {F.~J.}\ \bibnamefont
  {Llanes-Estrada}},\ }\href {\doibase 10.1103/PhysRevD.91.075017} {\bibfield
  {journal} {\bibinfo  {journal} {Phys. Rev.}\ }\textbf {\bibinfo {volume}
  {D91}},\ \bibinfo {pages} {075017} (\bibinfo {year} {2015}{\natexlab{b}})},\
  \Eprint {http://arxiv.org/abs/1502.04841} {arXiv:1502.04841 [hep-ph]}
  \BibitemShut {NoStop}%
\bibitem [{\citenamefont {Coleman}(1985)}]{Coleman:1985}%
  \BibitemOpen
  \bibfield  {author} {\bibinfo {author} {\bibfnamefont {S.}~\bibnamefont
  {Coleman}},\ }\href@noop {} {\emph {\bibinfo {title} {Dilatations}}},\
  \bibinfo {organization} {, in Aspects of Symmetry Symmetry: Selected Erice
  Lectures (pp. I-Vi). Cambridge University Press} (\bibinfo {year}
  {1985})\BibitemShut {NoStop}%
\bibitem [{\citenamefont {Halyo}(1993)}]{Halyo:1991pc}%
  \BibitemOpen
  \bibfield  {author} {\bibinfo {author} {\bibfnamefont {E.}~\bibnamefont
  {Halyo}},\ }\href {\doibase 10.1142/S0217732393000271} {\bibfield  {journal}
  {\bibinfo  {journal} {Mod. Phys. Lett.}\ }\textbf {\bibinfo {volume} {A8}},\
  \bibinfo {pages} {275} (\bibinfo {year} {1993})}\BibitemShut {NoStop}%
\bibitem [{\citenamefont {Goldberger}\ \emph {et~al.}(2008)\citenamefont
  {Goldberger}, \citenamefont {Grinstein},\ and\ \citenamefont
  {Skiba}}]{Goldberger:2008zz}%
  \BibitemOpen
  \bibfield  {author} {\bibinfo {author} {\bibfnamefont {W.~D.}\ \bibnamefont
  {Goldberger}}, \bibinfo {author} {\bibfnamefont {B.}~\bibnamefont
  {Grinstein}}, \ and\ \bibinfo {author} {\bibfnamefont {W.}~\bibnamefont
  {Skiba}},\ }\href {\doibase 10.1103/PhysRevLett.100.111802} {\bibfield
  {journal} {\bibinfo  {journal} {Phys. Rev. Lett.}\ }\textbf {\bibinfo
  {volume} {100}},\ \bibinfo {pages} {111802} (\bibinfo {year} {2008})},\
  \Eprint {http://arxiv.org/abs/0708.1463} {arXiv:0708.1463 [hep-ph]}
  \BibitemShut {NoStop}%
\bibitem [{\citenamefont {D'Ambrosio}\ and\ \citenamefont
  {Espriu}(2006)}]{DAmbrosio:2006xmn}%
  \BibitemOpen
  \bibfield  {author} {\bibinfo {author} {\bibfnamefont {G.}~\bibnamefont
  {D'Ambrosio}}\ and\ \bibinfo {author} {\bibfnamefont {D.}~\bibnamefont
  {Espriu}},\ }\href {\doibase 10.1016/j.physletb.2006.05.069} {\bibfield
  {journal} {\bibinfo  {journal} {Phys. Lett.}\ }\textbf {\bibinfo {volume}
  {B638}},\ \bibinfo {pages} {487} (\bibinfo {year} {2006})},\ \Eprint
  {http://arxiv.org/abs/hep-ph/0602008} {arXiv:hep-ph/0602008 [hep-ph]}
  \BibitemShut {NoStop}%
\bibitem [{\citenamefont {Froissart}(1961)}]{PhysRev.123.1053}%
  \BibitemOpen
  \bibfield  {author} {\bibinfo {author} {\bibfnamefont {M.}~\bibnamefont
  {Froissart}},\ }\href {\doibase 10.1103/PhysRev.123.1053} {\bibfield
  {journal} {\bibinfo  {journal} {Phys. Rev.}\ }\textbf {\bibinfo {volume}
  {123}},\ \bibinfo {pages} {1053} (\bibinfo {year} {1961})}\BibitemShut
  {NoStop}%
\bibitem [{\citenamefont {Ball}\ \emph {et~al.}(2013)\citenamefont {Ball},
  \citenamefont {Bertone}, \citenamefont {Carrazza}, \citenamefont
  {Del~Debbio}, \citenamefont {Forte}, \citenamefont {Guffanti}, \citenamefont
  {Hartland},\ and\ \citenamefont {Rojo}}]{Ball:2013hta}%
  \BibitemOpen
  \bibfield  {author} {\bibinfo {author} {\bibfnamefont {R.~D.}\ \bibnamefont
  {Ball}}, \bibinfo {author} {\bibfnamefont {V.}~\bibnamefont {Bertone}},
  \bibinfo {author} {\bibfnamefont {S.}~\bibnamefont {Carrazza}}, \bibinfo
  {author} {\bibfnamefont {L.}~\bibnamefont {Del~Debbio}}, \bibinfo {author}
  {\bibfnamefont {S.}~\bibnamefont {Forte}}, \bibinfo {author} {\bibfnamefont
  {A.}~\bibnamefont {Guffanti}}, \bibinfo {author} {\bibfnamefont {N.~P.}\
  \bibnamefont {Hartland}}, \ and\ \bibinfo {author} {\bibfnamefont
  {J.}~\bibnamefont {Rojo}} (\bibinfo {collaboration} {NNPDF}),\ }\href
  {\doibase 10.1016/j.nuclphysb.2013.10.010} {\bibfield  {journal} {\bibinfo
  {journal} {Nucl. Phys.}\ }\textbf {\bibinfo {volume} {B877}},\ \bibinfo
  {pages} {290} (\bibinfo {year} {2013})},\ \Eprint
  {http://arxiv.org/abs/1308.0598} {arXiv:1308.0598 [hep-ph]} \BibitemShut
  {NoStop}%
\bibitem [{\citenamefont {Barachetti}\ \emph {et~al.}(2016)\citenamefont
  {Barachetti}, \citenamefont {Rossi},\ and\ \citenamefont
  {Szeberenyi}}]{Barachetti:2120851}%
  \BibitemOpen
  \bibfield  {author} {\bibinfo {author} {\bibfnamefont {A.}~\bibnamefont
  {Barachetti}}, \bibinfo {author} {\bibfnamefont {L.}~\bibnamefont {Rossi}}, \
  and\ \bibinfo {author} {\bibfnamefont {A.}~\bibnamefont {Szeberenyi}},\
  }\href {http://cds.cern.ch/record/2120851} {\bibfield  {journal} {\bibinfo
  {journal} {CERN-ACC-2016-0007}\ } (\bibinfo {year} {2016})}\BibitemShut
  {NoStop}%
\bibitem [{\citenamefont {Khachatryan}\ \emph {et~al.}(2014)\citenamefont
  {Khachatryan} \emph {et~al.}}]{Khachatryan:2014hpa}%
  \BibitemOpen
  \bibfield  {author} {\bibinfo {author} {\bibfnamefont {V.}~\bibnamefont
  {Khachatryan}} \emph {et~al.} (\bibinfo {collaboration} {CMS}),\ }\href
  {\doibase 10.1007/JHEP08(2014)173} {\bibfield  {journal} {\bibinfo  {journal}
  {JHEP}\ }\textbf {\bibinfo {volume} {08}},\ \bibinfo {pages} {173} (\bibinfo
  {year} {2014})},\ \Eprint {http://arxiv.org/abs/1405.1994} {arXiv:1405.1994
  [hep-ex]} \BibitemShut {NoStop}%
\bibitem [{atl(2015)}]{atlas2015identification}%
  \BibitemOpen
  \href {https://cds.cern.ch/record/2041461} {\emph {\bibinfo {title}
  {{Identification of boosted, hadronically-decaying $W$ and $Z$ bosons in
  $\sqrt{s} = 13$ TeV Monte Carlo Simulations for ATLAS}}}},\ \bibinfo {type}
  {Tech. Rep.}\ \bibinfo {number} {ATL-PHYS-PUB-2015-033}\ (\bibinfo
  {institution} {CERN},\ \bibinfo {address} {Geneva},\ \bibinfo {year}
  {2015})\BibitemShut {NoStop}%
\bibitem [{\citenamefont {Aad}\ \emph {et~al.}(2015)\citenamefont {Aad} \emph
  {et~al.}}]{Aad:2015owa}%
  \BibitemOpen
  \bibfield  {author} {\bibinfo {author} {\bibfnamefont {G.}~\bibnamefont
  {Aad}} \emph {et~al.} (\bibinfo {collaboration} {ATLAS}),\ }\href {\doibase
  10.1007/JHEP12(2015)055} {\bibfield  {journal} {\bibinfo  {journal} {JHEP}\
  }\textbf {\bibinfo {volume} {12}},\ \bibinfo {pages} {055} (\bibinfo {year}
  {2015})},\ \Eprint {http://arxiv.org/abs/1506.00962} {arXiv:1506.00962
  [hep-ex]} \BibitemShut {NoStop}%
\bibitem [{\citenamefont {Heinrich}(2014)}]{heinrich2014reconstruction}%
  \BibitemOpen
  \bibfield  {author} {\bibinfo {author} {\bibfnamefont {J.~J.}\ \bibnamefont
  {Heinrich}},\ }\href {https://cds.cern.ch/record/1956424} {\enquote {\bibinfo
  {title} {{Reconstruction of boosted $W^{\pm}$ and $Z^{0}$ bosons from fat
  jets}},}\ } (\bibinfo {year} {2014}),\ \bibinfo {note}
  {{CERN-THESIS-2014-152. Thesis presented 12 Sep 2014}}\BibitemShut {NoStop}%
\bibitem [{\citenamefont {Alloul}\ \emph {et~al.}(2014)\citenamefont {Alloul},
  \citenamefont {Christensen}, \citenamefont {Degrande}, \citenamefont {Duhr},\
  and\ \citenamefont {Fuks}}]{Alloul:2013bka}%
  \BibitemOpen
  \bibfield  {author} {\bibinfo {author} {\bibfnamefont {A.}~\bibnamefont
  {Alloul}}, \bibinfo {author} {\bibfnamefont {N.~D.}\ \bibnamefont
  {Christensen}}, \bibinfo {author} {\bibfnamefont {C.}~\bibnamefont
  {Degrande}}, \bibinfo {author} {\bibfnamefont {C.}~\bibnamefont {Duhr}}, \
  and\ \bibinfo {author} {\bibfnamefont {B.}~\bibnamefont {Fuks}},\ }\href
  {\doibase 10.1016/j.cpc.2014.04.012} {\bibfield  {journal} {\bibinfo
  {journal} {Comput. Phys. Commun.}\ }\textbf {\bibinfo {volume} {185}},\
  \bibinfo {pages} {2250} (\bibinfo {year} {2014})},\ \Eprint
  {http://arxiv.org/abs/1310.1921} {arXiv:1310.1921 [hep-ph]} \BibitemShut
  {NoStop}%
\end{thebibliography}%


\end{document}